\PassOptionsToPackage{table}{xcolor}

\documentclass[11pt]{article}
\usepackage{jheppub}
\usepackage{bm}
\usepackage{physics}
\usepackage{slashed}
\usepackage{dsfont}
\usepackage{mathtools}
\usepackage{amsthm}
\usepackage{todonotes}
\usepackage{tikz,tikz-3dplot}
\usepackage{dsfont}
\usepackage{faktor}
\usepackage{pgfplots}
\pgfplotsset{compat=newest}
\usepackage{halloweenmath, mathrsfs}
\usepackage{multicol}
\usepackage{adjustbox}
\usepackage{soul}
\usepackage{subcaption}
\usepackage{centernot}

\pgfplotsset{compat=1.17}
\usetikzlibrary{shapes,arrows,cd,chains,decorations.markings,decorations.pathmorphing,calc}
\tikzset{
->-/.style args={#1rotate#2}{decoration={markings, mark=at position #1 with {\arrow[scale=3.5,rotate = #2 ]{stealth}}}, postaction={decorate}}
}
\definecolor{cof}{RGB}{219,144,71}
\definecolor{pur}{RGB}{186,146,162}
\definecolor{greeo}{RGB}{91,173,69}
\definecolor{greet}{RGB}{52,111,72}

\DeclareMathOperator{\bbZ}{\mathbb{Z}}
\DeclareMathOperator{\bbR}{\mathbb{R}}

\DeclareMathOperator{\calB}{\mathcal{B}}

\DeclareMathOperator{\calN}{\mathcal{N}}

\DeclareMathOperator{\calO}{\mathcal{O}}
\DeclareMathOperator{\calS}{\mathcal{S}}
\DeclareMathOperator{\calL}{\mathcal{L}}

\DeclareMathOperator{\calT}{\mathcal{T}}

\newcommand{\scarex}{``x^+"}
\newcommand{\scri}{\mathscr{I}}

\newmuskip\pFqskip
\mathchardef\pFcomma=\mathcode`, 

\theoremstyle{definition}

\usepackage{xcolor}
\usepackage{booktabs, array, makecell, multirow}

\usepackage{enumitem}

\setlist[itemize]{noitemsep}
\setlist[enumerate]{noitemsep}

\title{Do null defects dream of conformal symmetry?}
\abstract{We initiate the study of null line defects in Lorentzian conformal field theories in various dimensions. We show that null lines geometrically preserve a larger set of conformal isometries than their timelike and spacelike counterparts, explain a connection to non-relativistic systems, and constrain correlation functions using conformal Ward identities. We argue that having conformal symmetry, and especially maximal conformal symmetry, is extremely constraining -- nearly trivializing systems. We consider the (3+1)d scalar pinning field and null Wilson line examples in depth, compare their results to ultraboosted limits of timelike and spacelike systems, and argue that shockwave-type solutions are generic. A number of physical consistency conditions compel us to consider defect correlators as distributions on a restricted subspace of Schwartz test functions. Consequently, we provide a resolution to the longstanding problem of ultraboosted limits of gauge potentials in classical electromagnetism. We briefly analyze semi-infinite sources for the scalar in ($4-\epsilon$)-dimensions, consider solutions on the Lorentzian cylinder, and introduce the ``perfect null polygon'' which emerges for compatibility between Gauss' law and ultraboosted limits.}

\author[1]{Rajeev S. Erramilli,}
\author[2,3]{Justin Kulp,} 
\author[2,3]{and Fedor K. Popov}

\affiliation[1]{Institut des Hautes \'Etudes Scientifiques, 35 Route de Chartres, 91440 Bures-sur-Yvette, France }
\affiliation[2]{Simons Center for Geometry and Physics, Stony Brook University, Stony Brook, NY 11794, USA}
\affiliation[3]{Yang Institute for Theoretical Physics, Stony Brook University, Stony Brook, NY 11794, USA}

\begin{document}
\maketitle

\section{Introduction and Summary}
QFTs are largely characterized by their response to local deformations. Such deformations describe bounded observables and match neatly onto our physical intuition of measurements as coupling to external probe systems. For instance, adding an interaction term to a Lagrangian or Hamiltonian modifies the theory in all of spacetime, while adding interactions supported on loci of higher codimension describes defects and boundaries. The specific consideration of defects and impurities is fundamental to our understanding of symmetries \cite{Gaiotto:2014kfa}, the characterization of phase transitions \cite{Wilson:1974sk, Polyakov:1975rs, tHooft:1977nqb} and topological order \cite{Wen:1989iv}, and understanding dualities \cite{Kapustin:2005py}. Even more simply, massive point charges in the continuum and impurities in lattice models are described by defects at low-energies. 

Conformal Field Theories (CFTs) hold a preeminent position within the landscape of QFTs. The strong symmetry constraints on dynamics provide some of the rare non-perturbative, mathematically rigorous, and tractable examples of interacting QFTs which are nonetheless physically applicable and relevant: from strongly coupled gauge theories, to condensed matter systems near criticality, and even holographic descriptions of gravity. Just as CFTs provide an effective description for QFTs near the endpoint of RG flows, defect CFTs describe endpoints of RG flows in systems with impurities \cite{Cardy:1984bb, McAvity:1995zd, Billo:2016cpy} (see also \cite{Cuomo:2021rkm} and references within).

While QFTs and CFTs exist natively in both Lorentzian and Euclidean spacetimes, traditionally we Wick rotate Lorentzian theories to Euclidean signature to simplify important computations and conceptual concerns. This Euclidean perspective allows us to extract useful information about ground state properties, correlation functions, and static features of the Lorentzian theory. Indeed, general theorems guarantee the equivalence of Euclidean QFT to Lorentzian QFTs satisfying the Wightman axioms \cite{Osterwalder:1973dx, Osterwalder:1974tc, Glaser:1974hy} (see also \cite{Kravchuk:2020scc, Kravchuk:2021kwe, Dedushenko:2022zwd}). Therefore, even in the study of defect CFTs, a common strategy is to Wick rotate to Euclidean signature and use various techniques -- such as radial quantization -- to study the defects and their modification of the space of states \cite{Agmon:2020pde}. However, as we elaborate on shortly, there are various subtle features of Lorentzian theories that warrant holding on to the indefinite signature.

Although Lorentzian and Euclidean descriptions of QFTs are formally equivalent, the reconstruction of physically interesting real-time observables from Euclidean correlation functions is not guaranteed to be \textit{easy}. In particular, one of the most important differences between Lorentzian and Euclidean quantum field theories is the intrinsic causal structure of Lorentzian spacetimes and, relatedly, the ability for distinct points to be null-separated. Consequently, Lorentzian subtleties of OPE convergence regions, singularity structure, and the distributional behaviour of correlation functions are obscured in Euclidean signature \cite{Luscher:1974ez, Mack:1976pa, Kravchuk:2018htv, Qiao:2020bcs, Kravchuk:2020scc, Kravchuk:2021kwe}. Even if one is not interested in such Lorentzian observables directly, analytical tools such as the Lorentzian inversion formula for CFTs \cite{Caron-Huot:2017vep, Simmons-Duffin:2017nub} have shown that even Euclidean data for a theory can be better and more powerfully understood in Lorentzian spacetime. In other words, Wick rotation is not a one-way street: given the powerful assumption of analyticity, the Lorentzian perspective can bear fruit even for Euclidean physics.

Focusing on line defects, in Lorentzian signature, a line defect can be locally timelike, spacelike, or null. The last option, the null line defect, is intrinsically Lorentzian; in Euclidean signature, such an object would correspond to something degenerate at the origin with no meaningful extension. In Lorentzian signature, a null line defect is a fundamentally new type of object, interacting in subtle ways with the surrounding field theory. On physical grounds, we expect null-defects to be useful for describing impurities moving at the speed of sound in a material or charged particles moving at the speed of light. Although there are no fundamental massless charged particles in nature, the ability to approximate heavily boosted massive charged particles by null trajectories has played an important role in the study of cusp anomalous dimensions, related Bremsstrahlung and scattering problems in QCD and $\calN=4$ SYM, as well as factorization and the study of soft radiation \cite{Collins:1985ue, Korchemsky:1985xj, Korchemsky:1987wg, Korchemskaya:1992je, Alday:2007hr, Alday:2010ku, Correa:2012nk, Alday:2012hy, Henn:2013wfa, Fiol:2015spa} (see also \cite{collins2004factorizationhardprocessesqcd, Alday:2008yw, Mertens:2014yzd, Becher:2014oda} for useful reviews).

A useful idealization of the aforementioned finite deformations and/or bounded observables is provided by infinitesimal deformations. In CFTs -- or theories described by CFTs in the deep UV -- infinitesimal point-like deformations are precisely the ``local operators'' of the theory. Consequently, some of the most important questions in the study of CFTs (and QFT more generally) concern local operators. In Euclidean CFTs, many questions about the behaviour of a theory are addressed through the state-operator correspondence and the existence of a convergent OPE: any field configuration can, in principle, be arbitrarily well approximated by a sum of local operators acting on the vacuum. This also applies to defect systems, allowing us to understand defects as states and classify defect changing operators, e.g. Ishibashi states \cite{Cardy:1989ir, Cardy:2004hm}.

A similar object of increasing interest in recent years is the manifestly Lorentzian light ray operator and related ``detector'' operators, representing idealized and formalized calorimeters \cite{Hofman:2008ar, Caron-Huot:2022eqs} (for some historical references see \cite{Gross:1971wn, Anikin:1978tj, Balitsky:1987bk, Muller:1994ses} and \cite{Moult:2025nhu} for a modern review). In CFTs, light ray operators were realized from first principles in \cite{Kravchuk:2018htv} and have played an important role in understanding the Lorentzian inversion formula \cite{Caron-Huot:2017vep}, given new proofs of RG monotonicity theorems \cite{Hartman:2024xkw}, and have given a glimpse into the much larger landscape of observables on the null plane. At the simplest level, light ray operators are constructed via conformally invariant integrals of local operators along a null line. Thus, in the language above, they describe intrinsically Lorentzian extended infinitesimal deformations of a theory. Since infinitesimal deformations are idealizations of bounded observables, if light ray operators are ``physical,'' then we expect them to be linear approximations to a finite deformation -- a null line defect. Said the other way, we expect some subset of null conformal line defects to be ``exponentials of light ray operators.'' This analogy is stressed from both sides, e.g. by the subtle properties of light ray operators and their OPEs, questions of marginality and relevance of local operators, and even sensible notions of locality on null defects. Nevertheless, viewing null line defects as light-ray pinning fields opens a window into the interplay between locality, causality, and conformal symmetry in real time.\footnote{Twisting supersymmetric theories in Lorentzian signature also naturally leads to considering smeared local operators on lightlike trajectories by supersymmetric descent \cite{Argyres:2022npi}.}

\subsection{Outline, Summary, and Future Directions}
Despite the wide-ranging and deep results involving null line defects mentioned above, and interest in defect systems and Lorentzian CFTs more generally, a study of abstract null line defects has gone largely uninitiated to date. In this work we initiate some of the elementary steps in this program, using very simple examples and kinematic constraints to drive our investigations:
\begin{enumerate}
    \item[{\hyperref[sec:Kinematics]{\underline{$\S$.2.}}}] In Section \ref{sec:Kinematics} we discuss basic kinematic facts about null lines in Minkowski space. In particular, in Sections \ref{sec:NullLine} and \ref{sec:NullDefectAlgebra} we study the subalgebra of the Minkowski conformal algebra $\mathfrak{so}(2,d)$ preserving a geometric null line. Surprisingly, the geometric or ``maximal conformal symmetries'' are far larger than one might expect by analogy to timelike or spacelike lines, spanning the ``null defect algebra'':
    \begin{equation}
        \mathfrak{n}_d := (\mathfrak{sl}(2,\bbR) \times \mathbb{R} \times \mathfrak{so}(d-2)) \ltimes \mathfrak{h}_{d-2}\,,
    \end{equation}
    where $\mathfrak{h}_{d-2}$ is a ($d-2$)-dimensional Heisenberg algebra. In Section \ref{sec:Schrodinger} we comment on the relationship to the Schr\"odinger algebra and Lifshitz systems in one fewer dimension. This can be considered a conformal analogue of the famous appearance of Galilean symmetry in DLCQ and ultraboosted limits/lightcone quantization \cite{Weinberg:1966jm, Susskind:1967rg, Kogut:1969xa}. Finally, in Section \ref{sec:CSB} we discuss very briefly the conformal symmetries of a geometric codimension-1 null plane, as many fast moving particles actually single out an entire shockwave plane, playing an important role in the analysis of solutions.
    \vskip 0.25cm
    
    \item[{\hyperref[sec:GeneralConsiderations]{\underline{$\S$.3.}}}] The bulk of our results are presented in Section \ref{sec:GeneralConsiderations}. In order to understand null conformal line defects and their relation to the conformal symmetry algebra $\mathfrak{n}_d$, we start by considering the symmetries preserved by pinning field deformations in the UV in Section \ref{sec:PinningSymmetries}. Intriguingly, only the null Wilson line can preserve the full $\mathfrak{n}_d$ symmetry in the UV. In Section \ref{sec:WITariffs} we solve Ward identities for different subalgebras of $\mathfrak{n}_d$ to understand the space of allowed one-point functions in different scenarios. We find that null-conformal symmetry, and especially maximal symmetry $\mathfrak{n}_d$, is extremely kinematically constraining: \textit{in the maximally symmetric case, one-point functions must vanish and two-point functions are characterized by discontinuities along shockwave planes}.\\
    \vskip -0.25cm
    After these general results, we take an essential detour in Section \ref{sec:ExampleFreeScalar} to the (3+1)d free scalar, arriving at the causal and shockwave solutions:
    \begin{equation}
    \phi_{c}(x^+,x^-, x^\perp) 
        =  -\frac{h}{4\pi x^+} \,  \Theta\left(\scarex\right)\,,\quad
    \phi_{s}(x^+,x^-, x^\perp) = \frac{h}{4\pi} \delta(x^+) \log \mu^2 x_\perp^2\,.
    \end{equation}
    Besides foreshadowing the example of a massless charged particle, this example demonstrates the need to consider correlation functions as distributions on a restricted space of test functions $\calS_{\rm null}(\bbR^d) \subset \calS(\bbR^d)$ when null defects are inserted (a modification which is not necessary for timelike line defects), and revealing some of the deeper analytic structure of QFT correlation functions. This technical-seeming caveat turns out to have important physical consequences: preventing divergences, preventing arbitrary scales from emerging, and permitting the free scalar to possess all expected symmetries. We also discuss the importance of adiabatically turning on defects. In Section \ref{sec:limitsOfDefects} we study limits of timelike and spacelike defects, showing perfect agreement with our previous results after careful distributional limits, and in Section \ref{sec:shockwaves} we explain the generality of shockwave solutions and the appearance of a lightlike Kubo formula.\\
    \vskip -0.25cm
    Lastly, in Section \ref{sec:NullWilsonLine}, we consider the example of the Wilson line in pure Maxwell theory. The field strength one-point function turns out to be
    \begin{equation}\label{eq:introShock}
        F_{+i} = \frac{g}{2\pi} \frac{x_i}{|x_\perp|^2} \delta(x^+)\,,
    \end{equation}
    for \textit{both} causal and Feynman propagators. It is compatible with the aforementioned ultrarelativistic limits and Ward identities, but does not preserve maximal $\mathfrak{n}_d$ symmetry, but leads to the maximally symmetric ``perfect null polygon'' configuration instead. Moreover, by working in $\calS_{\rm null}(\bbR^4)$, we provide a resolution to a longstanding issue about the ultraboosted limits of Li\'enard-Wiechart gauge potentials in classical electromagnetism.
    \vskip 0.25cm
    
    \item[{\hyperref[sec:OtherResults]{\underline{$\S$.4.}}}]  In Section \ref{sec:OtherResults} we consider some additional results for null line defects. In Section \ref{sec:FiniteDefects} we briefly study symmetries of semi-infinite lines and use the freedom to study an interacting pinning field in $(4-\epsilon)$-dimensions. We find that the resulting correlation functions only see the defect creating operator, agreeing with the ultraboosted limit of a timelike defect. In Section \ref{sec:LorCyl} we turn to the Lorentzian cylinder. We solve for the scalar pinning field in timelike and lightlike scenarios and comment on the importance of adiabatically turning on the defect. In Section \ref{sec:ZZCC}, we again consider null Wilson lines. On the cylinder, we are forced to add additional charges for compatibility with Gauss' law. We contrast the case of two infinitely long null rays on the cylinder to the ultraboosted limit of two timelike lines and find the configurations are quite different. The ultraboosted limit leads to a ``perfect null polygon,'' a pair creation/annihilation process around the entire Poincar\'e patch of spacetime. This perfect null polygon preserves a larger set of symmetries than two infinitely long null rays and correctly matches \eqref{eq:introShock}.
\vskip 0.25cm

    \item[{\hyperref[sec:2dPaper]{\underline{$\S$.5.}}}] Finally, we end the paper in Section \ref{sec:2dPaper} by considering abstract maximally symmetric defects in (1+1)d. In Section \ref{sec:DefectPris} we discuss the (projective) unitary irreducible representations of $\mathfrak{n}_2$, potential positive energy conditions, and the branching of bulk lowest-weight representations to $\mathfrak{n}_2$. This gives us putative descriptions of potential defect local operators and a defect Hilbert space. Then, in Section \ref{eq:BulkWardIdentity2d} we write the Ward identities for bulk local operators, finding similar results to the higher dimensional cases, and speculate briefly on bulk-defect correlation functions.
\vskip 0.25cm
    
    \item[{\hyperref[app:Conventions]{\underline{$\mathscr{A}$.}}}] In Appendix \ref{app:Conventions} we present our conventions for the conformal algebra and lightcone coordinates throughout the paper.
\vskip 0.25cm

    \item[{\hyperref[app:Calculations]{\underline{$\mathscr{B}$.}}}] In Appendix \ref{app:Calculations} we provide more details for the calculations in Section \ref{sec:GeneralConsiderations}. In the process, we also explain the appearance of $\scarex$.
\end{enumerate}

\paragraph{Future Directions.} Despite learning a\href{https://www.youtube.com/watch?v=J6VjPM5CeWs}{\phantom{a}}number of things in our study of null line defects in CFTs, our results barely scratch the surface of this burgeoning subject. The following are some considerations that motivated us, as well as new questions from our studies:
\begin{itemize}
    \item \textbf{Defect Operators, OPE, and Hilbert Space.} Defects should support their own sets of local operators with defect OPEs and so on. While it is possible to classify various unitary irreducible representations of defect algebras (we do not record this here, except in the (1+1)d case in Section \ref{sec:2dPaper}), the meaning of a Hilbert space of defect local operators and their sufficiency in describing defect dynamics is not clear. For example, it is not clear how ``rigid'' a null line defect should be, i.e. whether it should admit a full-rank displacement operator. Likewise, the notion of OPE becomes quite difficult to define on null surfaces, as there is no obvious distance between any two points. See also the start of \ref{sec:ScalarPrimary}.
    
    \item \textbf{Non-Relativistic Systems and Lightfront Quantization.} One hope to answering the previous question could come via the connection to non-relativistic systems. As described in Section \ref{sec:Schrodinger}, there is a well-known relationship between non-relativistic systems, null-reduction, and lightfront quantization. However, many non-perturbative subtleties have plagued this relationship for decades. Given that our systems give a ``conformal enhancement'' of this story, it is plausible that the issues of null-reduction and lightfront quantization are resolvable in these conformal settings (see \cite{Fitzpatrick:2018ttk}), especially since we have good non-perturbative control over both CFTs and non-relativistic CFTs. Therefore, it is plausible that results from non-relativistic CFT (understanding of local operators, Hilbert space, defects etc.) can be lifted back to relativistic systems.

    \item \textbf{Holography of Various Types.} Much of our understanding of null line defects comes from the study of cusped Wilson lines in $\calN=4$ SYM. We might wonder about holographic dualities for our defect CFTs in more general situations. At least one kinematic insight comes from non-relativistic systems, where it is known that $pp$-wave deformations of $\mathrm{AdS}_{d+1}$ geometries have the same symmetries as a scalar pinning field. 
    
    Distinct from this, there is also the question of the relationship to flat space (or ``Carrollian'') holography, where scattering in asymptotically flat spacetimes is putatively described by ``CFTs'' at null infinity $\scri^{\pm}$. Our systems are directly related to Carrollian CFTs, indeed, a conformal map brings the $x^+ = 0$ plane to $\scri^{\pm}$. More generally, any understanding of locality or OPEs would be highly relevant to both problems.
    
    \item \textbf{Connection to SCET and Collider Physics.} Part of the motivation for studying light ray operators and null Wilson lines comes from collider physics. In Soft Collinear Effective Theory (SCET), one considers RG flows associated with chiral dilatations $J_0$ as opposed to usual dilatations $D$. Maximally symmetric null defects are in part distinguished by the preservation of both $J_0$ and $\bar{J}_0$ (equivalently, $D$ and $M_{+-}$). It is curious that this is both what we would expect from a conformal fixed point as well as a SCET fixed point -- making them special fixed points for both communities. It would be interesting to better understand the connection.
    
    \item \textbf{Existence of Maximally Symmetric Defects.} In Section \ref{sec:Tariff} we show, purely from Ward identity considerations, that maximally symmetric null defects do not change the kinematics of bulk correlation functions at all in $d>2$ dimensions except in discontinuities across shockwave planes. In non-CFTs, such discontinuities can be understood in perturbation theory -- it would be curious to know if they can be computed in CFTs or removed entirely. In (1+1)d, maximally symmetric defects describe lightlike conformal interfaces preserving $5$ out of $6$ generators of the conformal algebra. It would be interesting to know if we can constrain the space of maximally symmetric defects by considering, e.g. the analogue of Cardy conditions.
\end{itemize}

\acknowledgments We would like to thank Jacob Abajian, Omar Abdelghani, Jan Albert, Philip Argyres, Gabriel Cuomo, Jaume Gomis, Petr Kravchuk, Zohar Komargodski, Yue-Zhou Li, Juan Maldacena, Thibaud Raymond, Amit Sever, Adar Sharon, Philine van Vliet, Yifan Wang, and Mitch Weaver for helpful discussions. We also thank Zohar Komargodski and Adar Sharon for comments on the draft. JK thanks Sabrina Pasterski for early suggestions to study null conformal defects. The work of JK is supported by the NSERC PDF program.

\section{Kinematics for Null Lines}\label{sec:Kinematics}
In this section we briefly discuss kinematic aspects of our setup: the symmetries that preserve a null line and the connections to Schr\"odinger symmetry. Our conventions for conformal transformations and lightcone coordinates are described in Appendix \ref{app:Conventions}.

\subsection{Conformal Symmetries Preserving a Null Line}\label{sec:NullLine}
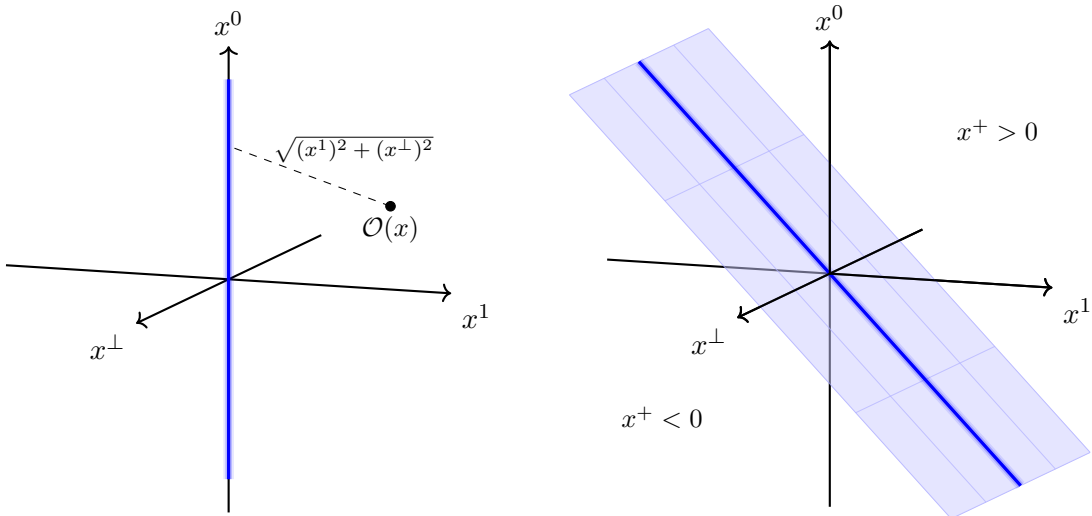
\begin{figure*}
	\begin{minipage}{0.49\textwidth}
        \centering
\tdplotsetmaincoords{80}{110}
\begin{tikzpicture}[tdplot_main_coords, scale=1.8]

  \draw[thick,->] (-2.0,0,0) -- (2.0,0,0) node[anchor=north east]{$x^\perp$};
  \draw[thick,->] (0,-1.75,0) -- (0,1.75,0) node[anchor=north west]{$x^1$};
  \draw[thick,->] (0,0,-1.75) -- (0,0,1.75) node[anchor=south]{$x^0$};

  \draw[line width=4pt, blue, opacity=0.1]
    (0,0,-1.5) -- (0,0,1.5);
  \draw[line width=2.5pt, blue, opacity=0.2]
    (0,0,-1.5) -- (0,0,1.5);
  \draw[line width=1pt, blue, opacity=0.5]
    (0,0,-1.5) -- (0,0,1.5);
  \draw[very thick, blue]
    (0,0,-1.5) -- (0,0,1.5);

  \coordinate (O) at (2.0,2.0,1.0); 
  \filldraw[black] (O) circle (1pt) node[below] {\small$\mathcal{O}(x)$};

  \coordinate (P) at (0,0,1.0);

  \draw[dashed] (O) -- (P);

  \node at (0.6,1.2,1.15) [] {\scriptsize$\sqrt{(x^1)^2 + (x^\perp)^2}$};

\end{tikzpicture}
	\end{minipage}
    \begin{minipage}{0.49\textwidth}
    \centering
    \tdplotsetmaincoords{80}{110} 
\begin{tikzpicture}[tdplot_main_coords, scale=1.8]

  \draw[thick,->] (-2.0,0,0) -- (2.0,0,0) node[anchor=north east]{$x^\perp$};
  \draw[thick,->] (0,-1.75,0) -- (0,1.75,0) node[anchor=north west]{$x^1$};
  \draw[thick,->] (0,0,-1.75) -- (0,0,1.75) node[anchor=south]{$x^0$};

  \def\L{1.5}

  \foreach \x in {-1.5, -0.75, 0, 0.75, 1.5} {
    \draw[blue!40, opacity=0.7]
      (\x,-1.5,1.5) -- (\x,1.5,-1.5);
  }
  \foreach \y in {-1.5, -0.75, 0, 0.75, 1.5} {
    \draw[blue!40, opacity=0.7]
      (-\L,\y,-\y) -- (\L,\y,-\y);
  }
  \fill[blue!20, opacity=0.5] 
    (-\L,-1.5,1.5) -- (\L,-1.5,1.5) -- (\L,1.5,-1.5) -- (-\L,1.5,-1.5) -- cycle;

\draw[line width=4pt, blue, opacity=0.1]
  (0,-1.5,1.5) -- (0,1.5,-1.5);
\draw[line width=2.5pt, blue, opacity=0.2]
  (0,-1.5,1.5) -- (0,1.5,-1.5);
\draw[line width=1pt, blue, opacity=0.5]
  (0,-1.5,1.5) -- (0,1.5,-1.5);
\draw[very thick, blue]
  (0,-1.5,1.5) -- (0,1.5,-1.5);

  \node at (0.5,1.5,1.25) {\small $x^+ > 0$};
  \node at (-0.5,-1.5,-1.25) {\small $x^+ < 0$};

  \draw[thick,->] (-2.0,0,0) -- (2.0,0,0);   
  \draw[thick,->] (0,0,0) -- (0,1.75,0);    
  \draw[thick,->] (0,0,0) -- (0,0,1.75);    

\end{tikzpicture}

    \end{minipage}
	\caption{Left, a timelike line  defect running up the $x^0$ axis. Correlation functions, such as one-point functions, depend on the distance of local operators to the line. Right, a null line defect along $x^+ = 0$ and $x^\perp = 0$. Geometrically, the line preserves far more symmetries than a timelike line. The defect splits space into two halves before ($x^+ < 0$) and after ($x^+>0$) the defect. Physics often depends on the distance to the shockwave plane $x^+ = 0$ or, conversely, is often confined to the shockwave plane.}
	\label{fig:nullSymmetries}
\end{figure*}

With our conventions in place, we can work out the subgroup of the conformal algebra preserving a null line. For the remainder of this paper, we will consider the case of a null line living along
\begin{equation}
    x^+ = 0 \quad \text{and} \quad x^\perp = 0\,.
\end{equation}
This is a null line running from the bottom right to top left in the light plane; see Figure \ref{fig:nullSymmetries}.

To compute the subgroup of the conformal group preserving this null line, we use our differential representations of the conformal algebra $\mathscr{D}_Q$. We divide them into three families:
\begin{enumerate}
    \item \textbf{Broken Transformations.} These are generators explicitly broken by the null line, of which there are $2d-3$:
    \begin{equation}
        P_+\,,\quad P_i\,,\quad M_{+i}\,.
    \end{equation}
    \item \textbf{Transverse Transformations.} These are generators which act trivially on the line (they leave $x^-$ fixed), of which there are $d(d-1)/2 +1$:
    \begin{equation}
        K_-\,,\quad K_i\,,\quad M_{-i}\,,\quad M_{ij}\,,\quad \bar{J}_0\,.
    \end{equation}
    \item \textbf{Preserved Transformations.} These are generators which act nontrivially on the line (they map it to itself but transform $x^-$), of which there are 3:
    \begin{equation}
        P_-\,,\quad K_+\,,\quad J_0\,. 
    \end{equation}
\end{enumerate}
It's useful to introduce the basis of generators
\begin{alignat}{5}
    J_{-1} &:= P_-\,,\quad
    && J_0 &&:= -\frac{1}{2}(D-\frac{1}{F^2}M_{+-})\,,\quad
    && J_{1} &&:= \frac{1}{2F^2}K_{+}\,,\label{eq:SL2Rm}\\
    \bar{J}_{-1} &:= P_+\,, \quad
    && \bar{J}_0 &&:= -\frac{1}{2}(D+\frac{1}{F^2}M_{+-})\,,\quad
    && \bar{J}_{1} &&:= \frac{1}{2F^2}K_{-}\,,\label{eq:SL2Rp}
\end{alignat}
which span the 2d conformal algebra acting on the light plane, $x^\perp = 0$. This basis will play an important role in the subsequent analysis as it manifests the following subalgebra:
\begin{equation}
    \mathfrak{sl}(2,\bbR)_- \times \mathfrak{sl}(2,\bbR)_+ \subset \mathfrak{so}(2,d)\,.
\end{equation}
$\bar{J}_0$ is called the ``collinear twist.'' The constant $F$ appears from our lightcone conventions, which are left arbitrary to ease the readers comparison to other references, see our conventions in Appendix \ref{app:Conventions}.

For a timelike or spacelike line, the line is preserved by the 1d conformal symmetry algebra $\mathfrak{g}_p = \mathfrak{sl}(2,\bbR)$ along the defect and the transverse symmetry algebras are $\mathfrak{g}_t = \mathfrak{so}(d-1)$ or $\mathfrak{g}_t = \mathfrak{so}(1,d-2)$ respectively. Critically, the two algebras commute so that the algebra of conformal symmetries preserving the line is simply the product $\mathfrak{g}_p \times \mathfrak{g}_t$. As a result, defect local operators can be understood as 1d CFT primaries $\hat{\calO}_{h,I}(x)$ with some scaling dimension $h$; transverse rotations amount to a flavour symmetry for the defect local operators.

In the case of a lightlike defect, the story is different. To start with, there are $d$ more generators $K_-$, $K_i$, and $\bar{J}_0$; we will provide some intuition for this in Section \ref{sec:extraGeneratorIntuition}. Also, the $\mathfrak{sl}(2,\bbR)_-$ and transverse algebra do not commute, leading to a richer structure for defect local primaries, which we will discuss in Section \ref{sec:NullDefectAlgebra}. We will call this maximal symmetry algebra the \textit{null-defect algebra}. As we will see in the next section, it takes the form:
\begin{equation}
    \mathfrak{n}_d := (\mathfrak{sl}(2,\bbR) \times \mathbb{R} \times \mathfrak{so}(d-2)) \ltimes \mathfrak{h}_{d-2}\,,
\end{equation}
where $\mathfrak{h}_{d-2}$ is a a ($d-2$)-dimensional Heisenberg algebra, we elaborate on this in more detail in Section \ref{sec:NullDefectAlgebra}.\footnote{We use notation of product and semi-direct product of groups, rather than sums, for our Lie algebras.}

So far, we consider no dynamics: this is purely a kinematical statement about the maximal subalgebra of the conformal algebra preserving a null line. When turning on a particular deformation along a line, e.g. by considering
\begin{equation}
    S_{\mathrm{CFT}} + \int dx^- \calO(0,x^-,0)\,,
\end{equation}
we will break some collection of the ``transverse'' and ``preserved'' transformations. A priori, we expect to flow to a conformal fixed point in the IR so that the full $\mathfrak{sl}(2,\bbR)_-$ emerges in the IR essentially by definition. However, there is no guarantee that the full null-defect algebra will actually emerge in the IR, or that the defect does not simply flow to something trivial. In the subsequent sections we explore the rigid constraints from such symmetries which -- in some cases -- almost entirely disallows RG flows to non-trivial defect CFTs or to defect CFTs with one-point functions.

\subsubsection{Some Intuition for Conformal Symmetries on the Lightcone}\label{sec:extraGeneratorIntuition}
The purpose of this subsection is to provide some more intuition for the finite conformal transformations preserving the null line. This will make the existence of the $d$ additional symmetries more clear.

\paragraph{Lightcone Boosts.} The differential operator generating the lightcone boosts $M_{\pm i}$ is
\begin{equation}\label{eq:lightconeBoost}
    \mathscr{M}_{\pm i} 
        = -i (x_{\pm} \partial_i - x_i \partial_{\pm})
        = i (F^2 x^{\mp} \partial_i + x^i \partial_{\pm})\,.
\end{equation}
Since the metric is anti-diagonal, the transformation shifts $\pm$ proportional to $x^i$ (as expected), but shifts the transverse coordinate $x^i$ proportional to the $x^{\mp}$ value and leaves $x^\mp$ unchanged. A finite transformation acts by
\begin{equation}
     \exp (\lambda M_{\pm i}): \begin{pmatrix}x^\pm\\x^\mp\\ x^i\end{pmatrix} \mapsto \begin{pmatrix}x^\pm + \lambda x^i + \frac{1}{2}\lambda^2 x^\mp\\ x^\mp\\x^i + \lambda x^\mp\end{pmatrix}\,.
\end{equation}
As a result, the boosts $M_{+i}$ are broken while the boosts $M_{-i}$ are transverse i.e. trivial on the null line, like the spatial rotations $M_{ij}$.

Also due to the anti-diagonal metric, we have $[M_{-i},M_{-j}] = 0$. Meanwhile the generators $M_{-i}$ transform as a vector of the transverse rotation algebra $\mathfrak{so}(d-2)$, so that
\begin{equation}\label{eq:isod2M}
    \langle M_{ij}, M_{-i} \rangle \cong \mathfrak{so}(d-2) \ltimes \bbR^{d-2} = \mathfrak{iso}(d-2)\,.
\end{equation}
This is compatible with the intuition from the timelike/spacelike defect lines, where the transverse rotations/boosts form an $\mathfrak{so}(d-1)$ or $\mathfrak{so}(1,d-2)$ algebra respectively. The two are related by Wick rotation, and the null case can be seen as an In\"on\"u-Wigner contraction and/or degenerate limit of the others. 

\paragraph{Special Conformal Transformations.} Next, we have the Special Conformal Transformations (SCTs), which are newly preserved in the null case. The finite action takes the form:
\begin{equation}
    y\cdot K: x^\mu \mapsto \frac{x^\mu-y^\mu x^2}{1-2(y\cdot x) + y^2 x^2}\,.
\end{equation}
This action is independent of metric. In Euclidean signature, SCTs fix the origin and move the point at infinity; this can be seen immediately by replacing $x^2=x^\mu=0$. In Lorentzian signature, $x^2=0 \centernot{\implies} x^\mu=0$, so the action of the SCT simplifies not just for the origin but for the future and past \emph{lightcones} of the origin as well. Namely, SCTs map the null cone to itself up to a local rescaling:
\begin{equation}\label{eq:SCTOnLightcone}
    y\cdot K: x^\mu \mapsto \frac{1}{1-2(y\cdot x)}x^\mu\,,\qquad \text{$x$ on lightcone}\,.
\end{equation}
This is the Lorentzian signature avatar of the fixed origin in Euclidean signature.

We can get another intuition for why SCTs map the null cone to itself by remembering the familiar fact that with inversion,
\begin{equation}
    I: x^\mu\mapsto \frac{x^\mu}{x^2}\,,
\end{equation}
SCTs are conjugate to ordinary translations. In Euclidean signature, inversion exchanges $0$ and infinity in Euclidean signature. Ordinary translations keep the point at infinity fixed but move the origin; conjugating by inversion, we see that SCTs move infinity and fix the origin, as we stated previously. In Lorentzian signature, inversion exchanges the null cone with timelike, spacelike, and null infinities (see also Section \ref{sec:LorCyl}). For a point on the null plane $(x^+, x^-, x^i = 0)$,
\begin{equation}
    I: x^{\pm} \mapsto \frac{x^\pm}{-x^\pm x^\mp} = -\frac{1}{x^\mp}\,,\qquad \text{for $x$ on the null plane}\,.
\end{equation}
When one of the coordinates goes to $0$, the conjugate coordinate diverges under inversion. The remaining finite coordinate can be interpreted as the position on null infinity $\mathscr{I}$; this position can be shifted by ordinary translations. (Note that the point at spacelike/timelike infinity is still fixed under translations, analogous to the Euclidean case.) Under inversion, this translation along $\mathscr{I}$ becomes nontrivial SCTs along a given direction on the null cone while still fixing the origin. In the same way that translations preserve null infinity, SCTs preserve the null cone and indeed the null defect line. More precisely, all $K_i$ and $K_-$ act trivially, while $K_+$ acts non-trivially.\footnote{The reader should note the opposite index $\pm$ in relation to translations.} (In the timelike/spacelike defect case, all $K_\mu$ are broken except the one directly along the defect line.)

As before, we note that the $K_i$ form a commuting $\mathfrak{so}(d-2)$ vector, so that
\begin{equation}\label{eq:isod2K}
    \langle M_{ij}, K_i \rangle \cong \mathfrak{iso}(d-2)\,.
\end{equation}

\paragraph{Chiral Light Plane Dilatations.} A distinguished role is played by the generators $J_0$ and $\bar{J}_0$. As in
\eqref{eq:lightconeBoost}, the anti-diagonal metric makes
\begin{equation}
    \mathscr{M}_{+-}= -i(x_+ \partial_- - x_- \partial_+) = iF^2 (x^- \partial_- - x^+ \partial_+)\,,
\end{equation}
and combined with the dilatation generator
\begin{equation}
    \mathscr{D}= -i(x^- \partial_- + x^+ \partial_+)\,,
\end{equation}
it is clear that the null-defect preserves the chiral $J_0$ and anti-chiral $\bar{J}_0$ dilatations. Importantly, their finite transformations are
\begin{align}
    J_0: (x^+, x^-, x^\perp)
        &\mapsto (x^+, \lambda^2 x^-, \lambda x^\perp)\,,\label{eq:J0Boost}\\
    \bar{J}_0: (x^+, x^-, x^\perp)
        &\mapsto (\lambda^2 x^+, x^-, \lambda x^\perp)\,.\label{eq:J0BarBoost}
\end{align}
That the chiral dilatations scale one coordinate with power $\lambda^2$ and the transverse coordinates with power $\lambda$ foreshadows the connection to Schr\"odinger-symmetric systems in Section \ref{sec:Schrodinger}.

\subsection{The Conformal Null-Defect Algebra}\label{sec:NullDefectAlgebra}
In this section, we will show that the maximal conformal symmetry algebra of the null line is the conformal null-defect algebra
\begin{equation}
    \mathfrak{n}_d := (\mathfrak{sl}(2,\bbR) \times \mathbb{R} \times \mathfrak{so}(d-2)) \ltimes \mathfrak{h}_{d-2}\,.
\end{equation}

The group of 1d conformal transformations which act non-trivially along the line is $\mathfrak{g}_p := \mathfrak{sl}(2,\bbR)_-$, defined in \eqref{eq:SL2Rm}. There is also the $\mathfrak{sl}(2,\bbR)_+$, defined in \eqref{eq:SL2Rp}, of which only $\bar{J}_0$ and $\bar{J}_1$ are preserved. Together, these two $\mathfrak{sl}(2,\bbR)$s generate the conformal transformations in the light plane. The generators satisfy the commutation relations:
\begin{equation}
    [J_m, J_n] = i(n-m) J_{m+n}\,,\quad
    [\bar{J}_m, \bar{J}_n] = i(n-m) \bar{J}_{m+n}\,,\quad
    [J_m, \bar{J}_n] = 0\,.
\end{equation}
With our choices of normalizations above, a primary $\calO_I$ satisfies:
\begin{align}
    [J_0,\calO_I(0)] 
        &= -\frac{i}{2}(\Delta + \tau) \calO_I(0)\\
    [\bar{J}_0,\calO_I(0)] 
        &= -\frac{i}{2}(\Delta - \tau) \calO_I(0)
\end{align}
where $\tau$ is \textit{minus} the projection along the $+-$ direction, i.e. the $T_{-\dots-}$ component of a traceless symmetric spin-$\ell$ tensor has $\tau=-\ell$. In particular, for a scalar primary $\calO(x)$ we have $\tau = 0$, and for a vector primary $A_\mu$ we have
\begin{equation}
    [J_0, A_{\pm}] 
        = -\frac{i}{2}(\Delta \mp 1)A_{\pm}\,,
    \quad
    [\bar{J}_0, A_{\pm}] 
        = -\frac{i}{2}(\Delta \pm 1)A_{\pm}\,,
\end{equation}
while the other components vanish.

Next we unpack the set of transverse transformations
\begin{equation}
    \mathfrak{g}_t 
        := \mathrm{span}\{K_i\,, M_{-i}\,, M_{ij}\,,\bar{J}_0\,,\bar{J}_1\} =: \mathfrak{g}_t' \cup \mathrm{span}\{\bar{J}_0\}\,.
\end{equation}
It is helpful to start by considering just $\mathfrak{g}_t'$. First, note that $\bar{J}_1 \propto K_-$ is completely central in $\mathfrak{g}_t'$. Indeed, the only non-trivial $\mathfrak{g}_t'$ commutators are the ones implied from \eqref{eq:isod2M} and \eqref{eq:isod2K}:
\begin{equation}
    [M_{ij}, K_{i}] = iK_j\,,\quad 
    [K_i, M_{-j}] = i\delta_{ij}K_-\,,\quad 
    [M_{ij}, M_{-i}] = iM_{-j}\,, 
\end{equation}
and the usual $M_{ij}$ rotations. This is just the Heisenberg algebra with $K_i$ and $M_{-i}$ playing the role of $\hat{x}_i$ and $\hat{p}_i$ respectively, and $K_-$ is $\hat{\mathds{1}}$.  There is also a natural $\mathfrak{so}(d-2)$ of rotations rotating the transverse directions into each other. Thus we have
\begin{equation}
    \mathfrak{g}_t' = \mathfrak{so}(d-2) \ltimes \mathfrak{h}_{d-2}\,.    
\end{equation}

There is a nice relationship between $\mathfrak{g}_p$ and $\mathfrak{g}_t'$. In particular, $SL(2,\bbR)$ is the outer automorphism group of $\mathfrak{h}_1$, which is neatly identified with our $\mathfrak{sl}(2,\bbR)_-$. If we take a look at $\mathfrak{g}_p$, we see that $\mathfrak{g}_t'$ decomposes into $\mathfrak{sl}(2,\bbR)$ modules, where each $\{K_i,M_{-i}\}$ pair transforms as an $\mathfrak{sl}(2,\bbR)$ 2-tuple:
\begin{alignat}{2}
    [J_{-1},K_i] &= -2iM_{-i}\,,\qquad
        &[J_{-1},M_{-i}] 
        &=0\,\nonumber\\
    [J_{0},K_i] &= \frac{i}{2}K_i\,,\qquad
        &[J_{0},M_{-i}] 
        &=-\frac{i}{2}M_{-i}\,\label{eq:outerAutomorphism}\\
    [J_{1},K_i] &= 0\,,\qquad
        &[J_{1},M_{-i}] 
        &=+\frac{i}{2}K_i\,.\nonumber
\end{alignat}
Of course the central $K_-$ and $\mathfrak{so}(d-2)$ rotations are $\mathfrak{sl}(2,\bbR)_-$ trivial. As we will discuss further in Section \ref{sec:Schrodinger}, together these form a Schr\"odinger algebra:
\begin{equation}
    \mathfrak{sch}_{d-2} := \mathfrak{g}_p \ltimes \mathfrak{g}_t'\,.
\end{equation}

Finally, we return to $\bar{J}_0$. By construction, $\bar{J}_0$ commutes with $\mathfrak{sl}(2,\bbR)_-$. In fact, the only non-trivial commutators of $\bar{J}_0$ are with the $\mathfrak{h}_{d-2}$. The elements transform with eigenvalues
\begin{equation}
    [\bar{J}_0, K_i] = \frac{i}{2}K_i\,,\quad
    [\bar{J}_0, \bar{J}_1] = i\bar{J}_1\,,\quad
    [\bar{J}_0, M_{-i}] = \frac{i}{2}M_{-i}\,.
\end{equation}
This is commensurate with our result in \eqref{eq:J0BarBoost}: $+$-direction objects boost with twice the eigenvalue of the transverse directions. In totality, we have the null-defect algebra
\begin{equation}
    \mathfrak{n}_d := (\mathfrak{sl}(2,\bbR) \times \mathbb{R} \times \mathfrak{so}(d-2)) \ltimes \mathfrak{h}_{d-2}\,.
\end{equation}

\subsection{Surprise Schr\"odinger Symmetry and Non-Relativistic CFTs}\label{sec:Schrodinger}
As mentioned in Section \ref{sec:NullDefectAlgebra}, the null defect algebra is the same as a Schr\"odinger algebra with one additional element $\bar{J}_0$, corresponding to our collinear twist. The Schr\"odinger group is the maximal group of kinematic symmetries of the free Schr\"odinger equation (see e.g. \cite{Henkel:1993sg, Henkel:2003pu, Henkel:2005dj} and \cite{Duval:2024eod} for a review). It is also the symmetry group of the non-relativistic Schr\"odinger CFTs. Such theories have dynamical critical exponent $z=2$ and have found a host of applications, especially in cold atom systems, where they describe heavy atoms, phonons, and/or vortices in a ``harmonic trap'' \cite{Son:2005rv, Mehen:2007dn, Nishida:2007pj, Bekaert:2011qd, Kravec:2019djc}. They also have a well-defined and convergent OPE \cite{Golkar:2014mwa, Goldberger:2014hca}, holographic correspondence \cite{Son:2008ye, Goldberger:2008vg, Balasubramanian:2008dm}, and interesting defect/impurities \cite{Raviv-Moshe:2024yzt}. See also \cite{Baiguera:2023fus} for an excellent review.\footnote{We emphasize that the non-relativistic Schr\"odinger CFTs have $z=2$ scaling in the time direction $D:(t,\vec{x})\mapsto (\lambda^2 t,\vec{x})$. These are \textit{not} the CFTs that would be invariant under the $c \to \infty$ In\"on\"u-Wigner contraction of the usual conformal algebra, which has $z=1$ scaling, called Galilean Conformal Field Theories. Many intuitions and issues are transferrable between them. See also \cite{Bagchi:2009my} and references within for further discussion.}

The Schr\"odinger algebra in $(d+1)$-dimensional spacetime is just the usual Galilean algebra, with time $P_0$ and space $P_i$ translations, Galilean boosts $K_i$ and rotations $M_{ij}$, and mass central extension $M$, as well as an $\mathfrak{sl}(2,\bbR)$ extension in the timelike direction only, with special conformal generator $C_0$ and $z=2$ dilatation operator $D$. In practice, the generators are:
\begin{equation}
    \mathfrak{sch}_d = \mathrm{span}\{P_0, C_0, D, M_{ij}, K_i, P_i, M\}\,,
\end{equation}
subject to the relations:
{\begin{alignat}{3}
    [D,P_0] 
        &= 2i P_0\,,\quad
    &[C_0,P_0] 
        &= i D\,,\quad
    &[D,C_0] 
        &= -2i C_0\,,\nonumber\\
    [D,P_i] 
        &= i P_i\,,\quad
    &[K_i,P_j] 
        &= i \delta_{ij} M\,,\quad
    &[D,K_i] 
        &= -i K_i\,,\nonumber\\
    [C_0,P_i] 
        &= i K_i\,,\quad
    &&&[P_0,K_i] 
        &= -i P_i\,,\\
    [M_{ij},P_k] 
        &= i(\delta_{ik} P_j-\delta_{jk}P_i)\,,\mkern-36mu
    &&&[M_{ij},K_k]
        &= i(\delta_{ik} K_j-\delta_{jk}K_i)\,,\nonumber
\end{alignat}
\vskip -2em
\begin{equation}
    \quad[ M_{ij}, M_{kl}]
        = -2i(\delta_{j[l} M_{k]i}-\delta_{i[l} M_{k]j})\,.\nonumber
\end{equation}}
The time components scaling with twice the power of the spatial components is what is meant by a $z=2$ system. To identify it explicitly with our generators, we can take
\begin{gather}
    \{P_0,D,C_0\} = \{J_{-1},-\tfrac{1}{2}J_0,J_1\}\,,\qquad
    M = -4\bar{J}_1\,,\nonumber\\
    P_i = -2 M_{-i}\,,\qquad
    K_i = K_i\,,\\
    M_{ij} = M_{ij}\,.\nonumber
\end{gather}
In this identification, one lightcone direction plays the role of Schr\"odinger time, while the transverse directions play the role of the Schr\"odinger position.

The connection to Schr\"odinger symmetry sheds light on our null defect problems. We also expect formal problems, such as the construction of a defect Hilbert space and the definition and convergence of the defect OPE, to be informed by Schr\"odinger CFTs. Conversely, problems in the null reduction of CFTs should be related to families of Schr\"odinger CFTs (see below). For example, Schr\"odinger primaries at $\calO(t=0,\vec{x}=0)$ are defined/found to satisfy \cite{Nishida:2007pj, Goldberger:2008vg}:
\begin{gather}
    [D, \calO(0)] = i\Delta \!\calO(0)\,,\quad
    [M, \calO(0)] = m \!\calO(0)\,,\\
    [C_0,\calO(t,\vec{x})] = [K_i,\calO(t,\vec{x})] = 0\,.
\end{gather}
The condition to be a null defect CFT primary should therefore be similar to the condition to be an $M=0$ Schr\"odinger primary.

Indeed, it has long been known that the null-reduction of $(d+2)$-dimensional Lorentzian systems leads to Schr\"odinger invariant systems in $(d+1)$-dimensional spacetimes \cite{Duval:1984cj, Banks:1996vh}; with our situation enriched by the presence of a conformal defect. In fact, the non-conformal analogue has been known even longer. In theories without conformal symmetry, working in lightcone quantization (aka the ``infinite momentum frame'') one finds a distinguished Galilean subgroup in the transverse directions and can consistently map between ultra-boosted relativistic mechanics and Schr\"odinger problems in one fewer dimension \cite{Weinberg:1966jm, Susskind:1967rg, Kogut:1969xa, Kogut:1972di, Venugopalan:1998zd}. In this picture, we take
\begin{equation}
    H_{\mathrm{QM}} := F P_-\,,\quad M := F P_+\,,
\end{equation}
so that the mass-shell equation
\begin{equation}
    -m_{\mathrm{rel.}}^2 = -(2/F^2)P_+ P_- + P_\perp^2
\end{equation}
becomes
\begin{equation}\label{eq:QMHammy}
    H_{\mathrm{QM}} = \frac{P_\perp^2}{2M} + V_0
\end{equation}
where (in this case) $V_0 = m_{\mathrm{rel.}}^2/(2M)$.

One notable result of Weinberg is that, \textit{in perturbation theory}, the vacuum structure in ultra-boosted frames (or on lightcone quantization surfaces) is trivial: there are no contributions from diagrams in which particles are created or destroyed from the vacuum \cite{Weinberg:1966jm}. In other words, the lightcone vacuum is perturbatively equivalent to the free vacuum. This plays an essential role in parton physics, for example, in justifying that multi-parton Fock states are eigenstates of the QCD Hamiltonian (see e.g. \cite{Venugopalan:1998zd}). The non-relativistic version of this statement is immediate: in non-relativistic theories, when $M$ is central, we do not have particle creation and annihilation. With \eqref{eq:QMHammy}, this also highlights an extremely important subtlety: essentially all of the non-perturbative information is hidden in the $M = P_- = 0$ sector (see also \cite{Fitzpatrick:2018ttk, Anand:2020gnn, NR0Mass}). We leave further study of these formal connections to future works.

\subsubsection{Null-Reduction and Delta-Function Potentials}\label{sec:NullDelta}
Consider, for simplicity, a scalar field $\phi$ in $d$ spacetime dimensions with action:
\begin{equation}
    S = \int dx^+ dx^- dx^\perp \big(\!-\!\!\tfrac{1}{2}(\partial_\mu \phi)^2 - V(\phi)\big)\,,
\end{equation}
then the equations of motion are
\begin{equation}
    \left(-\tfrac{2}{F^2}\partial_+\partial_- + \partial_i^2\right)\phi(x^+,x^-,x^\perp) = V'(\phi)\,.
\end{equation}
A Fourier transform in the $x^+$ direction gives, assuming $\lim_{x^+ \to \pm\infty} \phi(x^+,x^-,x^\perp) = 0$, or in some DLCQ setup, a Schr\"odinger equation with mass $m=\beta k_+/F^2$ and $x^-$ as the Schr\"odinger time:\footnote{
Our conventions for the Fourier transform are
\begin{align}
    \psi(k_+,x^-,x^\perp) 
        &= \alpha\int dx^+\, e^{i\beta k_+ x^+} \phi(x^+,x^-,x^\perp)\,,\\
    \phi(x^+,x^-,t) 
        &= \frac{\abs{\beta}}{2\pi\alpha} \int dk_+\, e^{-i\beta k_+ x^+}\psi(k_+,x^-,x^\perp)\,.
\end{align}
}
\begin{equation}\label{eq:SchroEoM}
    \left(2mi\partial_- + \partial_i^2\right)\psi(m,x^-,x^\perp) = V'_{*}(\psi)\,.
\end{equation}
Note that this is really a family of Schr\"odinger problems parametrized by $m \propto k_+$.

For example, we can consider a simple pinning-type line defect extended along the $x^-$ direction
\begin{equation}\label{eq:PinningPotential}
    V(\phi) = h \delta(x^+)\delta^{d-2}(x^\perp)\phi(x^+,x^-,x^\perp)\,.
\end{equation}
After a null-reduction, the corresponding Schr\"odinger problem is given by a Dirac-delta potential
\begin{equation}\label{eq:DiracDeltaPotential}
    i\partial_- \psi_m(x^-,x^\perp) = -\frac{1}{2m}\partial_i^2\psi_m(x^-,x^\perp) + h\delta^{d-2}(x^\perp)\,.
\end{equation}
Such $\delta$-function potentials are the non-relativistic limit of $\phi^4$ theory \cite{Beg:1984yh, Jackiw:1991je, Bergman:1991hf}. In $(2+1)$d, solutions for the Schr\"odinger equation with $\delta$-function source are $\sim\!\log(\abs{x_\perp})$ (up to scheme dependent pieces) for any fixed $m$. The classical (Feynman-causal) solution to \eqref{eq:SchroEoM} with potential \eqref{eq:PinningPotential} is a scalar shockwave:
\begin{equation}\label{eq:scalarShockwave}
    \phi_{s}(x^+,x^-,x^\perp) = h \delta(x^+) G_{F}^{(d-2)}(x^\perp)\,,
\end{equation}
where $G_F^{(d-2)}(x^\perp)$ is the Feynman propagator in $(d-2)$-dimensions; matching the log divergence of the Schr\"odinger solution for $d=4$.\footnote{The careful reader may notice that \eqref{eq:scalarShockwave} does not actually preserve the full null-defect algebra $\mathfrak{n}_d$. In fact, a straightforward calculation in Section \ref{sec:ScalarPrimary} shows that any scalar deformation on a line will never preserve the full null defect algebra in the UV.\label{footnote:nonMaximal}} As we will see, shockwave solutions are generic and often the only possible non-trivial field configurations compatible with symmetries.

The lesson from a potential like \eqref{eq:PinningPotential} is fairly generalizable. In particular, any Lagrangian pinning-type potential will give a Schr\"odinger (or more general Lifshitz) problem like \eqref{eq:DiracDeltaPotential} with a (possibly complicated) weighted $\delta$-function potential. Practically speaking, a $\delta$-function potential can define a defect (see \cite{Raviv-Moshe:2024yzt}) or, in many-body quantum mechanics, prescribes a particular singular behaviour for the wavefunction as particles collide.
So the fields content and OPEs of the null theory prescribe the scaling dimensions of local operators in the non-relativistic field theory.

\subsection{Symmetries of a Null-Plane and Conformal Symmetry Breaking}\label{sec:CSB}
As we will see in examples, explicit solutions and ultraboosted limits of conformal line defects often split spacetime into two halves before ($x^+ < 0$) and after ($x^+ > 0$) the defect. In other words, physics often depends not just on the distance to the line defect $x^+ = 0$ and $x^\perp = 0$, but distinguishes the whole null plane $x^+ = 0$.

A straightforward computation partitions the symmetries of the null plane into two families:
\begin{enumerate}
    \item \textbf{Broken Transformations.} There are $d$ generators explicitly broken by the null plane:
    \begin{equation}
        P_+\,,\quad K_+\,,\quad M_{+i}\,.
    \end{equation}
    \item \textbf{Preserved Transformations.} There are $d(d+1)/2+1$ generators preserved by the null plane:
    \begin{equation}
        P_-\,,\quad P_i\,, \quad K_-\,,\quad K_i\,,\quad D\,,\quad M_{+-}\,,\quad M_{-i}\,,\quad M_{ij}\,.
    \end{equation}
\end{enumerate}
The symmetries of the null-plane form a conformal Carollian algebra (equivalently, the Poincar\'e algebra see the Appendices of \cite{Chen:2021xkw, Kulp:2024scx}) with one additional generator $M_{+-}$ acting by outer automorphisms.\footnote{We relegate potential connections to flat space holography to future works.}

Here we focus on a simpler point. The SCT $K_+$, which maps the null line $x^+ = 0$ and $x^\perp = 0$ to itself, moves the $x^+ = 0$ plane. Specifically, a finite $K_+$ transformation maps a point in the $x^+ = 0$ plane to
\begin{equation}
    (0, x^-, x^\perp) \mapsto \frac{1}{1+2cF^2x^-}(-c x_\perp^2, x^-, x^\perp)\,,
\end{equation}
and so $K_+$ ``tilts'' the shockwave plane. 

This is not exotic: $K_0$ does not fix constant time slices, and a generic solution for a timelike defect one-point function is supported on all of spacetime with varying field values throughout spacetime. When $K_0$ acts, the solution maps back to itself up to spin and scaling terms. The real difference for null defects is the flattening of solutions into shockwaves, supported \textit{only} on $x^+ = 0$. Then, since $K_+$ does not leave the $x^+ = 0$ plane invariant, there is essentially no hope for conformal symmetry to be preserved. i.e. pure shockwave solutions break $K_+$ symmetry. In interacting examples, other effects may be able to turn on field strengths away from the shockwave plane and prevent spontaneous breaking of conformal symmetry (see Section \ref{sec:shockwaves}).

\section{General Considerations for Infinite Line Defects}\label{sec:GeneralConsiderations}
A natural question now is: which subalgebras of the null-defect algebra survive in examples with non-trivial defects? In this section we start to address this question by proving some general results about abstract null conformal line defects. As we will see, infinitely long defects are quite strongly constrained. We will discuss some other interesting situations in Section \ref{sec:OtherResults}. We mainly focus on $d>2$ spacetime dimensions here, leaving $d=2$ to Section \ref{sec:2dPaper}.

In answering our question, it is important to keep track of which particular effect breaks or preserves different symmetries. This gives us a rough hierarchy of considerations:
\begin{enumerate}
    \item \textbf{Classical UV Action}. If we start with a UV CFT and add a pinning-type deformation
    \begin{equation}
        S_{\mathrm{CFT}} + h\int dx^{-} \cdot \calO(0,x^-,0)\,,
    \end{equation}
    the interaction itself may break some additional symmetries at the level of the classical action. We will see this abstractly in Section \ref{sec:PinningSymmetries} and explicitly with the free scalar in Section \ref{sec:ExampleFreeScalar}.
    \item \textbf{Symmetry Breaking}. Studying some simple examples, we find that just considering the UV symmetries associated to the classical action is too naive. Sometimes there are additional symmetry breaking solutions, and sometimes the only non-trivial solutions are ones breaking some of the UV symmetries. 
    \item \textbf{Quantum Mechanical Effects}. In an interacting quantum theory, we can have quantum effects which break symmetries, e.g. by giving an anomalous dimension to operators. This is further enriched by the fact that we have different types of flows we can consider e.g. $J_0$ flows or $D$ flows (or even a mixture).
    \item \textbf{Emergent IR Symmetries}. Finally, there can be emergent IR symmetries, which can only be studied if we know the actual IR fixed point.
\end{enumerate}

Depending on the asymptotic regions in question, combinations of the above broken and emergent symmetries could result in different effective theories describing an IR defect fixed point. 
To say we have a null conformal defect, the defect dynamics should preserve some $\mathfrak{sl}(2,\bbR)$ that maps the null-line to itself. As we will see, there are also natural examples that preserve some scale-like transformations inside the $\mathfrak{sl}(2,\bbR) \times \bbR$ group. Before considering examples, we note some physical expectations and very strong constraints from the kinematics in Section \ref{sec:NullDefectAlgebra}:
\begin{itemize}
    \item Generically, we expect all of the $M_{ij}$ to be preserved. On general grounds, it would be strange if there was a breaking of rotational symmetry in the transverse directions to the light plane.\footnote{Monodromy defects are an explicit counterexample to this, but they are, by their very nature, relative to some higher surface defect.}
    \item Since $\mathfrak{sl}(2,\bbR)$ acts by outer automorphisms on the $\mathfrak{h}_{d-2}$ (recall \eqref{eq:outerAutomorphism}) it is hard to have any part of the transverse group without having \textit{all} of it. Specifically, having just translation symmetry $P_-$ and transverse SCTs $K_i$ leads to all of $\mathfrak{h}_{d-2}$. Likewise, having just special conformal symmetry $K_+$ and lightcone boosts $M_{-i}$ leads to all of $\mathfrak{h}_{d-2}$.
    \item If we consider power counting in $M_{+-}$-boost weight, i.e. $J_0 - \bar{J}_0$ weight, we generically expect the coupling $h$ to transform non-trivially under boosts. Thus we expect any $h$ dependence of correlation functions to scale suitably in the lightlike limit and cause generic correlation functions to vanish or diverge unless there is a compensating scaling.
    \item The very large symmetry algebra of $\mathfrak{n}_d$ leads one to guess that there may actually be no non-trivial maximally symmetric null conformal line defects.
\end{itemize}

We proceed as follows. Since we don't yet know a good definition of an abstract null conformal line defect, we must study some examples. Thus, we start by considering pinning field type defects in Section \ref{sec:PinningSymmetries}, and find a connection to general Lifshitz symmetric systems. In Section \ref{sec:WITariffs} we solve general Ward identities for one-point functions and argue that interesting null defects cannot exist without breaking many transverse symmetries; we use similar arguments for two-point functions to infer that maximally symmetric null defects are trivial up to a discontinuity in the shockwave plane. 

Next we consider the (3+1)d free scalar in Section \ref{sec:ExampleFreeScalar}, it has multiple natural solutions (depending on boundary conditions) which we compare to the UV symmetries. In this example, we argue that correlation functions are actually distributions on a restricted space of test functions $\calS_{\rm null}(\bbR^d) \subset \calS(\bbR^d)$, unique to null line defects, and then consider the matching with UV symmetries. In Section \ref{sec:limitsOfDefects}, we consider ultrarelativistic limits of timelike and spacelike line defects and the constraints on null defects as ultraboosted limits. In Section \ref{sec:shockwaves} we discuss the prevalence of shockwave solutions, especially in perturbation theory. 

Finally, in Section \ref{sec:NullWilsonLine} we consider the massless charged particle or infinite null Wilson, as well as their comparison to different limits. Using our modified space of test functions, we offer a resolution to the non-convergence of ultraboosted limits of gauge potentials in classical electromagnetism.

\subsection{Pinning Defects and Symmetries of the Action}\label{sec:PinningSymmetries}
Let $\calO_I$ be a primary operator of scaling dimension $\Delta$, transforming with some spin indices $I$. We can integrate it along the $x^+ = 0$ ray, turning on a pinning-type deformation of the schematic form
\begin{equation}
    S_{\mathrm{pin}} = h \int dx^- \cdot \calO(0,x^-,0)\,.
\end{equation}
The differential action of $\mathfrak{n}_d$ on $\calO(0,x^-,0)$ is given by
\begin{equation*}
\begin{aligned}
    [P_-,\calO_{I}] 
        &= i\partial_-\calO_I\,,
    &[K_i,\calO_{I}] &= -2{(S_{-i})_I}^J x^- \calO_J\,,
    \\
    [D,\calO_{I}]
        &= i(x^-\partial_-+\Delta)\calO_{I}\,,
        &[M_{-i},\calO_{I}] &= -{(S_{-i})_{I}}^J \calO_J\,,
        \\
    [M_{+-},\calO_{I}]
        &= -iF^2 x^-\partial_- \calO_{I}-{(S_{+-})_I}^J \calO_J\,,
     &[M_{ij},\calO_{I}] &= -{(S_{ij})_{I}}^J \calO_J\,,
    \\
    [K_+,\calO_I] 
        &= i 2F^2 ((x^-)^2\partial_- +\Delta x^-)\calO_I + 2{(S_{+-})_I}^J x^- \calO_J\,,
    &[K_-,\calO_{I}] &= 0\,.
    \\
\end{aligned}
\end{equation*}
We can consider the variation of the action under a general element of $\mathfrak{n}_d$, say with parameters $p_-, k_+, \dots, m_{-i}, m_{ij}$, denoted $\delta_\mathfrak{n}$, then
\begin{equation}
\begin{aligned}
    \delta_\mathfrak{n} \calO_I(0,x^-,0)
        &= -i\Delta(d+2F^2k_+ x^-)\calO_I(0,x^-,0)\\
        &- i(p_- + (d-F^2 m_{+-})x^- + 2F^2k_+ (x^-)^2)\partial_- \calO_{I}(0,x^-,0)\\
        &+(m_{-i}+2k_{i}x^-){(S_{-i})_I}^J \calO_J(0,x^-,0)\\
        &+(m_{+-}-2k_{+}x^-){(S_{+-})_I}^J \calO_J(0,x^-,0)
        +m_{ij}{(S_{ij})_I}^J \calO_J(0,x^-,0)\,.
\end{aligned}
\end{equation}
We can integrate by parts along $x^-$, with boundaries $(x_i^-,x_f^-)$, modifying the expression for $S_{\rm pin}$ to
\begin{equation}\label{eq:generalVariation}
\begin{aligned}
    \delta_\mathfrak{n} S_{\rm pin} 
        &= \int dx^-\left[ -i(d(\Delta-1)+F^2(2k_+(\Delta-2)x^- +m_{+-}))\calO_I(0,x^-,0)\right.\\
        &+(m_{-i}+2k_{i}x^-){(S_{-i})_I}^J \calO_J(0,x^-,0)\\
        &\left.+(m_{+-}-2k_{+}x^-){(S_{+-})_I}^J \calO_J(0,x^-,0)
        +m_{ij}{(S_{ij})_I}^J \calO_J(0,x^-,0)\right]\\
        &- i(p_- + (d-F^2 m_{+-})x^- + 2F^2k_+ (x^-)^2) \calO_{I}(0,x^-,0)\bigg\vert_{x_i^{-}}^{x_f^{-}}\,.
\end{aligned}
\end{equation}
Of course, any finite length defect generically breaks $P_- \sim J_{-1}$, $J_0$, and $K_+ \sim J_1$ symmetry, and so we will assume that the defect infinite is in extent with vanishing boundary terms. We will return to finite (or semi-infinite) length defects in Section \ref{sec:FiniteDefects}. Using this, we can consider the subalgebra of UV symmetries preserved by the action in a few different special cases. The results of this section are summarized in the Table \ref{tab:null-line-summary} for primary operators integrated along the line.

\begin{table}
\centering
\small
\begin{tabular}{
    >{\centering\arraybackslash}m{2.0cm} 
    >{\centering\arraybackslash}m{2.0cm} 
    >{\centering\arraybackslash}m{5.0cm} 
    >{\centering\arraybackslash}m{5.0cm}}
\toprule
\thead{Case} & \thead{Action} & \thead{Always Preserved} & \thead{Extra Symmetries} \\
\midrule

Scalar primary
  & $h\!\int dx^-\,\mathcal{O}$
  & $P_-,\,M_{ij},\,K_i,\,K_-,\,M_{-i}$
  & \makecell[l]{
      $\Delta=1\!:\;D$\\
      $\Delta=2\!:\;K_+,\,J_0$\\
      $\Delta\neq1,2\!:\;J' = D - \tfrac{\Delta-1}{F^2} M_{+-}$
    } \\

\addlinespace

Spin‑1 primary
  & $g\!\int dx^-\,A_-$
  & $P_-,\,M_{+-},\,M_{ij},\,K_i,\,K_-,\,M_{-i}$
  & \makecell[l]{
      $\Delta=1\!:\;D,\,K_+$\\
      $\Delta\neq1\!: \,\,\text{---}$
    } \\

\addlinespace

Spin‑$\ell$ primary
  & $h\!\int dx^-\,T_{-\dots-}$
  & $P_-,\,M_{ij},\,K_i,\,K_-,\,M_{-i}$
  & \makecell[l]{
      $\Delta=2-\ell\!:\;K_+,\,J_0$\\
      $\Delta\neq2-\ell\!:\;J' = D + \tfrac{\Delta-1}{(\ell-1)F^2} M_{+-}$
    } \\

\bottomrule
\end{tabular}
\caption{Summary of pinning field defects and associated UV symmetries.}
\label{tab:null-line-summary}
\end{table}

\subsubsection{Scalar Primary and General Covariance}\label{sec:ScalarPrimary}
In the scalar case, but not the subsequent spinning primaries, we need some care in defining our pinning field defect. i.e. we must be less schematic in our notation $dx^- \cdot \calO$. To write a ``local defect'' compatible with general covariance, $S$ should be a functional of the trajectory
\begin{equation}\label{eq:generalCovScalar}
    S_{\rm{pin}}[h, x] = \int_{x(\sigma)} d\sigma\, h(\sigma) \calO(x(\sigma))\,.
\end{equation}
Here $\sigma$ is a worldline time coordinate and $h(\sigma)$ is a local density of the scalar $\mathcal{O}$ on the line. This density behaves like an einbein so that the action $S$ is invariant under 1d diffeomorphisms on the worldline i.e. $\sigma \mapsto \sigma'(\sigma)$ and $h(\sigma')\to h(\sigma) \frac{d\sigma}{d\sigma'}$. In the case of a timelike line, we can make a canonical choice and set $h(\sigma) = h \sqrt{-\frac{dx}{d\sigma}\frac{dx}{d\sigma}}$, to make it compatible with the symmetries of the bulk theory, leaving an integral over the ``proper time'' along the worldline
\begin{equation}
    S_{\rm{pin}}[x] =  h \int_{x(\tau)} d\tau\, \calO(x(\tau))\,.
\end{equation}

In the null case, we want to again integrate a scalar primary $\calO$ along the ray parametrized by $x^\mu = \sigma v^\mu$ where $v=\tfrac{1}{2}(1,-1,0^\perp)$ with $\sigma \in (-\infty,\infty)$, in which case $\dot{x}^2 = 0$. At this point, we can either leave the density $h(\sigma)$ as is and consider a generally covariant action, as in \eqref{eq:generalCovScalar}, or we can again choose a particular density, so that it preserves as many of the bulk conformal symmetries as possible for a scalar pinning field. 

We choose to fix $h(\sigma)= h = {\rm const}$, so that our action is
\begin{equation}
    S_{\mathrm{pin}} = h \int dx^- \calO(0,x^-,0)\,.
\end{equation}
Our choice of $h(\sigma)=h={\rm const}$ is useful for comparison to the null Wilson line in Section \ref{sec:NullWilsonLine} (which is automatically generally covariant), and more closely resembles the light-ray integral transform of a primary operator in Section \ref{sec:shockwaves}. A consequence of our choice is that our defect has fewer local degrees of freedom: displacement operators only exist in the ($d-2$)-transverse directions $x^\perp$. Moreover, some rescalings of the free coupling must be taken into account under boosts in \ref{sec:limitsOfDefects}, but then our choice correctly reproduces ultrarelativistic limits.

Now we can consider the variation of the action $\delta_{\mathfrak{n}}S_{\mathrm{pin}}$ under a general conformal transformation preserving the null line. In this case, \eqref{eq:generalVariation} simplifies dramatically, with no boundary terms (by assumption) and no spin terms. The variation of the integrand is (dropping the integral and $\calO$):
\begin{align}
    \delta_{\mathfrak{n}}\calL_{\mathrm{pin}} 
        &\propto ((\Delta-1)d + F^2 m_{+-}) + 2F^2 x^- k_+(\Delta-2)\,.
\end{align}
Consequently, the classical action always preserves $\{P_-, M_{ij}, K_i, K_-, M_{-i}\}$ and the following additional symmetries based on the scaling dimension of $\calO$:
\begin{enumerate}
    \item $\mathbf{\Delta = 1}$. Preserves scaling $D$, as we would expect from standard powercounting. $M_{+-}$ and $K_+$ are necessarily broken.
    \item $\mathbf{\Delta = 2}$. Preserves $K_+ \sim J_1$ and $J_0$. Collinear twist $\bar{J}_0$ is necessarily broken.
    \item \textbf{Generic $\mathbf{\Delta}$}. Generically, $J_1 \sim K_+$ is broken, but the action still preserves
    \begin{equation}
        J' := D-\frac{(\Delta-1)}{F^2}M_{+-}\,.
    \end{equation}
\end{enumerate}
The last $J'$ reduces to the other $D$ and $J_0$ in the special cases $\Delta = 1, 2$. The surviving algebra with $J'$ closes
    \begin{equation}
    \begin{aligned}
        [J',P_-] = i\Delta P_-\,,\quad
        [J',K_i] = -i K_i\,,\quad
        [J',K_-] = i(\Delta-2)K_-\,,\quad
        [J',M_{-i}] = i(\Delta-1)M_{-i}\,.
    \end{aligned}
    \end{equation}
We immediately recognize this family of symmetries, foreshadowed in Section \ref{sec:Schrodinger}. In this case, we see that (up to sign/labeling conventions), the subalgebra of the conformal group preserved by a scalar with scaling dimension $\Delta$ is the Lifshitz algebra with $z = \Delta$.

As explained in Section \ref{sec:Schrodinger}, a null-reduction of our $d$-dimensional defect CFTs should relate them to ($d-1$)-dimensional non-relativistic systems with Lifshitz symmetry. Such systems are believed to be dual to certain pp-wave deformations of $\mathrm{AdS}_{d+1}$ spacetimes \cite{Son:2008ye, Goldberger:2008vg, Balasubramanian:2008dm, Kachru:2008yh}. Consequently, we imagine a heavy or non-dynamical null defect in a Lorentzian CFT could deform spacetime to one of these Lifshitz geometries. We leave holographic interpretations of our results to other works.

\subsubsection{Spin-1 Primary}\label{sec:WLKinematics}
Let $A_\mu$ be a vector/1-form primary integrated along $x^\mu = \sigma v^\mu$ where $v=\tfrac{1}{2}(1,-1,0,0)$ with $\sigma \in (-\infty,\infty)$ as before
\begin{equation}\label{eq:nullWilsonLine}
    S_{\mathrm{WL}} = g \int_\Gamma dx^\mu A_\mu(x) = g \int_{-\infty}^{\infty} dx^{-}\, A_-(0,x^-,0)\,.
\end{equation}
Like before, we can consider the variation of the action under a general conformal transformation. The terms not-depending on spin matrices are exactly as in the scalar case, so the only extra work comes from the spin terms. Using our identities from \eqref{eq:SIdentities}, the expression is proportional to $A_{-}$ with coefficient:
\begin{align}
    \delta_{\mathfrak{n}} \calL_{\mathrm{WL}} \propto (\Delta-1)(d+2F^2k_+x^-)\,.
\end{align}
As a result, the classical action always preserves $\{P_-, M_{+-}, M_{ij}, K_i, K_-, M_{-i}\}$ and the following additional symmetries based on the scaling dimension of $A_\mu$:
\begin{enumerate}
    \item $\mathbf{\Delta = 1}$. Preserves $D$ and $K_+$, this is the entire null defect algebra $\mathfrak{n}_d$.
    \item \textbf{Generic $\mathbf{\Delta}$}. No additional symmetries are preserved, $D$ and $K_+$ are broken.
\end{enumerate}

Consider a 1-form/vector operator $A_\mu(x)$ which is primary with weights $\Delta = \ell = 1$. Any operator with those conformal weights is necessarily defined only up to gauge transformations and thus is necessarily a ``gauge field,'' (see (4.86) of \cite{Karateev:2018oml} for a precise discussion). Thus it's a bit odd to consider it as a ``primary'' in the sense of being a part of the CFT data of some physical CFT. Moreover, for $d>2$ it is below the unitarity bound $\Delta\geq d+\ell-2$, so any CFT it's a part of would be non-unitary. However, any such operator necessarily has a 2-form primary-descendant:
\begin{equation}
   F_{\mu\nu} := (dA)_{[\mu\nu]}\,,\quad [K_\lambda, F_{\mu\nu}] = 0\,.
\end{equation}
When $2<d\leq 4$ the local operator $F_{\mu\nu}$ is compatible with unitarity and can be discussed as part of the CFT data without reference to $A_\mu$ at all. In unitary theories, $A_\mu$ can and should be excised completely from the space of local operators -- it is only a trick for defining correlation functions of $F_{\mu\nu}$ (see also \cite{El-Showk:2011xbs} for a systematic discussion). On the other hand, it is still useful to consider interactions (which are not the same as local operators) and other finite observables built from $A_\mu$, like the null Wilson line
\begin{equation}
    W[\Gamma] = \mathcal{P}\exp(g\int_\Gamma dx^\mu A_\mu(x))\,.
\end{equation}
Thus we claim that the null Wilson line preserves all of $\mathfrak{n}_d$ at the level of UV symmetries. Likewise, a supersymmetric null Wilson line preserves the entire null conformal algebra, half of the supercharges, and any $R$-symmetries \cite{GomisNull}. Indeed, the scalar components of the usual supersymmetric $1/2$-BPS Wilson drop out on null-trajectories, reducing to the standard Wilson line \cite{Alday:2007he}, and giving a holographic interpretation in $\calN=4$ SYM. We will return to null Wilson lines in more detail in Section \ref{sec:NullWilsonLine}.

\subsubsection{Collinear Projection of Spin-\texorpdfstring{$\ell$}{ell} Primary}
Finally, we consider the example that we integrate the $T_{\mu - \dots -}$ component of a symmetric traceless rank-$\ell$ tensor:
\begin{equation}
    S_{\mathrm{ANEC}} = h \int_\Gamma dx^\mu\, T_{\mu-\dots -}(x) = h \int_{-\infty}^{\infty} dx^{-}\, T_{-\dots-}(0,x^-,0)\,.
\end{equation}
All of the spin matrices drop out from the action except the $(S_{+-}^{(\ell)})$ which simply sends $T_{-\dots -} \mapsto i\ell F^2 T_{-\dots -}$. The variations give:
\begin{align}
    \delta_{\mathfrak{n}}\calL_{\mathrm{ANEC}}
        &\propto d(\Delta-1)-F^2m_{+-}(\ell-1)+2F^2 k_+ x^- (\ell+\Delta-2)\,.
\end{align}
The classical action therefore always preserves $\{P_-, M_{ij}, K_i, K_-, M_{-i}\}$ and the following additional symmetries based on the scaling dimension and spin $\ell$
\begin{enumerate}
    \item $\mathbf{\Delta = 2-\ell}$. Preserves $K_+$ and also $J_0$, but breaks $\bar{J}_0$. This is strongly constrained by unitarity bounds (as in the previous example).
    \item \textbf{Generic $\mathbf{\Delta}$ and $\mathbf{\ell}$}. Generally $J_1\sim K_+$ is broken, but the action still preserves
    \begin{equation}
        J' := D+\frac{1}{F^2}\frac{(\Delta-1)}{(\ell-1)}M_{+-}\,.
    \end{equation}
\end{enumerate}
This last case technically subsumes all previous examples.\footnote{Due to strong constraints from unitarity bounds, we generically will never add a relevant deformation by exponentiating (a component of) a spinning primary. In principle, we can still do so -- regulating by shifting slightly in the transverse plane -- at the cost of needing infinitely many counterterms. It would be curious to know if the counterterms could be constrained by the large number of symmetries, possibly making the exponential of the ANEC operator, or something similar, an interesting observable.}

\subsection{Ward Identities and Constraints on Maximal Conformal Defects}\label{sec:WITariffs}
Next we understand how symmetries constrain the general structure of correlation functions in the presence of an infinite null line defect by directly solving the Ward identities. We restrict our attention to $d>2$ and repeat this exercise in the $d=2$ case in Section \ref{sec:2dPaper}.

We will be conservative and treat the defect as a deformation of the Hamiltonian. In this case, a defect correlation function is
\begin{gather}
    \expval*{\mathcal{O}_1(x_1) \cdots \mathcal{O} _n(x_n)}_{\mathrm{D}} 
        = \expval*{U_{\mathrm{D}}^\dagger(t_1,t_0) \mathcal{O}_1(x_1) U_{\mathrm{D}}(t_1,t_0) \cdots U_{\mathrm{D}}^\dagger(t_n,t_0) \mathcal{O}_1(x_n) U_{\mathrm{D}}(t_n,t_0)}\,, \notag\\
    U_{\mathrm{D}}(t_2, t_1) = T \exp\left(-i \int^{x^-_2}_{x^-_1} \!\!dx^-\, \mathcal{O}(0,x^-,0)  \right),
\end{gather}
where we have used the interaction picture to define how operators evolve in time. The time $t_0$ signals when the defect is turned on. In this section it is infinite, but we will return to semi-infinite and finite defects in Section \ref{sec:FiniteDefects}.\footnote{One consequence of considering the defect this way, is that it is \textit{not} viewed as an operator inserted in a correlation function -- this only really makes sense in $x^+$-time lightcone quantization anyway. Thus we will not generally consider Wightman correlators with an operator $\hat{D}$ inserted with different orderings relative to operators e.g. $\mel*{\Omega}{\calO_1(x_1)\hat{D}\calO_2(x_2)}{\Omega}$ or permutations.}

\subsubsection{One-Point Function Ward Identities}
Here we consider the symmetry constraints on the one-point functions of primary operators. The general form of allowed one-point functions is highly dependent, of course, on what the symmetry structure is at the fixed point. Here we demonstrate, by way of example, that conformal endpoints of RG flows are not expected to support non-trivial one-point functions (unless many symmetries are broken).

Consider a scalar primary operator $\calO$ of dimension $\delta$ in the presence of a null defect. Below, we enumerate the constraints on potential one-point functions of $\calO$. Our calculations are detailed in Appendix \ref{app:scalarCalc}. In addition to power-law solutions, we also allow for shockwave solutions, proportional to $\delta(x^+)$. Note: normally we do not care about contact terms in discussing correlation functions, since we only care about operators at separated points, but here we expect potential shockwave terms $\propto\delta(x^+)$ and must be careful not to throw them away.\footnote{Recall we defined the generator
\begin{equation}
    J_{\Delta}' := D-\frac{(\Delta-1)}{F^2}M_{+-}\,,
\end{equation}
which arose from studying the UV symmetries of pinning field defects in Section \ref{sec:PinningSymmetries}. Here we are not, a priori, studying UV symmetries of pinning field defects, but find it useful to still use the generator $J_\Delta'$ in discussing our algebras.}
\begin{enumerate}
    \item \textbf{Minimal Conformal Symmetry.} The defect only preserves $\{P_-, J_{\Delta}', K_+, M_{ij}\}$, where $J'_{\Delta}$ is a chiral dilatation symmetry. Then a one-point function necessarily takes the form:
    \begin{equation}
    \expval*{\calO(x^+,x^-,x^\perp)}_{\rm{D}} = \frac{c}{|x_\perp|^\delta} + \tilde{c}\,\delta_{\Delta,2}\delta(x^+)\delta(|x_\perp|^2)\,.
    \end{equation}
    In particular, we note the curious side effect that any local operator only depends on the distance in the transverse directions $|x_\perp|$ and not on the separation from the defect in the $x^+ = 0$ plane. The second term is a genuine contact term and can be ignored.
    
    \item \textbf{$\mathbf{\Delta = 1}$ Pinning Symmetries.} The defect preserves $\{P_-, M_{ij}, K_i, K_-, M_{-i}\}$ and $D = J'_1$. Note that it is \textit{not} conformal (there is no $K_+$) symmetry. Then a scalar primary one-point function takes the form:
    \begin{equation}
        \expval*{\calO(x^+,x^-, x^\perp)}_{\rm{D}} = \frac{c}{|x^+|^\delta} + \delta_{\delta,\Delta}\delta(x^+)h(|x_\perp|)\,,
    \end{equation}
    where $h$ is a homogeneous function of ``degree $0$.'' This will eventually match our result in Section \ref{sec:ExampleFreeScalar}, with a ``degree 0'' homogeneous function $\log(\mu^2 x_\perp^2)$. Here we note that a scale, but not conformal, invariant one-point function leads to $|x^+|$, not $|x_\perp|$, dependence.
    
    \item \textbf{$\mathbf{\Delta = 2}$ Pinning Symmetries.} The defect preserves $\{P_-, M_{ij}, K_i, K_-, M_{-i}\}$ and $J_0 = J'_{\Delta=2}$ and $K_+$, so is conformal. In this case, only pure contact terms on the defect exist:
    \begin{equation}
        \expval*{\calO(x^+,x^-, x^\perp)}_{\rm{D}} = c\,\delta(x^+)\delta(|x^\perp|^2)\,.
    \end{equation}
    We note, by comparison to the previous two cases, that the addition of $K_+$ to a full conformal symmetry has essentially forced the solution to be trivial.
    
    \item \textbf{Generic $\mathbf{\Delta}$ Pinning Symmetries.} The defect preserves $\{P_-, M_{ij}, K_i, K_-, M_{-i}\}$ and $J'_\Delta$. For generic $\Delta$, a shockwave solution exists
    \begin{equation}\label{eq:GenericUVPinning}
        \expval*{\calO(x^+,x^-, x^\perp)}_{\rm{D}} = \delta_{\delta,\Delta}\delta(x^+)h(|x_\perp|)\,,
    \end{equation}
    where $h$ is a homogeneous function of degree $2-2\Delta$. This is the generalization of the previous free scalar result to allow for more general ``codimension 1 CFT correlators'' along the shockwave plane.
    \item \textbf{Maximal Conformal Symmetry $\mathbf{\mathfrak{n}_d}$.} All non-trivial one-point functions are pure contact terms.
\end{enumerate}

In general, we can consider RG flows in dilatations $D$ or more general twisted dilatations $J_{\Delta}'$, terminating on a $D$ or $J_{\Delta}'$ invariant fixed point. If null conformal line defects exist at endpoints of RG flows, then we are most interested in the minimal and maximal conformal symmetries. From the preceding analysis, we see that one-point functions become trivial or essentially trivial in the IR unless copius numbers of symmetries are broken along the RG flow. The intermediate $\Delta\neq 2$ ``pinning symmetry'' cases are not conformal, and describe the symmetries of nearly maximally symmetric scale invariant fixed points. 

In free cases, where the UV symmetries are the IR symmetries, we can confirm that known one-point functions match our abstract forms above. e.g. the $\Delta=1$ case exactly matches the (3+1)d free scalar solution in Section \ref{sec:ExampleFreeScalar}. For the $\Delta = 2$ case, we can consider 
a (5+1)d free scalar with equation of motion
\begin{equation}
    \partial^2 \phi = h \delta(x^+)\delta^4(x^\perp)\,.
\end{equation}
Similar to the (3+1)d case, this PDE admits a shockwave solution
\begin{equation}
    \phi_s(x^+,x^-,x^\perp) = h\delta(x^+)|x_\perp|^{-2}\,,
\end{equation}
but this solution spontaneously breaks the $K_+$ symmetry of the action since the shockwave plane is not $K_+$ invariant. We can instead try to explicitly solve the PDE by using the 6d retarded propagator $G_R(x) = \alpha \Theta(x^0)\delta'(x^2)$, in which case we see that the final answer is compatible with our ansatz.

Vanishing one-point functions do not mean that the defect is trivial. Naively, given a collection of local operators in the presence of a defect, one could repeatedly apply the OPE to produce a single local operator, then all bulk correlation functions would be determined by the bulk OPE coefficients and the one-point functions. However, in Lorentzian signature this is not true. Instead, we must first separate local operators into those ``before'' ($x^+ < 0$) and ``after'' ($x^+ > 0$) the defect, as well as those at $x^+ = 0$. In each region, we can use the OPE to reduce correlation functions of local operators into configurations with a single operator before and a single operator after the defect.

Intuitively, this makes sense in comparison to light-ray operators -- in as much as some null defects can be considered as exponentials of light-ray operators. A generic light-ray operator $\mathbb{O}(x)$ annihilates the vacuum $\Omega$, and thus must always be sandwiched between at least two local operators (or computed in a non-trivial state) for a non-zero correlation function $\mel*{\Omega}{\calO_1(x_1) \mathbb{O}(x) \calO_2(x_2)}{\Omega}$. From this perspective, it might be expected that one-point functions vanish (or be $\delta$-supported), and that defects are characterized by configurations with a local operator before and after (although, care should be taken not to conflate ordering in a Wightman function with time ordering when being more precise).

\paragraph{Maximal Symmetry.} Let's turn again to the maximally symmetric conformal defects, and lets move on from scalars. In the maximally symmetric case (or maximally symmetric without $\bar{J}_0$ symmetry), all of the SCTs are preserved. Under inversion, an SCT becomes a translation and a primary transforms like
\begin{equation}
    \calO_I(x) \mapsto \calO_I'(\tilde{x}) = x^{2\Delta} \mathcal{I}_I^{J}(x)\calO_J(x)
\end{equation}
where $\tilde{x} = x^\mu/x^2$ and $\mathcal{I}$ is the inversion tensor for the appropriate spin rep. If a one-point function is invariant under all SCTs, then the inverted one-point function must be a constant $c_J$, hence we have:
\begin{equation}
    \expval*{\calO_{I}(x)}_{\rm{D}} = \frac{c_J}{x^{2\Delta}}\mathcal{I}_{I}^J(x)\,.
\end{equation}
However, a maximally symmetric defect should also be dilatation invariant, and this constrained form is already not compatible with dilatations unless $\Delta = 0$. Consequently, all non-trivial one-point functions  (spinning or not) in a maximally symmetric theory are $0$.\footnote{In principle we could also have $\delta$-function contact terms on the lightcone $x^2=0$. However, the remaining symmetries will kill all $\delta$-functions except those supported exactly on the geometric null line, i.e. those contact terms like $\delta(x^+)\delta^{d-2}(x^\perp)$, which we have already seen.}


\subsubsection{Two-Point Functions and the Tariff Function}\label{sec:Tariff}
Next we argue that two-point functions are nearly trivial by considering symmetry constraints. For simplicity, consider two scalar operators $\mathcal{O}_{1}$ and $\mathcal{O}_{2}$. As in the one-point function case, we use symmetry with respect to all $K_\mu$'s to constrain the two-point function to the form:
\begin{gather}
 \expval*{\calO_{\Delta_1}(x_1)\calO_{\Delta_2}(x_2)}_{\rm D} = \frac{1}{x_1^{2\Delta_1 } x_2^{2\Delta_2}}f\left(\frac{x_1}{x_1^2} - \frac{x_2}{x_2^2}\right)\,.
\end{gather}
By $M_{+-}$, $M_{-i}$, and $M_{ij}$ invariance, this becomes
\begin{gather}
\expval*{\calO_{\Delta_1}(x_1)\calO_{\Delta_2}(x_2)}_{\rm D} =  \frac{t(s_1, s_2)}{x_1^{2\Delta_1 } x_2^{2\Delta_2}} f\left(\frac{|x_{12}|^2}{x_1^2 x_2^2}\right)\,.
\end{gather}
Here, $s_i = \mathrm{sign}(x_i^+)$, stating whether it is before ($s_i = -1$) or after ($s_i = +1$) the defect, which is an invariant of these rotations and boosts; we will say $s_i = 0$ if $x_i^+ =0$.\footnote{The defect itself cuts the shockwave plane, so one may also need to consider which side of the defect an operator is on in the shockwave plane.} This means there are 9 inequivalent configurations corresponding to whether each of the two operators is in the past, future, or on the same null plane as the defect. We keep track of this with an as yet undetermined coefficient $t(s_1,s_2)$ that depends on $s_i$ in this expression. We can determine the function $f$ by applying dilatations $D$, which simplifies the two-point function to
\begin{gather}
\expval*{\calO_{\Delta_1}(x_1)\calO_{\Delta_2}(x_2)}_{\rm D} = 
 \frac{t(s_1,s_2)}{x_1^{2\Delta_1 } x_2^{2\Delta_2}}  \left\vert\frac{x_1 x_2}{x_{12}}\right\vert^{\Delta_1 + \Delta_2}\,.
\end{gather}
Finally, the remaining $P_-$ forces $\Delta_1 = \Delta_2$, so that:
\begin{gather}\label{eq:2ptFunctionTariff}
\expval*{\calO_{\Delta_1}(x_1)\calO_{\Delta_2}(x_2)}_{\rm D} = 
 \delta_{\Delta_1, \Delta_2}\frac{t(s_1, s_2)}{|x_1 - x_2|^{2\Delta_1}}\,.
\end{gather}

The final answer \eqref{eq:2ptFunctionTariff} for the two-point function in the presence of the defect is kinematically constrained by maximal conformal symmetry to take the form of a standard two-point function, up to an unknown ``tariff function'' $t(s_1, s_2)$. The tariff function represents a potential hidden tax on operators crossing the $x^+ = 0$ border between the $x^+ < 0$ and $x^+ > 0$ regions. This discontinuity in propagators across shockwave planes is well-known from studies of scattering in shockwave backgrounds (see e.g. Section 4.2 of \cite{Raj:2025hse}). With care, this derivation can also be extended to spinning operators.

Let us consider a simple case to get a sense of our tariff function. In principle, if the two operators are on the same side of the defect (for instance $x^+>0$), then we could use the OPE to expand them into a sum of local operators. With maximal symmetry, all non-trivial one-point functions vanish, so we expect that $t(\pm 1,\pm 1) = 0$. This argument also strongly constrains any higher point functions, since we can use the OPE to reduce any collection of operators on either side of the shockwave plane into a sum of local operators. Thus we reduce all non-trivial defect physics to the non-triviality of the tariff function.

\subsection{Example: (3+1)d Free Scalar + \texorpdfstring{$\phi$}{phi}}\label{sec:ExampleFreeScalar}
We now turn to one of our most important and instructive examples for the remainder of the paper. Consider the (3+1)d free scalar with a line defect stretched along $x^+ = 0$, obtained by integrating $\phi(x^+,x^-,x^\perp)$ along $x^+ = 0$ and $x^\perp = 0$. The action is:
\begin{gather}\label{eq:freeScalarAction}
    S = \frac{1}{2}\int d^{4} x \, (\partial_\mu \phi)^2 + h  \int dx^-\, \phi(0,x^-,0)\,.
\end{gather}
The bulk equations of motion for this theory with these boundary conditions are just the wave equation with a source:
\begin{equation}\label{eq:scalarEoM}
    \partial^2 \phi = h \delta(x^+)\delta^2(x^\perp)\,.
\end{equation}
To understand the one-point function, we want to find the field configuration created by the source
\begin{equation}\label{eq:nullSource}
    J(x^+,x^\perp) :=  h\, \delta(x^+)\delta^2(x^\perp) = \sqrt{2}hF\, \delta(x^0+x^1)\delta^2(x^\perp)\,.
\end{equation}

For a physical solution, we should solve the PDE using the causal/retarded Green's function. Recall that the retarded Green's function for a free scalar in four dimensions is
\begin{gather}\label{eq:4dGF}
    G_R(x) = -\frac{\delta(x^0 - r)}{4\pi r}\,,
\end{gather}
where $r^2 = (x^1)^2 + (x^\perp)^2$. The details of the calculation can be found in Appendix \ref{app:freeScalarCalcs}, the final answer is:

\begin{equation}\label{eq:scalarSolution}
    \phi_{c}(x^+,x^-, x^\perp) 
        =  -\frac{h}{4\pi x^+} \,  \Theta\left(\scarex\right)\,.
\end{equation}
Here we introduce $\scarex$ in scare-quotes where
\begin{equation}
    \scarex := \frac{F}{\sqrt{2}} x^+ + \frac{x_\perp^2}{2\sqrt{2}F x^+}\,.
\end{equation}
$\scarex$ arises naturally in integrating the Green's function \eqref{eq:4dGF} against the source \eqref{eq:nullSource}. As we explain in detail in Appendix \ref{app:xpIdentities}, $x^+ > 0$ is the same region as $\scarex > 0$; the coordinates differ in their behaviour on the $x^+ = 0$ plane. Roughly, $\scarex$ ensures that $\phi_c$ has the correct singular behaviour as $x^+ \to 0$, thereby satisfying the equations of motion.\footnote{In earlier works, a similar looking expression is obtained from limiting procedures \cite{Aichelburg:1970dh, Jackiw:1991ck} (see \cite{Azzurli:2014lha} around (2.24)), with $x^+$ instead of $\scarex$. There, it is correctly noted that the naive solution using $\Theta(x^+)$ is not a solution of the equations of motion. Section \ref{footnote:Excuse} provides further insight into this difference, coming from a restricted space of test functions.} As we explain next in Section \ref{footnote:Excuse}, all solutions should be understood as distributions on some (suitably restricted) space of test functions.

We see that the field exists and is finite in the region $\scarex>0$; we remind the reader that this is the same region as $x^+>0$. This is commensurate with our lightlike source and physical setup (no incoming radiation from infinities). The field falls off at large distances $\sim (x^+)^{-1}$ and is singular on the shockwave plane $x^+ = 0$. Switching to retarded Bondi coordinates, $(r,u,\theta,\varphi)$, the radiation profile along a generic generator of $\scri^+$ is
\begin{equation}\label{eq:radiation}
    r^\Delta \phi_{c}(r,u,\theta,\varphi)\Big\vert_{\scri^+} = -\frac{\sqrt{2}h F}{4\pi(1+\cos\theta)}\sum_{n=0}^\infty \frac{(-1)^n}{(1+\cos\theta)^n}\left(\frac{u}{r}\right)^n\,.
\end{equation}
Along $\theta = \pi$, the field profile is independent of $r$, i.e. $\phi_c(r,u,\pi,\varphi) = -\tfrac{\sqrt{2}hF}{4\pi}\Theta(u)\frac{1}{u}$.

We can also solve with the time-ordered/Feynman propagator. This is the solution that would appear in ``in-out'' formalism matrix elements or by taking a straightforward Fourier transform. In this case, we obtain the shockwave solution
\begin{equation}\label{eq:shockwaveScalar}
    \phi_{s}(x^+,x^-,x^\perp) = h \delta(x^+)G_{F}^{(d-2)}(x^\perp)\,,
\end{equation}
where $G_{F}^{(d-2)}(x^\perp)$ is the massless Euclidean Feynman propagator in ($d-2$)-dimensions (recall also Section \ref{sec:NullDelta}).\footnote{In 2d we have, for some scale $\mu$, the Green's function:
\begin{equation}
    G_F^{(2)}(x_\perp) = \frac{1}{4\pi}\log \mu^2 x_\perp^2\,,
\end{equation}
satisfying $\Box G_F^{(2)}(x_\perp) = \delta^{(2)}(x_\perp)$.} Unlike the causal solution, the shockwave solution is supported only along the shockwave plane $x^+ = 0$.

Finally, there is an acausal solution obtained by using the advanced propagator. Of course, it matches the retarded solution with a support flip
\begin{equation}
    \phi_{a}(x^+,x^-, x^\perp) 
        =  \frac{h}{4\pi x^+} \,  \Theta\left(-\scarex\right)\,.
\end{equation}
The causal and acausal solutions can be seen to satisfy the antipodal matching condition $\phi_c^{(1)}(-\infty,\theta,\varphi)=\phi_a^{(1)}(\infty,\pi-\theta,\varphi+\pi)$.

To emphasize, all of the preceding solutions to the equations of motion \eqref{eq:scalarEoM} are non-unique. In all cases, we can augment the solutions by adding a homogeneous term $\phi_h$ satisfying $\partial^2 \phi_{h} = 0$, so boundary conditions are essential. For example, $\phi_c$ is the unique causal solution only once we have further asserted no incoming radiation from infinity. 

\subsubsection{Spaces of Test Functions and the Conformal Anomaly}\label{footnote:Excuse}
The solutions $\phi_c$, $\phi_s$, $\phi_a$, and all of our subsequent solutions, should be understood in a distributional sense on a restricted space of Schwartz test functions $\calS_{\rm null}(\bbR^d) \subset \calS(\bbR^d)$. In particular, defect correlators need a restricted space of test functions to resolve pathological behaviours of integrals near the null defect and at infinity.\footnote{This is not completely unfamiliar. While tempered distributions $\calS'$ are argued to be sufficient to describe correlation functions of local operators in CFTs \cite{Kravchuk:2021kwe}, it is known that slightly larger spaces of distributions can still be interesting \cite{Jaffe:1967nb}, or even spaces as large as hyperfunctions \cite{Kontsevich:2021dmb} (see also \cite{wells1981hyperfunction}). Our restrictions are far more mild, and just involve shrinking the space of Schwartz functions to remove divergences -- a similar procedure removes IR divergences from zero modes in the (1+1)d free boson \cite{Streater:1989vi, Derezi_ski}.} Here we find this space, focusing on the causal solution for simplicity.

We want to define $\phi_c$ as the solution to:
\begin{gather}
    \Box_R \phi_c(x) = h \delta(x^+) \delta^{d-2}(x_i)\,,
\end{gather}
where $\Box_R$ is the d'Alembertian. The subscript reminds us that we consider the causal solution. Our goal is to make sense of $\phi_c(x)$ the same way we make sense of $\delta(x)$ -- as a ``component'' of the distribution $\delta$ satisfying $(\delta,f) = f(0)$. That is, we define $\phi_c$ so that for any Schwartz function $f \in \mathcal{S}(\bbR^d)$ we have
\begin{equation}\label{eq:distDefn}
    (\phi_c, f) = h \int dx^- (\Box_R^{-1}f)(x^-,0,0)\,.
\end{equation}

Since $f$ is a Schwartz test function, the inverse $\Box_R^{-1}f$ exists and actually belongs to $C^\infty(\bbR^d)$. In particular, we can consider $f$ integrated along the null ray $x^+ = x^\perp = 0$. Focusing on the special case when $d=4$, then
\begin{align}
    (\Box_R^{-1}f)(x^-,0,0) 
        &= -\frac{1}{2\pi} \int d^4y \, \delta(-2F^2(y^-\!-\!x^-)y^+ + y_\perp^2) \Theta(x^-\!-\!y^-\!-\!y^+) f(y)\,,\\
        &= -\frac{1}{2\pi} \int \frac{dy^- d^2y^\perp}{2F^2 |x^-\!-\!y^-|} \, \Theta(x^-\!-\!y^-\!-\!y_{*}^{+}) f(y^-,\tfrac{y_\perp^2}{2F^2 (y^-\!-\!x^-)}, y^\perp)\,.
\end{align}
For $x^- \gg 0$, we can write this as\footnote{Subleading contributions decay like $(x^-)^{-2}$ so it is sufficient to consider convergence for this leading $(x^-)^{-1}$ term.}
\begin{equation}
    (\Box_R^{-1}f)(x^-,0,0)
        \approx -\frac{1}{4\pi F^2 |x^-|} \int d^4y \, \delta(y^+) f(y^-,y^+, y^\perp)\,.
\end{equation}
Thus, for \eqref{eq:distDefn} to converge, we demand that
\begin{equation}
    \int d^4y \, \delta(y^+) f(y) = 0\,;
\end{equation}
in Fourier space this is $\int dk^- \hat{f}(k^-,0,0) = 0$. This restricts us to a smaller space of test functions, which we call $\calS_0(\bbR^4) \subset \calS(\bbR^4)$, and changes the space of distributions so that $c \delta(x^+)$ is now zero distributionally for any constant $c$. Morally speaking, ``$\calS_0(\bbR^4) = \ker(\delta(x^+))$,'' although this is not actually well-defined.

An important consequence of $c\delta(x^+)$ being zero is that the shockwave solution no longer introduces an artificial scale. In (3+1)d, shockwave solutions are of the form
\begin{equation}
    \phi_s(x) = \frac{h}{4\pi} \delta(x^+) \log \mu^2 x_\perp^2\,,
\end{equation}
for some scale $\mu^2$. Up to now, we may have been worried about the physical meaning of the scale $\mu^2$, but we see that
\begin{equation}
    \mu\frac{d}{d\mu} \phi_s(x) = \frac{h}{2\pi} \delta(x^+)\,,
\end{equation}
which is now zero distributionally. Thus, no scale is introduced. 

A similar phenomenon happens in usual CFT without defects. In momentum space, correlation functions can carry an explicit logarithmic cutoff: in even dimensions $\expval*{\calO(p)\calO(-p)} \sim p^d \log(\mu^2/p^2)$, see \cite{Gomis:2015yaa}. This does not violate scale invariance because the change of scale is proportional to a contact term, but does lead to a conformal anomaly, detectable on a non-trivial background. In our case, we have a similar conformal anomaly
\begin{gather}
    \expval{T^\mu_\mu(x) \phi(y)}_{\rm D} = \frac{h}{2\pi} \delta(y^+) \delta^{(d)}(x - y)\,.
\end{gather}
For comparison, a timelike defect would necessarily have $\expval*{T^\mu_\mu(x) \calO(y)}_{\rm D} = 0$.


We can repeat the same exercise with the test functions for a timelike defect to emphasize where the difference lies:
\begin{align}
    (\Box_R^{-1}f)(x^0) 
        &= -\int \frac{d^4y}{4\pi|\vec{y}|}\, \delta(x^0-y^0-|\vec{y}|)f(y)\\
        &= -\int \frac{d^3\vec{y}}{4\pi|\vec{y}|}\, f(x^0-|\vec{y}|,\vec{y})\,.
\end{align}
Since $f$ is Schwartz, it falls off quickly for large $|\vec{y}|$ and the integral converges automatically. Interpreted physically, a point $x^0$ in the far future is in causal contact with a large past-oriented cone, and any Schwartz function falls off quickly at large spatial distances, so the integral is manifestly zero. Conversely, in the lightlike case, the null-integral requires additional constraints because the Schwartz condition is not strong enough to guarantee vanishing.

Now we must make a further refinement of our space of test functions so that it interacts well with conformal symmetries. In particular, any vector field with polynomial coefficients will map a Schwartz function to a Schwartz function, so $\calS$ is invariant under conformal vector fields. The putative space of null defect test functions $\calS_0(\bbR^4)$ is not itself invariant under the null defect algebra. 

As a simple test, we can consider the transformation of $\delta(x^+)$ under the null defect algebra (since $\calS_0$ is morally the kernel of $\delta(x^+)$). A straightforward analysis shows that the space is invariant under all transformations except $K_i$ and $K_+$. For example, if $\delta(x^+)$ transforms with scaling weight $\gamma$, then:
\begin{equation}
    K_i: \delta(x^+) \mapsto x_i (\gamma-1)\delta(x^+)\,.
\end{equation}
We can equally consider successive powers of $K_i$ to similar results. To ensure that the action of $K_i$ doesn't spoil the convergence of our integrals, for generic scaling weight, we must further demand our space of test functions satisfy
\begin{gather}
   \int d^4 x\, \delta(x^+) x_2^{k_2} x_3^{k_3} \,f(x)= 0\,.
\end{gather}

$K_+$ invariance is a bit less straightforward. In this case, we can consider a finite $K_+$ transformation by an amount $\alpha$:
\begin{equation}
    \delta(x^+) \mapsto \frac{1}{(1+2\alpha x^-)^\gamma}\delta\left(x^+ - \alpha \frac{x_\perp^2}{F^2(1+2\alpha x^-))}\right)\,.
\end{equation}
We can expand this in series in $\alpha$ to obtain new conditions which must hold order-by-order on the restricted space of test functions. These conditions generally mix derivatives of $\delta$-functions at different orders. Enforcing $K_+$ invariance by making all of these conditions distibutionally zero defines our space of null-defect test-functions. We can write this schematically as:
\begin{equation}
    \calS_{\rm null}(\bbR^4) := \calS_{0}(\bbR^4)^{\mathfrak{n}_4}\,.
\end{equation}

\subsubsection{Switching on the Null Line}\label{sec:SwitchingOn}
To resolve some mysteries associated with the previous solution, let us understand how it changes when we include a weight function $h(x^-)$ on the current:
\begin{gather}
    \Box_R \phi_{c}(x) = h(x^-) \delta(x^+) \delta^{2}(x_i)\,. \label{eq:switchon}
\end{gather}
We also consider the adiabatic switching on of a null defect on the cylinder in Section \ref{sec:LorCyl}.

If we repeat the manipulations of the previous section and Appendix \ref{app:freeScalarCalcs} for this weighted current, we easily arrive at the solution
\begin{equation}
    \phi_c(x) = -\frac{h\left(\frac{-x^2}{2F^2x^+}\right)}{4\pi x^+}\Theta(x^+)\,.
\end{equation}
If we act with the Laplacian, we find
\begin{equation}\label{eq:switchon2}
    \Box_R \phi_c(x) = \frac{1}{2\pi F^2 x^+}h'\left(\frac{-x^2}{2F^2x^+}\right)\delta(x^+)\,.
\end{equation}
This looks very different from the expected RHS of \eqref{eq:switchon}. However, as distributions on the space of test functions $\mathcal{S}_{\rm null}(\mathbb{R}^4)$, they are actually the same.

To see this, integrate a test function $f \in \mathcal{S}_{\rm null}(\mathbb{R}^4)$ against the RHS of \eqref{eq:switchon2}, then
\begin{align}
    \int d^4x \, \Box_R \phi_c(x) f(x)
    &=\int dx^+ dx^- d^2x^\perp\, f(x^+,x^-,x^\perp) \frac{1}{2\pi F^2 x^+}h'\left(\frac{-x^2}{2F^2x^+}\right)\delta(x^+)\\
    &=\frac{1}{\pi}\int dx^+ dx^- d^2x^\perp \, f(x^+,x^-,\sqrt{2 x^+} F x^\perp) h'(x^--x_{\perp}^2)\delta(x^+)\\
    &=\int dx^- \, f(0,x^-,0) \left(h(x^-) - h(-\infty)\right)\,.
\end{align}
So it appears that
\begin{equation}
    \Box_R \phi_c(x) = \left(h(x^-) - h(-\infty)\right)\delta(x^+)\delta^2(x^\perp)\,.
\end{equation}
For our purported solution to be consistent, it means that $h(-\infty)$ must be $0$. We can think of the $\Theta(\scarex)$ appearing in the static solution as representing this fact: \textit{$\Theta(\scarex)$ is the same as $\Theta(x^+)$ with the understanding that it becomes $0$ when $-x^2/x^+$ goes to $-\infty$.} Alternatively, it means that our infinite static defect has strength $h=0$ at infinity.

\subsubsection{Symmetries of the Free Scalar Solutions}
Now we can explicitly consider the symmetries of the scalar solutions we found. Based on the analysis of Section \ref{sec:ScalarPrimary}, we expect a $\Delta=1$ scalar primary deformation to preserve $P_-$, $D$, $M_{ij}$, $K_i$, $K_-$ and $M_{-i}$ and break $K_+$ and $M_{+-}$. Since the theory is free, no classical or quantum interactions will spoil this.

Our causal solution, in regulated form is:
\begin{gather}
    \phi_c(x) = -\frac{1}{4\pi x^+}\frac{\Theta(x^+)}{\sqrt{1 + \delta^2/(\scarex)^2}}\,.
\end{gather}
$\phi_c$ should transform covariantly like a 1-point function under the above generators. This function is clearly symmetric with respect to $P_-$, and $M_{ij}$, and it clearly breaks $M_{+-}$ symmetry. We can check that the solution is invariant with respect to dilatations by a direct application of the differential operator
\begin{equation}
    \lim_{\delta \to 0}\,(\mathscr{D}_D\phi_c(x) + i \phi_c(x)) = -\frac{i}{4\pi}\delta(x^+)\,.
\end{equation}
Such a solution is $0$ on the space $\mathcal{S}_{\rm null}(\bbR^4)$ but not for general $f \in \mathcal{S}(\bbR^4)$.

It is in fact even easier to use the results of the previous section, representing $\phi_c$ with a general switching function:
\begin{equation}
    \phi_c(x) = -\frac{h\left(\frac{-x^2}{2F^2x^+}\right)}{4\pi x^+}\Theta(x^+)\,.
\end{equation}
Then we can apply the differential operators and replace $h$ by a constant. If we do this, we find that
\begin{align}
    [D,\phi_c(x)] 
        &= (\mathscr{D}_D + i)\phi_c(x) = -\frac{i h}{4\pi}\delta(x^+)\,,\\ 
    [K_i,\phi_c(x)] 
        &= (\mathscr{D}_{K_i} - 2 i x_i)\phi_c(x) = \frac{i h}{2\pi}x_i\delta(x^+)\,,\\ 
    [M_{-i},\phi_c(x)] 
        &= \mathscr{D}_{M_{-i}}\phi_c(x) = 0\,,\\ 
    [K_{-},\phi_c(x)] 
        &= (\mathscr{D}_{K_i} + 2 i F^2 x^+)\phi_c(x)  = 0\,,
\end{align}
so all solutions are zero or distributionally zero on the space $\mathcal{S}_{\rm null}(\bbR^4)$. By comparison, it is easy to confirm that the solution is not $K_+$ invariant by functional methods, as opposed to the shockplane and action arguments used earlier.

\subsection{Limits of Timelike and Spacelike Defects}\label{sec:limitsOfDefects}
Part of the rationale for studying null defects is to model ultra-boosted limits of (often charged) particles. As is well-known from e.g. studies of cusp anomalous dimension \cite{Correa:2012nk, Henn:2013wfa}, considering the null limit from both timelike and spacelike solutions, and analytic continuations, also generally reveals different features of the physics.

Consider the previous example of a free scalar in (3+1)d, but with a timelike source
\begin{equation}
    \partial^2\phi = h \delta(x^1)\delta^2(x^\perp)\,.
\end{equation}
The one-point function/classical solution in the presence of such a timelike defect is of course
\begin{equation}
    \phi_{\mathrm{timel.}}(x) = -\frac{h}{{4\pi}(x_1^2 + x_\perp^2)^{1/2}}\,.
\end{equation}

Now we boost in the $01$-plane, so that the particle is moving at finite speed $0 < u < 1$, and
\begin{equation}\label{eq:timelikeScalar}
    \phi_{u}(x) = -\frac{h}{4\pi(\gamma^2(x^1 + u x^0) + x_\perp^2)^{1/2}} = -\frac{h}{4\pi}\frac{\sqrt{1-u^2}}{R_u}\,,
\end{equation}
where $R_u^2:= (ux^0 + x^1)^2 + (1-u^2) x_\perp^2$ and $\gamma$ is the Lorentz factor. If we consider the infinite boost of the timelike solution, then we have
\begin{equation}
    \lim_{u\to 1} \phi_u(x) = -\frac{h}{4\pi\abs{x^\perp}}\delta_{x^+,0}\,,
\end{equation}
where $\delta_{x^+,0}$ is the Kronecker delta in $x^+$ (not Dirac delta), and so is distributionally zero. We have recorded some useful identities relevant for the ultrarelativistic limits in Appendix \ref{app:URLimits}.

We can repeat the same exercise, starting with a spacelike source:
\begin{equation}
    \partial^2 \phi = h \delta(x^0)\delta^2(x^\perp)\,.
\end{equation}
The solution is
\begin{equation}
    \phi_{\mathrm{spacel.}}(x) = -\frac{h}{2\pi((x^0)^2-x_\perp^2)^{1/2}} \Theta(x^0 - |x^\perp|)\,.
\end{equation}
Note the factor of $2\pi$ instead of $4\pi$, owing to the fact that there are two points on the defect in causal contact with any point in the future of the defect.

If we boost in the $01$-plane, so that the source approaches the lightcone, then
\begin{align}
    \phi_{u}(x) 
        &= -\frac{h}{2\pi(\gamma^2(x^0 + u x^1)-x_\perp^2)^{1/2}} \Theta\left(\tfrac{x^0 + u x^1}{\sqrt{1-u^2}}-|x^\perp|\right)\\
        &= -\frac{h}{2\pi} \frac{\sqrt{1-u^2}}{T_u} \Theta\left(\tfrac{x^0 + u x^1}{\sqrt{1-u^2}}-|x^\perp|\right)\,,
\end{align}
where $T_u^2 = (x^0 + ux^1)^2-(1-u^2)x_\perp^2$. If we consider the infinite boost of the spacelike solution to the lightcone, $\lim\limits_{u\to 1} \phi_{u}(x)$, a similar limit to the timelike case shows that the result is a Kronecker delta and thus distributionally zero. All of the distributional limits should be considered in the space $S'_{\rm null}(\mathbb{R}^4)$.

Physically, the reason these two solutions vanish in the ultra boosted limit can be understood from the fact that $h$ effectively scales under boosts like $\sqrt{1-u^2}$. If instead we scale the coupling $h \mapsto \tfrac{1}{\sqrt{2} F\sqrt{1-u^2}}h$, then the limit of the timelike solution becomes\footnote{The rescaling of $h$ turning Kronecker deltas into Dirac deltas can be understood mathematically by recalling the regulated forms of Dirac and Kronecker deltas e.g. in a heat-kernel regularization. Up to overall finite normalizations, the difference between the Dirac and Kronecker delta in the limit $\epsilon \to 0$ is the value at $0$. The Dirac delta diverges because the heat kernel regulator $\sim \epsilon^{-1/2}$. In our case, since $h$ effectively transforms with a power of $\sqrt{1-u^2} \sim \sqrt{\epsilon}$, it turns would-be Dirac deltas into Kronecker deltas.}
\begin{align}
    -\frac{h}{4\pi\sqrt{2}F} \lim_{u \to 1} \frac{1}{R_u}
        &= \frac{h}{4\pi}\left(\log(\mu^2 x_\perp^2) \delta(x^+) - \frac{1}{|x^+|}\right)\\
        &= h\delta(x^+)G_F^{(2)}(x^\perp) - \frac{h}{4\pi|x^+|}\,.
\end{align}
Rescaling the spacelike solution gives:
\begin{align}
    -\frac{h \sqrt{2} F}{2\pi} \lim_{u \to 1} \frac{1}{T_u}\Theta\left(\tfrac{x^0 + u x^1}{\sqrt{1-u^2}}-|x^\perp|\right) 
        &= -\frac{h}{4\pi} \left(\log(\mu^2 x_\perp^2) \delta(x^+)+ \frac{1}{|x^+|}\right)\Theta\left(\scarex\right)\\
        &= h\delta(x^+) G_F^{(2)}(x^\perp) - \frac{h}{4\pi|x^+|}\Theta\left(\scarex\right)\,.
\end{align}
In taking the limit, we used the ultrarelativistic limits in Appendix \ref{app:URLimits}.

We note that the limits produce terms which look like both our causal solution in \eqref{eq:scalarSolution} and our scalar shockwave in \eqref{eq:shockwaveScalar}. However, while the limit of the purely timelike solution correctly reproduces the shockwave, it produces ``both'' the causal and acausal solutions with a $|x^+|^{-1}$ term. This is to be expected: any point in spacetime is in the causal past and causal future of at least one point on any infinite timelike defect. Only in the strict $u \to 1$ limit does spacetime become split into two halves (plus the shockwave plane). Thus it is not surprising that the limit of the causally-indifferent regulator gives causally ambiguous results. On the other hand, the spacelike defect limits only to the causal solution because the regulator itself already involves a causal choice. We could just as easily have used the spacelike defect to produce the advanced solution.

In more general theories with, say, a timelike conformal line defect, the coupling $h$ can be replaced by the (measurable) one-point function coefficient that characterizes the conformal defect. If we boost the defect to obtain a null defect, we generically expect the limit of the one-point function to vanish (or diverge), since $h$ will effectively go to $0$ under the boost. As seen explicitly in Section \ref{sec:PinningSymmetries}, one exception to this expectation is the null Wilson line, which is completely $J_0$ and $\bar{J}_0$ invariant. On the other hand, we have also seen that if $h$ lives in some (conformal) family, then it is possible to rescale $h$ with the boost and have a one-point function survive. Thus we arrive at the following physical expectations:
\vskip 1em
\noindent \textit{Any ultra boosted conformal defect either}:
\begin{enumerate}
    \item \textit{has vanishing one-point functions,}
    \item \textit{or, is part of a defect conformal manifold (e.g. like the free theory; see below),}
    \item \textit{or, is a null Wilson line.}
\end{enumerate}
Some comments are in order. First, we do not preclude the existence of null line defects which are not obtainable from the limit of ultraboosted non-null line defects, but we have already seen in Section \ref{sec:WITariffs}, and next in Section \ref{sec:shockwaves}, that they are very strongly constrained. Second, statement 3 is obviously scheme dependent, so it should be understood as ``something that is kinematically like a Wilson line,'' as discussed in Section \ref{sec:WLKinematics} and references within. Third, we note that the ultraboosted limit of a timelike \textit{conformal} line defect may not even be conformal. The careful reader may have noticed this already in the free scalar example, since the (3+1)d free scalar solutions do not preserve $K_+$ symmetry, and hence are not even invariant under the $\mathfrak{sl}(2,\bbR)_-$ acting along the line. Instead there is only a scale invariance along the line. The loss of conformality in limiting solutions is to be expected from electrodynamics, where field strengths are known to ``pancake'' orthogonal to the motion of a charged particle, until they become supported entirely on the shockwave plane in the limit.

Finally, we might consider boosting other solutions. If we consider a timelike conformal defect with one bulk local operator $\calO_{\Delta}$ and one defect local operator $\hat{\calO}_{\hat{\Delta}}$, then the expectation value takes the form (up to $i\epsilon$-prescription)
\begin{gather}
    \expval*{\hat{\calO}_{\hat{\Delta}}(\tau,0) \calO_{\Delta}(t,\vec{x})} = \frac{1}{(-(t-\tau)^2+\vec{x}^{\,2})^{\hat{\Delta}} |\vec{x}|^{\Delta - \hat{\Delta}}  }\,,
\end{gather}
and vanishes under boosts. When $\Delta \neq \hat{\Delta}$, the two-point function vanishes in the ultra-boosted limit unless it can be rescaled, as before. When $\Delta = \hat{\Delta}$, the two-point function is Lorentz invariant and survives the boost without any scaling. So the only two-point functions which are non-trivial are those where the bulk and defect primary are ``the same operator.'' At this point, we might consider using ultra-boosted limits of timelike correlation functions to start defining defect correlation functions and primaries. Unfortunately, this is not so straightforward, since the rescalings of the defect primaries $\hat{\calO}_{\hat{\Delta}}$ would depend on the probe scaling/spin (as above). It would be interesting to make sense of this in future work.

\subsection{1-Point Functions Start as Shockwaves}\label{sec:shockwaves}
To find a one-point function in the presence of a defect, classically, we first solve the Green's problem for the equations of motion, then integrate the Green's function against the source to get our specific solution. We did this in the case of the free scalar in (3+1)d in Section \ref{sec:ExampleFreeScalar}. While the free scalar example was a useful warmup, we might ask how generic the causal, acausal, and shockwave solutions were.

Consider some general free field theory $\Phi(x)$ sourced by $J(x) = h \delta(x^+)\delta^{d-2}(x^\perp)$, then the causal and acausal solutions are obtained by integrating
\begin{equation}
    \Phi_{c/a}(x) = \int d^d\tilde{x}\,  G_{R/A}(x-\tilde{x})J(\tilde{x})\,.
\end{equation}
The difference in the two solutions is
\begin{align}
    \Phi_c(x) - \Phi_a(x) 
        &= \int d^d\tilde{x}\, (G_R(x-\tilde{x})-G_A(x-\tilde{x}) J(\tilde{x})\\ 
        &= -ih\int d\tilde{x}^-\, \mel*{0}{[\Phi(x),\Phi(0,\tilde{x}^-,0)]}{0}\label{eq:solutionDiff1}\\
        &= -ih\mel*{0}{\,[\Phi(x),\mathbb{L}[\Phi]]\,}{0}\label{eq:solutionDiff2}\,.
\end{align}
In passing from \eqref{eq:solutionDiff1} to \eqref{eq:solutionDiff2} we have identified the integral of $\Phi(0,\tilde{x}^-,0)$ with the light ray operator $\mathbb{L}[\Phi]$. When $\Delta + J > 1$, the light ray transform converges and $\mathbb{L}[\Phi]$ annihilates the vacuum on the left and on the right \cite{Kravchuk:2018htv}.\footnote{To be more precise, each piece in the commutator contributes the same delta- function contact term, which is cancelled in the commutator.} Thus we have
\begin{equation}
    \Phi_c(x) = \Phi_a(x)\,,\quad \text{when $\Delta+ J > 1$}\,.
\end{equation}
The only way that $\Phi_c$ and $\Phi_a$ can be equal and be non-zero is if they are $\delta$-supported on their common shockwave plane $x^+ = 0$. Thus when $\Delta + J > 1$, we expect free-field solutions to be shockwaves.

We can actually consider a more general source term and bulk local operator. Start by integrating $\calO$ along $x^+ = 0$ and $x^\perp = 0$, then our interaction Hamiltonian is
\begin{equation}
    H_{\mathrm{int}} = \int d^dx\, \delta(x^+) \delta^2(x^\perp) \calO(x^+,x^-,x^\perp) = \int dx^- \calO(0,x^-,0)\,.
\end{equation}
In $x^+$-ordering, this can be viewed as a ``time-dependent'' perturbation to the free Hamiltonian, where the entire Hamiltonian changes at time $x^+ = 0$. So we expect some Kubo-like formula to describe the perturbation.

For any bulk local operator $\calB$, we can compute the causal and acausal one-point functions in a state $\Psi$ using a Schwinger-Keldysh prescription:
\begin{align}
    \expval{\calB(x)}_{\Psi,c} 
        &:= \mel*{\Psi}{\bar{\calT}(e^{i h \int H_{\mathrm{int}}})\calT(\calB e^{-i h \int H_{\mathrm{int}}})}{\Psi}\,,\label{eq:causalKubo}\\
    \expval{\calB(x)}_{\Psi,a} 
        &:= \mel*{\Psi}{\calT(e^{-i h \int H_{\mathrm{int}}})\bar{\calT}(e^{i h \int H_{\mathrm{int}}}\calB )}{\Psi}\,\label{eq:acausalKubo}.
\end{align}

Let us expand \eqref{eq:causalKubo} in powers of $h$. To order $O(h^2)$, we have (dropping the outer $\bra{\Psi}$ and $\ket{\Psi}$ for brevity):
\begin{align}
    \expval{\calB(x)}_{\Psi,c}
        &= \calB(x) + i h \int dy_1^- \left(\calO(y_1^-)\!\calB(x)-\calT(\calB(x)\!\calO(y_1^-)\right)\nonumber\\
        &-\frac{h^2}{2} \int dy_1^- dy_2^- \left(\calT(\calB(x)\!\calO(y_1^-)\!\calO(y_2^-))\right.\\
        &\hphantom{-\frac{h^2}{2} \int dy_1^- dy_2^-}\left.-2\calO(y_1^-)\!\calT(\calB(x)\!\calO(y_2^-))+\bar{\calT}(\calO(y_1^-)\!\calO(y_2^-))\calB(x)\right) \nonumber\,.
\end{align}
As a sanity check, note that the $O(h^1)$ term does simplify to something resembling (the integral of) a retarded propagator
\begin{equation}
    -i h \int dy_1^- \, [\calB(x),\calO(y_1^-)]\, \Theta(x^0-y_1^0)\,.
\end{equation}
The first two terms in the expansion give
the Kubo formula for the expectation value of an operator with a time-dependent perturbation in quantum mechanics \cite{Kubo:1957mj}. A similar calculation for the acausal one-point function \eqref{eq:acausalKubo}, gives:
\begin{align}
    \expval{\calB(x)}_{\Psi,a}
        &= \calB(x) - i h \int dy_1^- \left(\calO(y_1^-)\!\calB(x)-\bar{\calT}(\calO(y_1^-)\!\calB(x)\right)\\
        &-\frac{h^2}{2} \int dy_1^- dy_2^- \left(\bar{\calT}(\calO(y_1^-)\!\calO(y_2^-)\!\calB(x))\right.\\
        &\hphantom{-\frac{h^2}{2} \int dy_1^- dy_2^-}\left.-2\calO(y_1^-)\bar{\calT}(\calO(y_2^-)\!\calB(x))+\calT(\calO(y_1^-)\!\calO(y_2^-))\calB(x)\right) \nonumber\,.
\end{align}

As before, we can now consider their difference (dropping the outer states again for simplicity):
\begin{gather}
    \expval*{\calB(x)}_{\Psi,c} - \expval*{\calB(x)}_{\Psi,a}
        = - i h \int dy_1^-\, [\calB(x),\calO(y_1^-)]\\
        - \frac{h^2}{2} \int dy_1^- dy_2^- \left(\calT(\calB(x)\!\calO(y_1^-)\!\calO(y_2^-)) - \bar{\calT}(\calO(y_1^-)\!\calO(y_2^-)\!\calB(x)) -2\calO(y_1^-)\calT(\calB(x)\!\calO(y_2^-)) \right. \nonumber\\
        \hphantom{- \frac{h^2}{2} \int dy_1^- dy_2^-}\left.+2\calO(y_1^-)\bar{\calT}(\calO(y_2^-)\!\calB(x))+\calT(\calO(y_1^-)\!\calO(y_2^-))\calB(x)+\bar{\calT}(\calO(y_1^-)\!\calO(y_2^-))\calB(x)\right) \nonumber
\end{gather}
By the same arguments as in the free field case, the first term is a light ray correlator in the state $\Psi$. In states killed by the light ray operator, like the vacuum $\Psi = \Omega$, we thus expect the leading term in $h$ to be a shockwave solution. Higher order interaction terms do generically lead to corrections on top of the shockwave solution, but as we saw in Section \ref{sec:WITariffs}, if the effects of interactions also preserve symmetries, they are quite strongly constrained.

\subsection{Example: Li\'enard-Wiechart and the Null Wilson Line}\label{sec:NullWilsonLine}
Perhaps the most important and physical example is the massless charged particle. In principle, just as a standard Wilson line can be considered as the trajectory of a very heavy massive particle, we can similarly consider a null Wilson line to be the trajectory of a charged massless particle (with some conflict with the intuition of it being ``very heavy''). In particular, pure (3+1)d Maxwell theory is conformal \cite{El-Showk:2011xbs}, and so we expect that the null Wilson line should be an example of a maximally symmetric conformal line defect based on Section \ref{sec:PinningSymmetries}.

Consider a charged particle moving along the infinite left-moving straightline trajectory $x^\mu = \sigma v^\mu$. Then we are interested in the action
\begin{equation}
    S = \int d^4x\, \left(-\frac{1}{4} F_{\mu\nu} F^{\mu\nu} - A_\mu J^\mu\right)\,,\quad
    J^\mu = g F^2 v^\mu \delta(x\cdot \bar{v})\delta^{2}(x^\perp)\,,
\end{equation}
where we introduce the velocities
\begin{equation}
    v := (1,-u,0,0)\,,\quad
    \bar{v} := (-u,1,0,0)\,,
\end{equation}
where $0\leq u<1$. Clearly the static charge corresponds to $u=0$ and the massless charge corresponds to the limit $u = 1$. In these important extremes, the only non-trivial components of $J^\mu$ are:
\begin{alignat}{2}
    u &= 0:\qquad 
        &J_{0}^0 &= g F^2 \delta(x^1)\delta^2(x^\perp)\,,\\
    u &= 1:\qquad 
        &J_{1}^- &= g \delta(x^+)\delta^2(x^\perp)\,,
\end{alignat}
where the subscripts denote the values of $u$.

The simplest gauge-invariant observable we can compute is $F_{\mu\nu}$, corresponding to the one-point function $\expval*{F_{\mu\nu}}$ in the quantum theory. It is instructive to compare the direct solution of the massless ($u=1$) problem to the one obtained by solving $0\leq u<1$ and taking the limit as $u\to 1$.

\subsubsection{The Ultraboosted Limit of a Charged Particle}
We start by considering the infinite boost limit of a relativistic charged particle. The textbook solution is obtained by working in Lorenz gauge $\partial_\mu A^\mu_u = 0$ and solving the corresponding Green's problem. In Lorenz gauge, the potential satisfies the equation of motion $\partial^2 A_u^\mu = J_u^\mu$, and a straightforward calculation with the retarded propagator gives the well-known Li\'enard-Wiechart solution:\footnote{One can also reconsider the formal procedure in Appendix \ref{app:freeScalarCalcs} and see exactly where this case differs from a lightlike source: when solving for $\tilde{x}^{0*}$.}
\begin{equation}
    A_u^\mu = -\frac{g F^2}{4\pi} \frac{v^\mu}{R_u}\,,
\end{equation}
where $R_u^2:= (ux^0 + x^1)^2 + (1-u)^2 x_\perp^2$.

We can compute the corresponding field strength $F_{\mu\nu} = \partial_\mu A_\nu - \partial_\nu A_\mu$, it gives:
\begin{equation}\label{eq:massiveChargedF}
    F^u_{\mu\nu} = -\frac{gF^2}{4\pi}\frac{1-u^2}{R_u^3}(v_\mu x_\nu - x_\mu v_\nu)\,.
\end{equation}
Famously, the electric and magnetic fields pancake outwards in the orthogonal directions to the line of charge.

Next we consider the limit as $u \to 1$. In key distinction to the scalar in \eqref{eq:timelikeScalar}, the boost weight of $F_{\mu\nu}$ means the limit survives and is a Dirac delta function (not Kronecker delta) without any rescaling of the coupling $g$. Using the ultrarelativistic limits in Appendix \ref{app:URLimits}, we find that the limiting solution is a shockwave:
\begin{equation}\label{eq:massiveChargedFLimit}
    F_{+i} = \frac{g F^2}{2\pi} \frac{x_i}{|x_\perp|^2} \delta(x^+)\,.
\end{equation}
As seen in previous examples, this does not preserve conformal symmetry since the shockwave solution necessarily breaks $K_+$.

If we consider the limit of the gauge potential, we have \cite{Jackiw:1991ck}:
\begin{equation}\label{eq:ALimit}
    A_1^- = \frac{gF^2}{2\pi}\left(\log(\mu^2 x_\perp^2) \delta(x^+) - \frac{1}{|x^+|}\right)\,.
\end{equation}
Taking the field strength, the first term of \eqref{eq:ALimit} reproduces the shockwave solution \eqref{eq:massiveChargedFLimit}. The second piece is pure gauge and does not contribute to the field strength. This is in contradistinction to the limit of the spacelike particle, where the regulator comes naturally equipped with $\Theta(\scarex)$, which does change the field strength (as we see in the next section).

Now we note a crucial difference between timelike and null defects. For timelike defects, we expect our space of test functions to coincide with the Schwartz functions $\mathcal{S}(\mathbb{R}^4)$. But, as explained in Section \ref{footnote:Excuse}, null defects demand a smaller space of test functions $\calS_{\rm null}(\bbR^4)$, so that the limiting gauge potential $A_1^-$ is a distribution in $\calS'_{\rm null}(\bbR^4)$. Thus, we claim that \textit{$A_1^{-}$ has no arbitrary scale dependence on $\mu^2$.} Moreover, a direct check of the symmetries of $F_{\mu\nu}$ shows it satisfies all Ward identities except $K_i$ and $K_+$. The violation of $K_i$ Ward identities is up to terms proportional to $\delta(x^+)$, but it is $A_{\mu}$ which is a distribution on $\calS_{\rm null}(\bbR^4)$, not $F_{\mu\nu}$, so this is sensible.  

\subsubsection{Direct Solution for the Massless Charged Particle}
We can try to study the massless charged particle by working directly with the current
\begin{equation}
    J_{1}^- = g \delta(x^+)\delta^2(x^\perp)\,.
\end{equation}
If we work in a Lorenz gauge $\partial^2 A_1^\mu = J_1^\mu$, then the only non-trivial equation is $\partial^2 A_1^- = J_1^-$. The solution using the retarded propagator is
\begin{equation}
    A_{1,c}^-(x^+,x^-,x^\perp) = -\frac{g}{4\pi x^+}\Theta(\scarex)\,,
\end{equation}
or using the Feynman propagator
\begin{equation}
    A_{1,s}^-(x^+,x^-,x^\perp) = g\delta(x^+)G_F^{(2)}(x^\perp)\,.
\end{equation}

Our $A_{1,c}^-$ is almost the solution written in \cite{Jackiw:1991ck} and (2.24) of \cite{Azzurli:2014lha} (see also \cite{Aichelburg:1970dh}). Previous authors obtain $\Theta(x^+)$, instead of $\Theta(\scarex)$, then correctly note that their result is not a solution of the equations of motion/Lorenz gauge condition, and the issues are chalked-up to the non-existence of certain distributional limits. Our solution does solve the Lorenz gauge condition and, as seen from the scalar analysis, matches limits of different solutions. A major difference is that we consider the restricted space of test functions $\calS_{\rm null}(\bbR^4)$, where other authors consider the full Schwartz space $\calS(\bbR^4)$. Thus we claim that the introduction of at least $\calS_{\rm null}(\bbR^4)$, and possibly even more restrictive spaces of test functions, is necessary for understanding defect theories and, in particular, certain limits of test particles in classical electromagnetism.

The shockwave solution is identical to the gauge potential in the previous example, so we focus on the new causal solution $A_{1,c}^-(x^+,x^-,x^\perp)$ and drop the subscripts $1$ and $c$ for brevity. Unlike the previous case, the causal solution is not obviously pure gauge because of the $\Theta(\scarex)$. In this case, the only non-trivial gauge field component is $A_+ = -F^2A^-$.

Let us compute the field strength corresponding to $A$ by using the smeared coupling, as we did for the scalar in \eqref{eq:switchon}. In that case, we assume that the charge of the particle varies in $x^-$ so that
\begin{equation}
    A^- 
        = -\frac{1}{4\pi x^+}g\left(\frac{-x^2}{2F^2x^+}\right)\Theta(x^+)\,.
\end{equation}
In this case, we have
\begin{align}
    F_{-+} 
        &= \frac{F^2}{4\pi x^+}g'\left(\frac{-x^2}{2F^2x^+}\right)\Theta(x^+)
        = \frac{F^4 x^+}{2\pi x_\perp^2} \partial_+ g\left(\frac{-x^2}{2F^2x^+}\right)\Theta(x^+)\,,
        \\
    F_{+i} 
        &= \frac{x^i}{4\pi (x^+)^2}g'\left(\frac{-x^2}{2F^2x^+}\right)\Theta(x^+)
        = \frac{F^2 x_i}{2\pi x_\perp^2}\partial_+ g\left(\frac{-x^2}{2F^2x^+}\right)\Theta(x^+)\,.
\end{align}

Now we can pick a particular switching function, representing it by:
\begin{equation}
    A^- 
        = \lim_{\lambda \to -\infty}\frac{-g}{4\pi x^+}\Theta\left(\frac{-x^2}{2F^2x^+}-\lambda \right)\Theta(x^+)\,.
\end{equation}
With this switching function, we have that
\begin{equation}
    \partial_+ g\left(\frac{-x^2}{2F^2x^+}\right)\Theta(x^+) = g \, \delta\left(x^+ - \frac{x_\perp^2}{2F^2(x^- - \lambda)} \right)\,.
\end{equation}
And, in the limit that the defect switches on in the infinite past, $\lambda \to -\infty$, the field strengths become
\begin{align}
    F_{-+} 
        &= \frac{F^4 x^+}{2\pi x_\perp^2} \partial_+ g\left(\frac{-x^2}{2F^2x^+}\right)\Theta(x^+)
        = \frac{g F^4 x^+}{2\pi x_\perp^2} \delta(x^+)
        =0\,,\\
    F_{+i} 
        &= \frac{F^2 x_i}{2\pi x_\perp^2}\partial_+ g\left(\frac{-x^2}{2F^2x^+}\right)\Theta(x^+)
        = \frac{g F^2}{2\pi}\frac{x_i}{|x_\perp|^2}\delta(x^+)\,,
\end{align}
which exactly matches the previous shockwave solution. It may seem somewhat surprising that the causal gauge potential produces the shockwave field strength. Mathematically, the causal gauge potential is supported on $x^+ > 0$ and so produces a non-trivial function at the shockwave plane under taking $\partial_+$, but physically the reason is still opaque and warrants a better explanation.

Since the solutions match, they must be gauge equivalent $A^\mu_c = A^\mu_s + \partial^\mu \varepsilon$. Specifically, as they both satisfy the Lorenz gauge, the gauge parameter must be a harmonic function $\Box \varepsilon = 0$. In \eqref{eq:radiation} we gave the asymptotic expansion of $A_c$ along $\scri^+$, and the two asymptotic gauge fields clearly differ: for the causal solution the field is non-zero at infinity, while in the shockwave it is supported on the ray $x^+ = 0$ (i.e. $\theta = \pi$). Thus the harmonic function $\varepsilon$ is not just the Fourier transform of the PDE $\Box \varepsilon = 0$, which privileges $r^{-1}$ fall-offs (see also Section 2.12 of \cite{Strominger:2017zoo} for relevant discussions). On the other hand, at least for generic generators/angles, the fields naively appear to have the same (zero) memory, possibly with a special exception along $\theta = \pi$.

\begin{figure}[t]
  \centering
  \begin{minipage}{0.48\textwidth}
        \centering
  \begin{tikzpicture}[scale=2]
    \pgfmathsetmacro{\piVal}{3.14159}
    \pgfmathsetmacro{\qual}{10}

    \draw[gray!50,thin] (-\piVal/2,0)--(\piVal/2,0);

    \def\timeList{-10,-4, 0, 4,10}
    \foreach \t in \timeList {
      \draw[gray!50,thin,domain=-40:40,samples=15*\qual,variable=\x]
        plot (
          { ( atan(\t + \x) - atan(\t - \x) )/2 * \piVal/180 },
          { ( atan(\t + \x) + atan(\t - \x) )/2 * \piVal/180 }
        );
    }
    \def\timeList{-2.,-1.34,1.34, 2.}
    \foreach \t in \timeList {
      \draw[gray!50,thin,domain=-30:30,samples=30*\qual,variable=\x]
        plot (
          { ( atan(\t + \x) - atan(\t - \x) )/2 * \piVal/180 },
          { ( atan(\t + \x) + atan(\t - \x) )/2 * \piVal/180 }
        );
    }
    \def\timeList{-0.83, -0.4, 0.4, 0.83}
    \foreach \t in \timeList {
      \draw[gray!50,thin,domain=-20:20,samples=40*\qual,variable=\x]
        plot (
          { ( atan(\t + \x) - atan(\t - \x) )/2 * \piVal/180 },
          { ( atan(\t + \x) + atan(\t - \x) )/2 * \piVal/180 }
        );
    }

    \begin{scope}[rotate=90]
      \draw[gray!50,thin] (-\piVal/2,0)--(\piVal/2,0);
      \def\timeList{-10,-4,4,10}
      \foreach \t in \timeList {
        \draw[gray!50,thin,domain=-40:40,samples=15*\qual,variable=\x]
          plot (
            { ( atan(\t + \x) - atan(\t - \x) )/2 * \piVal/180 },
            { ( atan(\t + \x) + atan(\t - \x) )/2 * \piVal/180 }
          );
      }
      \def\timeList{-2.,-1.34,1.34,2.}
      \foreach \t in \timeList {
        \draw[gray!50,thin,domain=-30:30,samples=30*\qual,variable=\x]
          plot (
            { ( atan(\t + \x) - atan(\t - \x) )/2 * \piVal/180 },
            { ( atan(\t + \x) + atan(\t - \x) )/2 * \piVal/180 }
          );
      }
      \def\timeList{-0.83,-0.4,0.4,0.83}
      \foreach \t in \timeList {
        \draw[gray!50,thin,domain=-20:20,samples=40*\qual,variable=\x]
          plot (
            { ( atan(\t + \x) - atan(\t - \x) )/2 * \piVal/180 },
            { ( atan(\t + \x) + atan(\t - \x) )/2 * \piVal/180 }
          );
      }
    \end{scope}

\def\smax{15}
\def\alphaListLow{0,-0.75} 

\foreach \alpha in \alphaListLow {
  \foreach \lw/\op in {3.0/0.1, 2.5/0.2, 1/1} {
    \draw[red, line width=\lw pt, opacity=\op, domain=-\smax:\smax, samples=200, variable=\s]
      plot (
        { ( atan( (\s*cosh(\alpha) + \s*sinh(\alpha)) )
            - atan( (\s*cosh(\alpha) - \s*sinh(\alpha)) )
          )/2 * \piVal/180 },
        { ( atan( (\s*cosh(\alpha) + \s*sinh(\alpha)) )
            + atan( (\s*cosh(\alpha) - \s*sinh(\alpha)) )
          )/2 * \piVal/180 }
      );
  }

  \draw[very thick, red, domain=-\smax:\smax, samples=30*\qual, variable=\s]
    plot (
      { ( atan( (\s*cosh(\alpha) + \s*sinh(\alpha)) )
          - atan( (\s*cosh(\alpha) - \s*sinh(\alpha)) )
        )/2 * \piVal/180 },
      { ( atan( (\s*cosh(\alpha) + \s*sinh(\alpha)) )
          + atan( (\s*cosh(\alpha) - \s*sinh(\alpha)) )
        )/2 * \piVal/180 }
    );

  \foreach \lw/\op in {3.0/0.1, 2.5/0.2, 1/1} {
    \draw[red, line width=\lw pt, opacity=\op]
      ({ (atan(\smax*cosh(\alpha)+\smax*sinh(\alpha)) - atan(\smax*cosh(\alpha)-\smax*sinh(\alpha)))/2 * \piVal/180 },
       { (atan(\smax*cosh(\alpha)+\smax*sinh(\alpha)) + atan(\smax*cosh(\alpha)-\smax*sinh(\alpha)))/2 * \piVal/180 })
      -- (0,\piVal/2);
    \draw[red, line width=\lw pt, opacity=\op]
      ({ (atan(-\smax*cosh(\alpha)-\smax*sinh(\alpha)) - atan(-\smax*cosh(\alpha)+\smax*sinh(\alpha)))/2 * \piVal/180 },
       { (atan(-\smax*cosh(\alpha)-\smax*sinh(\alpha)) + atan(-\smax*cosh(\alpha)+\smax*sinh(\alpha)))/2 * \piVal/180 })
      -- (0,-\piVal/2);
  }
}

\draw[very thick] (-\piVal/2,0) -- (0,\piVal/2) -- (\piVal/2,0) -- (0,-\piVal/2) -- cycle;

\def\alphaHigh{-2.0}
\foreach \lw/\op in {3.0/0.1, 2.5/0.2, 1/1} {
  \draw[red, line width=\lw pt, opacity=\op, domain=-\smax:\smax, samples=200, variable=\s]
    plot (
      { ( atan( (\s*cosh(\alphaHigh) + \s*sinh(\alphaHigh)) )
          - atan( (\s*cosh(\alphaHigh) - \s*sinh(\alphaHigh)) )
        )/2 * \piVal/180 },
      { ( atan( (\s*cosh(\alphaHigh) + \s*sinh(\alphaHigh)) )
          + atan( (\s*cosh(\alphaHigh) - \s*sinh(\alphaHigh)) )
        )/2 * \piVal/180 }
    );
}

\draw[very thick, red, domain=-\smax:\smax, samples=30*\qual, variable=\s]
  plot (
    { ( atan( (\s*cosh(\alphaHigh) + \s*sinh(\alphaHigh)) )
        - atan( (\s*cosh(\alphaHigh) - \s*sinh(\alphaHigh)) )
      )/2 * \piVal/180 },
    { ( atan( (\s*cosh(\alphaHigh) + \s*sinh(\alphaHigh)) )
        + atan( (\s*cosh(\alphaHigh) - \s*sinh(\alphaHigh)) )
      )/2 * \piVal/180 }
  );

\foreach \lw/\op in {3.0/0.1, 2.5/0.2, 1/1} {
  \draw[red, line width=\lw pt, opacity=\op]
    ({ (atan(\smax*cosh(\alphaHigh)+\smax*sinh(\alphaHigh)) - atan(\smax*cosh(\alphaHigh)-\smax*sinh(\alphaHigh)))/2 * \piVal/180 },
     { (atan(\smax*cosh(\alphaHigh)+\smax*sinh(\alphaHigh)) + atan(\smax*cosh(\alphaHigh)-\smax*sinh(\alphaHigh)))/2 * \piVal/180 })
    -- (0,\piVal/2);
  \draw[red, line width=\lw pt, opacity=\op]
    ({ (atan(-\smax*cosh(\alphaHigh)-\smax*sinh(\alphaHigh)) - atan(-\smax*cosh(\alphaHigh)+\smax*sinh(\alphaHigh)))/2 * \piVal/180 },
     { (atan(-\smax*cosh(\alphaHigh)-\smax*sinh(\alphaHigh)) + atan(-\smax*cosh(\alphaHigh)+\smax*sinh(\alphaHigh)))/2 * \piVal/180 })
    -- (0,-\piVal/2);
}

\def\squeak{0.015}
\draw[very thick, white] (-\piVal/2-\squeak,\squeak) -- (-\squeak,\piVal/2+\squeak);
\draw[very thick, white] (\squeak,-\piVal/2-\squeak) -- (\piVal/2+\squeak,-\squeak);

    \node at (0,\piVal/2 + 0.1) {$i^+$};
    \node at (0,-\piVal/2 - 0.1) {$i^-$};
    \node at (0.95,0.95) {$\mathscr{I}^+$};
    \node at (0.9,-0.9) {$\mathscr{I}^-$};
    \node at (\piVal/2+0.1,0) {$i^0$};

\foreach \scale/\op in {0.7/0.1, 0.6/0.5, 0.4/1} {
  \fill[blue, opacity=\op] (-\piVal/2,0) circle[radius=\scale*0.05];
}

  \end{tikzpicture}
  \end{minipage}
\begin{minipage}{0.48\textwidth}
        \centering
  \begin{tikzpicture}[scale=2]
    \pgfmathsetmacro{\piVal}{3.141592653589793}
    \pgfmathsetmacro{\qual}{10}

    \draw[gray!50,thin] (-\piVal/2,0)--(\piVal/2,0);

    \def\timeList{-10,-4, 0, 4,10}
    \foreach \t in \timeList {
      \draw[gray!50,thin,domain=-40:40,samples=15*\qual,variable=\x]
        plot (
          { ( atan(\t + \x) - atan(\t - \x) )/2 * \piVal/180 },
          { ( atan(\t + \x) + atan(\t - \x) )/2 * \piVal/180 }
        );
    }
    \def\timeList{-2.,-1.34,1.34, 2.}
    \foreach \t in \timeList {
      \draw[gray!50,thin,domain=-30:30,samples=30*\qual,variable=\x]
        plot (
          { ( atan(\t + \x) - atan(\t - \x) )/2 * \piVal/180 },
          { ( atan(\t + \x) + atan(\t - \x) )/2 * \piVal/180 }
        );
    }
    \def\timeList{-0.83, -0.4, 0.4, 0.83}
    \foreach \t in \timeList {
      \draw[gray!50,thin,domain=-20:20,samples=40*\qual,variable=\x]
        plot (
          { ( atan(\t + \x) - atan(\t - \x) )/2 * \piVal/180 },
          { ( atan(\t + \x) + atan(\t - \x) )/2 * \piVal/180 }
        );
    }

    \begin{scope}[rotate=90]
      \draw[gray!50,thin] (-\piVal/2,0)--(\piVal/2,0);
      \def\timeList{-10,-4,4,10}
      \foreach \t in \timeList {
        \draw[gray!50,thin,domain=-40:40,samples=15*\qual,variable=\x]
          plot (
            { ( atan(\t + \x) - atan(\t - \x) )/2 * \piVal/180 },
            { ( atan(\t + \x) + atan(\t - \x) )/2 * \piVal/180 }
          );
      }
      \def\timeList{-2.,-1.34,1.34,2.}
      \foreach \t in \timeList {
        \draw[gray!50,thin,domain=-30:30,samples=30*\qual,variable=\x]
          plot (
            { ( atan(\t + \x) - atan(\t - \x) )/2 * \piVal/180 },
            { ( atan(\t + \x) + atan(\t - \x) )/2 * \piVal/180 }
          );
      }
      \def\timeList{-0.83,-0.4,0.4,0.83}
      \foreach \t in \timeList {
        \draw[gray!50,thin,domain=-20:20,samples=40*\qual,variable=\x]
          plot (
            { ( atan(\t + \x) - atan(\t - \x) )/2 * \piVal/180 },
            { ( atan(\t + \x) + atan(\t - \x) )/2 * \piVal/180 }
          );
      }
    \end{scope}

\draw[very thick] (-\piVal/2,0) -- (0,\piVal/2) -- (\piVal/2,0) -- (0,-\piVal/2) -- cycle;
\draw[very thick] (-\piVal/2,-\piVal/2) -- (-\piVal/2,\piVal/2);
\draw[very thick] (\piVal/2,-\piVal/2) -- (\piVal/2,\piVal/2);

\foreach \lw/\op in {3.0/0.1, 2.5/0.2, 1/1} {
  \draw[red, line width=\lw pt, opacity=\op] (0,-\piVal/2) -- (\piVal/4,-\piVal/4);
  \draw[blue, line width=\lw pt, opacity=\op] (\piVal/2,0) -- (\piVal/4,-\piVal/4);
  \draw[red, line width=\lw pt, opacity=\op] (-\piVal/4,\piVal/4) -- (\piVal/4,-\piVal/4);
  \draw[blue, line width=\lw pt, opacity=\op] (-\piVal/2,0) -- (-\piVal/4,\piVal/4);
  \draw[red, line width=\lw pt, opacity=\op] (-\piVal/4,\piVal/4) -- (0,\piVal/2);
  \draw[blue, line width=\lw pt, opacity=\op] (-\piVal/4,\piVal/4) -- (-\piVal/2,\piVal/2);
  \draw[blue, line width=\lw pt, opacity=\op] (\piVal/4,-\piVal/4) -- (\piVal/2,-\piVal/2);
}

    \node at (0,\piVal/2 + 0.1) {$i^+$};
    \node at (0,-\piVal/2 - 0.1) {$i^-$};
    \node at (0.95,0.95) {$\mathscr{I}^+$};
    \node at (0.75,-1) {$\mathscr{I}^-$};
    \node at (\piVal/2+0.1,0) {$i^0$};

  \end{tikzpicture}
  \end{minipage}
  \caption{Left, the worldline (red) of static and various boosted massive charged particles. As the boost increases, the charged particle bends along an approximately null trajectory, but still ends at timelike infinities $i^\mp$, forcing segments to lie along null infinity $\scri^{\pm}$ and form a ``zig-zag'' pattern. Right, Minkowski space embeds as a patch on the Lorentzian cylinder, an oppositely charged particle (blue) zig-zags under boost and forms a null square in the limit.}
  \label{fig:superBoost}
\end{figure}
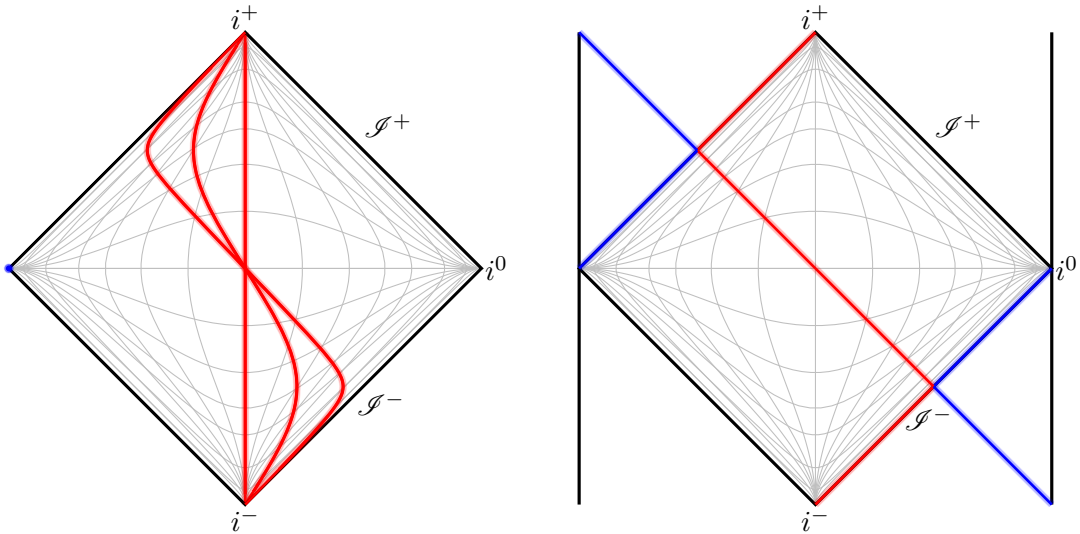

The fact that the field profiles appear different at infinity is perhaps expected. Roughly speaking, when we draw the worldline of a timelike particle and boost it, it bends as in Figure \ref{fig:superBoost}. As we boost $u \to 1$, we recover a lightlike source, but with parts of the original timelike worldline pushed to $\scri^{\pm}$ in the limit. There is also a second related conceptual issue: with our source in Minkowski space, there is nowhere for flux lines to end. For general theories, this is not a problem, and charge flux can escape to infinity. But Minkowski space can be conformally mapped to the Lorentzian cylinder $\bbR \times S^{d-1}$, with compact spatial slices. In this case, there should be another charge for flux lines to end on. In the case of two timelike lines, two oppositely charged particles would travel antipodally up the cylinder, and the worldline of the charge sink would be at spacelike infinity $i^0$ in the conformal compactification of the original Minkowski spacetime (blue in Figure \ref{fig:superBoost}). In the ultraboosted limit, the configuration describes the creation and annihilation of massless charged particles along null horizons. We discuss this in more detail in Section \ref{sec:ZZCC}.

\section{Semi-Infinite Defects, Candy Canes, and the Perfect Null Polygon}\label{sec:OtherResults}
In this section we consider some extensions of the previous results to more complicated scenarios. In Section \ref{sec:FiniteDefects}, we consider semi-infinite line defects. This is reasonable since any infinite defect is an idealization, realistically, a defect must start and end at some point. Moreover, the re-summation of soft radiation into null Wilson lines produces only semi-infinite lines, not infinite lines. In Section \ref{sec:FiniteDefects}, we consider semi-infinite lines and try to solve the scalar equations of motion in $4-\epsilon$ dimensions. Then, in Sections \ref{sec:LorCyl} and \ref{sec:ZZCC} we turn to consider defects on the cylinder. In Section \ref{sec:LorCyl} we give an exact solution for the conformally coupled scalar and comment on the importance of representing our idealized infinite defects by adiabatically turning on solutions. In Section \ref{sec:ZZCC} we discuss null Wilson lines on the cylinder and show that the ultraboosted limit of the charged particle naturally corresponds to a maximally symmetric ``perfect null polygon'' around the boundary of Poincar\'e patches on the cylinder.

\subsection{Semi-Infinite Defects and Interacting Fixed Points}\label{sec:FiniteDefects}
We consider a semi-infinite future-pointing line defect, stretched along $x^+ = 0$, $x^\perp = 0$, and with $x^- > 0$. The semi-infinite line explicitly breaks $P_-$ and $K_+$ symmetry in the maximal conformal algebra $\mathfrak{n}_d$. Thus we have that the maximally symmetric conformal semi-line preserves: $M_{+-}$, $M_{ij}$, $K_i$, $K_-$, $M_{-i}$, and $D$. We might try to see if there are interesting correlation functions in the presence of a semi-infinite line.

A straightforward exercise, mimicking Section \ref{sec:WITariffs}, reveals that even the semi-line conformal symmetries are still extremely constraining for one-point functions. For example, demanding $D$, $M_{+-}$ and $K_-$ symmetries immediately forces scalar correlation functions to be functions of $|x_\perp|$, and anything more makes them trivial. We will not repeat the same exhaustive exercise from before, instead we discuss the simple case of a scalar primary in the presence of a ``minimal'' scale-invariant defect.

From the UV definition of the pinning null defect we expect that some form of scaling $J_\Delta'$, transverse SCTs $K_i$, transverse rotations $M_{ij}$ and boosts $M_{i-}$ will be respected. As we see shortly, if we try to preserve $K_i$ along the flow, we find trivial correlation functions, so any interacting example must still sacrifice some symmetries. Actually, this is expected from studying loop corrections, where we inevitably find UV and IR divergences that can only be resolved by an appropriate regularization of the defect and thus breaking of defect symmetries.

From the most minimal point of view we could try to preserve twisted scaling $J_\Delta'$ and transverse rotations $M_{ij}$. This constrains the one-point function of a scalar field of dimension $\delta$ to take the form
\begin{align}\label{eq:ansatzPhi4}
    \Delta &\neq 2\,, \quad 
        (x^+)^{-\frac{\delta}{2-\Delta}} f\left(\frac{(x^-)^{2-\Delta}}{(x^+)^\Delta}, \frac{|x_\perp|^{2-\Delta}}{x^+}\right)\,,\\
    \Delta &= 2\,,\quad
    (x^-)^{-\frac{\delta}{2}}f\left(x^+, \frac{|x_\perp|^2}{x^+}\right)\,.
\end{align}
Now we consider what other symmetric solutions we would like. For example, we might also demand that our solution preserves the transverse Lorentz boosts $M_{-i}$. If we do that, some straightforward manipulations confirm that a general $M_{ij}$, $J_\Delta'$ and $M_{-i}$ symmetric ansatz is of the form
\begin{align}\label{eq:ansatzPhi41}
    \Delta &\neq 2\,, \quad 
        (x^+)^{-\frac{\delta}{2-\Delta}} f((x^+)^{\frac{2}{\Delta-2}}(x^2))\,,\\
    \Delta &= 2\,,\quad
    (x^2)^{-\frac{\delta}{2}} f(x^+)\,,\label{eq:ansatzPhi42}
\end{align}
where $x^2$ is the usual Lorentz-invariant distance. 


\subsubsection{Example: Interacting Scalar in \texorpdfstring{$d=4-\epsilon$}{d=4-epsilon}}\label{sec:phi4Theory}
Now let us turn specifically to the semi-infinite defect in the Wilson-Fisher theory in $d=4-\epsilon$ dimensions. The action is
\begin{gather}
    S = \int d^d x \left[\frac12 \left(\partial_\mu \phi\right)^2 + \frac{\lambda_*}{4!} \phi^4\right] + h \int^\infty_0 dx^- \phi(0,x^-,0)\,, \quad \lambda_* = \frac{16\pi^2 \epsilon}{3}\,.
\end{gather}
In $d$-dimensions, the solution of the free field equation is 
\begin{gather}\label{eq:freeWF}
    \phi_0 = \frac{h \Gamma\left(\frac{d-2}{2}\right)}{4\pi^\frac{d-2}{2}}\frac{\Theta(-x^2) \Theta(x^+)\left(-x^2\right)^\frac{4-d}{2}}{x^+}\,, \quad \Box \phi_0 = h \delta(x^+) \delta^{d-2}(x^\perp) \Theta(x^-)\,.
\end{gather}
Now we can solve order by order in the Wilson-Fisher theory in $d=4-\epsilon$. We note that this free solution is compatible with the ansatz \eqref{eq:ansatzPhi41}.


At first order in perturbation theory, we have
\begin{equation}
    \Box_{R,\epsilon}\phi_1 = \lambda \phi_0^3\,,
\end{equation}
where $\lambda = -\frac{16\pi^2 \epsilon}{3}$. As a brute force approach, we can try to solve
\begin{gather}
    \phi_1(x) 
        \propto \lambda \int d^d\tilde{x}\, G_R(x-\tilde{x}) [\phi_0(\tilde{x})]^3\\
    \propto -\lambda h^3 \int d\tilde{x}^+ d\tilde{x}^- d^{d-2}\tilde{x}^\perp\, \frac{\left(-\tilde{x}^2\right)^{3\frac{4-d}{2}}}{(\tilde{x}^+)^3 \left((x -\tilde{x})^2\right)^\frac{d-2}{2}} \Theta(x^0-\tilde{x}^0)\Theta(\tilde{x}^+)\Theta(\tilde{x}^-)\Theta(-\tilde{x}^2)\,.
\end{gather}
This last integral is UV divergent. Indeed, even the free solution is singular near $x = 0$. We can resolve the issue by regularizing the source, but any (unitary) regularization scheme will break some symmetries of the idealized source. We will choose to keep our solution boost invariant (i.e. $M_{ij}$ and $M_{-i}$). Thus, the resulting one point function in the first order of perturbation theory has the following form
\begin{equation}
\begin{aligned}
    \phi(x) 
        &= \frac{h \Gamma\left(\frac{d-2}{2}\right)}{4\pi^\frac{d-2}{2}} \frac{\left(-x^2\right)^\frac{4-d}{2} \Theta(-x^2)\Theta(x^+)}{x^+}\\ 
        &+ \frac{\lambda}{2(d-3)(3d-14)} \left(\frac{h \Gamma\left(\frac{d-2}{2}\right)}{4\pi^\frac{d-2}{2}}\right)^3 \frac{(-x^2)^\frac{14-3d}{2}\Theta(-x^2)\Theta(x^+)}{\left(x^+\right)^3 }\,.
\end{aligned}
\end{equation}

We can now improve our perturbation theory by considering the limit
\begin{gather}
    \epsilon\to 0\,,\quad 
    h\to\infty\,, \quad\text{keeping} \quad 
    \lambda_*h^2 = \rm{fixed}\,.
\end{gather}
After resummation, we expect that our one point function is a function of the form 
\begin{gather}
    \phi(x) = \frac{h(-x^2)^\frac{4-d}{2}}{x^+}\Phi(z)\,,\quad
    z := \lambda_* h^2\frac{(-x^2)^{5-d}}{(x^+)^2}\,.
\end{gather}
Note that this ansatz is different from the one derived in the previous subsection. This is because we are working in perturbation theory in the leading $\epsilon$ expansion; it's only when we have the non-perturbative resummed contributions would the preceding ansatz be valid. After this, we can resum leading divergences to find solve the following equation for $\Phi(z)$
\begin{gather}
 2 (d-3) \left(2 (d-5) z \Phi ''(z)+(3 d-14) \Phi '(z)\right)+\Phi (z)^3 = 0 \,.
\end{gather}
In the leading $\epsilon\to 0$ limit, we get the following equation
\begin{gather}
    4 z \Phi''(z) + 4 \Phi'(z) - \Phi^3(z)= 0\,, \quad \Phi(0) = 1\,.
\end{gather}
The same equation actually arises in the case of timelike defect in $d=4-\epsilon$ dimensions, and the solution of this equation is 
\begin{gather}
    \Phi(z) \sim \frac{1}{\sqrt{z}}\,, \quad z\to\infty\,.
\end{gather}
That leads to the answer of the form
\begin{gather}
    \phi(x) = \frac{1}{\sqrt{- \lambda_* x^2}}\,,
\end{gather}
meaning that the field $\phi$ only sees the start of the defect.

\subsection{An Adiabatic Defect on the Lorentzian Cylinder}\label{sec:LorCyl}
CFTs in Lorentzian signature naturally live on the Lorentzian cylinder $\bbR \times S^{d-1}$, obtained as the universal cover of the conformal compactification of Minkowski spacetime. We use the conventions of \cite{Luscher:1974ez, Kravchuk:2018htv}, where $(\tau, \hat{e})$ are global coordinates on the Lorentzian cylinder, with sphere of unit radius, and usual Minkowski space with coordinates $\{x^0,x^1,\dots, x^{d-1}\}$ is embedded in a Poincar\'e patch by
\begin{equation}
    x^0 = \frac{\sin\tau}{\cos\tau + e^d}\,,\quad x^i = \frac{e^i}{\cos\tau + e^d}\,.
\end{equation}
This is similar to the usual plane-cylinder relation in Euclidean space, except that Minkowski space only covers a patch of the Lorentzian cylinder. Indeed, the usual Minkowski metric is
\begin{equation}
    ds^2_{\bbR^{1,d-1}} = \frac{1}{(\cos\tau + e^d)^2}(-d\tau^2 + d\Omega_{d-1}^2) = \frac{1}{(\cos\tau + e^d)^2}ds_{\bbR \times S^{d-1}}^2\,,
\end{equation}
and the Wightman functions on the plane and the cylinder are related by
\begin{equation}
    \mel*{\Omega}{\calO_1(x_1) \cdots \calO_n(x_n)}{\Omega}_{\bbR^{1,d-1}}
        = \prod_{i=1}^n (\cos \tau_i+ e_i^d)^{\Delta_i} \mel*{\Omega}{\calO_1(\tau_1,\hat{e}_1) \cdots \calO_n(\tau_n,\hat{e}_n)}{\Omega}_{\bbR \times S^{d-1}}\,.
\end{equation}

\paragraph{A Timelike Scalar Defect on $\bbR\times S^{d-1}$.} Let's consider a conformally coupled scalar on the cylinder. The action of a conformally coupled scalar is
\begin{equation}
    S = \frac{1}{2} \int d^dx \sqrt{-g}((\partial_A\phi)^2 + \tfrac{d-2}{4(d-1)} R \phi^2)\,.
\end{equation}
On our unit cylinder, the equations of motion become
\begin{equation}
    \left(-\partial_\tau^2 + \Delta_{S^{d-1}} - \tfrac{(d-2)^2}{4}\right)\phi(\tau,\Omega) = 0\,,
\end{equation}
where $\Delta_{S^{d-1}}$ is the spherical Laplacian. The eigenfunctions are $Y_{\ell,\vec{m}}(\Omega)$ satisfying $\Delta_{S^{d-1}} Y_{\ell, \vec{m}} = - \ell(\ell+d-2) Y_{\ell,\vec{m}}$. See \cite{Nagano:2021tbu} for more helpful formulas.

As a warm-up, let's start by considering a timelike defect on $\bbR \times S^{d-1}$. The equation of motion is
\begin{equation}
    \left(-\partial_\tau^2 + \Delta_{S^{d-1}} - \tfrac{(d-2)^2}{4}\right)\phi(\tau,\Omega) = -h(\tau) \delta_{S^{d-1}}(\Omega,\rm{NP})\,,
\end{equation}
where $\delta_{S^{d-1}}(\Omega,0)$ is the sphere delta-function, fixing $\Omega$ to the North Pole. Here, we have allowed $h(\tau)$ to depend on time to represent the switching time of the defect. We can write a solution on the sphere $\phi(\tau,\Omega) = \phi_{\ell,\vec{m}}(\tau) Y_{\ell,\vec{m}}(\Omega)$, so the PDE becomes
\begin{equation}
    \left(\partial_\tau^2 + M_\ell^2\right)\phi_{\ell,\vec{m}}(\tau) = h(\tau)Y_{\ell,\vec{m}}^*(\rm{NP})\,,
\end{equation}
where $M_{\ell} = \tfrac{d-2}{2}+\ell$ and $Y_{\ell,\vec{m}}^*({\rm{NP}}) = c \delta_{\vec{m},0}$. Plugging in an infinitely long static current, $h(\tau) = h$, the corresponding static solution is
\begin{equation}
    \phi_{\ell,\vec{m}}^{\mathrm{static}}(\tau) = \frac{h}{M_{\ell}^2} c \delta_{\vec{m},0}\,.
\end{equation}
More generally, the causal solution for a general switching function is formally:
\begin{equation}
    \phi_{\ell,\vec{m}}^{c}(\tau) = \frac{1}{M_{\ell}}\int_{-\infty}^{\tau} \!\!d\tau'\, \sin(M_\ell (\tau-\tau'))Y_{\ell,\vec{m}}^*(\Omega(\tau'))h(\tau')\,.
\end{equation}

In a realistic scenario, we cannot create an infinitely long defect: we must turn it on at some time. Indeed, we have already seen that properly turning on defects fixes many issues already issues in Section \ref{sec:SwitchingOn}. Setting $h(\tau') = h \Theta(\tau')$ above, the general solution gives:
\begin{equation}
    \phi_{\ell,\vec{m}}^{c}(\tau) = \frac{h}{M_\ell^2}(1-\cos(M_\ell \tau))\Theta(\tau) c\delta_{\vec{m},0}\,.
\end{equation}
Thus creating the defect instantaneously in time leads to non-dissipating ripples that propagate forever into the future.  
\begin{figure}[t]
\centering




\includegraphics[width=\textwidth]{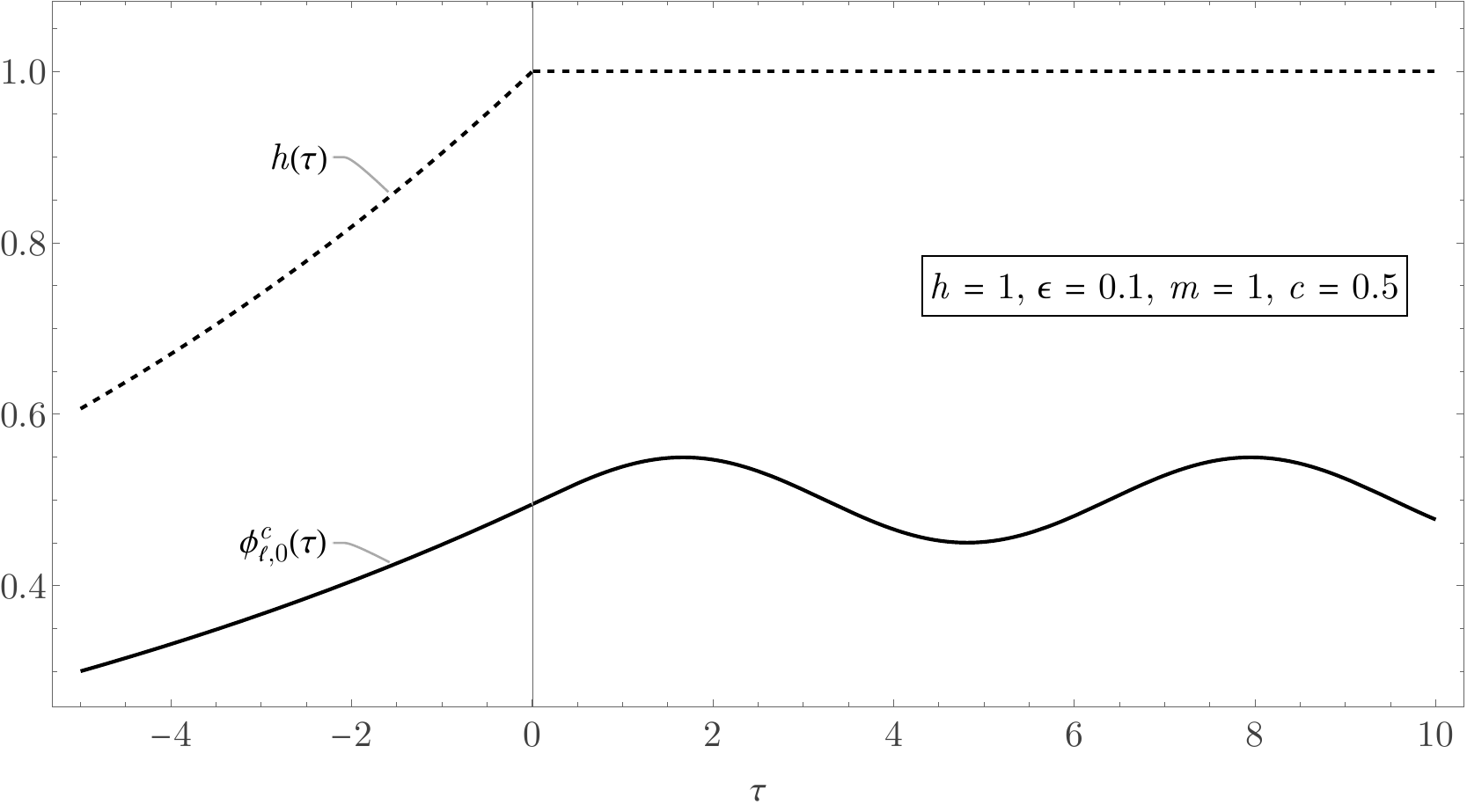}
\caption{Plots of $h(\tau)$ and $\phi_{\ell,0}^c(\tau)$ for an adiabatic defect with a specific choice of parameters.}
\label{fig:adiabatic}
\end{figure}

To cure this, we must smoothly turn on the defect. For example, if we use the interpolating current
\begin{equation}
    h(\tau) = h\Theta(\tau) + h e^{\epsilon \tau} \Theta(-\tau)\,,
\end{equation}
then the solution is
\begin{equation}
    \phi_{\ell,0}^c(\tau) = \frac{hc \,e^{\epsilon \tau}}{M_{\ell}^2+\epsilon^2} \Theta(-\tau) + \frac{hc}{M_{\ell}^2+\epsilon^2}\left(1+\frac{\epsilon}{M_{\ell}^2}(M_{\ell} \sin(M_{\ell} \tau) + \epsilon (1-\cos(M_{\ell}\tau))\right)\Theta(\tau)\,,
\end{equation}
as shown in Figure \ref{fig:adiabatic}. In the limit as $\epsilon \to 0$, this solution limits to the static solution, with small oscillations of order $\epsilon$ from the static solution. The key take away is that a stable defect should be created by adiabatically turning on the source from $0$ strength.

\paragraph{Moving Trajectories on $\bbR \times S^3$.} Now we specialize to (3+1)d on the cylinder $\bbR \times S^3$ and consider an angled trajectory $\Omega_0(\tau)$
\begin{equation}
    \left(\partial_\tau^2 - \Delta_{S^{3}} + 1\right)\phi(\tau,\Omega) = h(\tau) \delta_{S^3}(\Omega,\Omega_0(\tau))\,.
\end{equation}
Instead of hyperspherical harmonics, let us use Wigner D-functions as our basis for $L^2(S^3)$ and switch to group notation for $S^3 \cong SU(2)$. 

The Wigner D-functions give an orthonormal basis for $L^2(SU(2))$ \cite{tung1985group}. They are functions of $g \in SU(2)$
\begin{equation}
    D_{m_1m_2}^{j}(g)\,,\quad |m_i| \leq j\,,\quad j \in \tfrac{1}{2}\bbZ_{\geq 0}\,,
\end{equation}
satisfying the orthonormality conditions 
\begin{equation}
    \int dg\, \overline{D^{j'}_{m'_1 m'_2}(g)} D^{j}_{m_1 m_2}(g) = \frac{1}{2j+1} \delta^{jj'}\delta_{m_1 m_1'}\delta_{m_2 m_2'}\,.
\end{equation}
We have used a normalized Haar measure on $SU(2)$ i.e. $\int_{SU(2)} dg = 1$ so that the $\delta$-function satisfies $\int dg' \delta(g,g') f(g') = f(g)$. By unitarity, $\overline{D^{j}_{m_1 m_2}(g)} = D^j_{m_2,m_1}(g^{-1})$. As standard harmonic analysis would suggest, they are eigenfunctions of the Laplacian
\begin{equation}
    \Delta_{S^3} D_{m_1 m_2}^{j}(g) = -4j(j+1)D_{m_1m_2}^j(g)\,,
\end{equation}
and give a resolution of identity on $SU(2)$
\begin{equation}
    \delta(g,g') = \sum_{j} (2j+1) \sum_{m_1,m_2} \overline{D^{j}_{m_1 m_2}(g')} D^{j}_{m_1 m_2}(g)\,.
\end{equation}

On $\bbR\times S^3$, the cylinder equation of motion is now
\begin{equation}
    \left(\partial_\tau^2 - \Delta_{SU(2)} + 1\right)\phi(\tau,g) = h(\tau) \delta(g,g_0(\tau))\,.
\end{equation}
Expanding $\phi(\tau,g)$ in terms of D-functions, we have
\begin{equation}
    \phi(\tau,g) = \sum_{j} (2j+1) \sum_{m_1,m_2} \phi_{m_1 m_2}^{j}(\tau) D^{j}_{m_1m_2}(g)\,,
\end{equation}
so that the ODE for a particular mode $D_{m_1m_2}^j$ is
\begin{equation}
    (\partial_\tau^2 + M_{j}^2)\phi_{m_1m_2}^{j}(\tau) = h(\tau) D_{m_2m_1}^j(g_0^{-1}(\tau))
\end{equation}
where $M_j^2 = 4j(j+1)+1$. We again use the retarded Green's function to find the general solution
\begin{equation}
    \phi_{m_1 m_2}^j(\tau) = \frac{1}{M_j} \int_{-\infty}^{\tau} \!\!d\tau'\,\sin(M_j (\tau-\tau')) D_{m_2 m_1}^{j}(g_{0}^{-1}(\tau')) h(\tau')\,.
\end{equation}
We warn the reader that $\tau$ is not the proper time of the defect.

A lightlike trajectory on the Lorentzian cylinder $\bbR \times S^3$ can be realized by wrapping a Hopf fibre of $S^3 \cong S^2 \times S^1$ in time, positioned over the North poles of an $S^2$ \cite{Alday:2007mf}
\begin{equation}
    g_0(\tau) 
        = \begin{pmatrix}
            e^{i\tau} & 0\\
            0 & e^{-i\tau}
        \end{pmatrix}\,,\quad
    D_{m_2 m_1}^j(g_0^{-1}(\tau)) 
        = \delta_{m_1 m_2} e^{-2i m_1 \tau} \,.
\end{equation}
In this case, the only non-trivial components, for any switching function, will be the diagonal modes $\phi_{mm}^j(\tau)$, i.e.\footnote{If we parametrize an arbitrary $SU(2)$ group element
\begin{equation}
    g = 
    \begin{pmatrix}
        a & b\\
        -\bar{b} & \bar{a}
    \end{pmatrix}
    = 
    \begin{pmatrix}
        e^{-i(\alpha+\gamma)/2}\cos(\beta/2) & -e^{-i(\alpha-\gamma)/2}\sin(\beta/2)\\
        e^{i(\alpha-\gamma)/2}\sin(\beta/2) &e^{i(\alpha+\gamma)/2}\cos(\beta/2)
    \end{pmatrix}\,,
\end{equation}
where $\alpha,\gamma \in [0,2\pi)$ and $\beta \in [0,\pi]$, then it is helpful to know that
\begin{equation}
    D_{mm}^j(g) = a^{m+\abs{m}}\bar{a}^{\abs{m}-m}P_{j-\abs{m}}^{(0,2\abs{m})}(\abs{a}^2 - \abs{b}^2) = e^{-im(\alpha+\gamma)}\cos(\tfrac{\beta}{2})^{2\abs{m}}P_{j-\abs{m}}^{(0,2\abs{m})}(\cos\beta)\,,
\end{equation}
where $P^{(\alpha,\beta)}_n$ is a Jacobi polynomial. 
}
\begin{equation}
    \phi(\tau,g) 
        = \sum_j (2j+1) \sum_m \phi_{mm}^j(\tau) D^j_{mm}(g)\,.
\end{equation}

For the infinitely long static current $h(\tau)=h$ and/or solution with the adiabatic switch, the modes are:
\begin{equation}
    \phi_{mm}^j(\tau) = \frac{h}{M_j^2 - 4m^2}e^{-2im\tau}\,.
\end{equation}
The solution looks potentially resonant for some values of $m$, but such a resonance can never actually occur. We can write down an exact solution
\begin{align}
    \phi(\tau,g) 
        = \sum_{j\in \frac{1}{2}\mathbb{N}^0} (2j+1) \sum_{m=-j}^{j}\frac{h}{M_j^2-4m^2} e^{im(\alpha+\gamma-2\tau)}\cos(\frac{\beta}{2})^{2\abs{m}}P_{j-\abs{m}}^{(0,2\abs{m})}(\cos\beta)\,.
\end{align}
The solution converges, but not absolutely. This is obvious numerically, but can also be seen analytically. Consider, for example, the special points exactly opposite the defect, by taking $(\alpha,\beta,\gamma) = (\alpha,\pi,\gamma)$. In this case, $D^j_{mm}(\alpha,\pi,\gamma) = (-1)^j\delta_{m0}$ and the sum simplifies to 
\begin{equation}
    \phi(\tau,\alpha,\pi,\gamma) = h\sum_{j\in \mathbb{N}^0}(2j+1) \frac{(-1)^j}{M_j^2} = h\frac{\pi}{4}\,.
\end{equation}
However, the sum diverges absolutely like the harmonic series.


\subsection{Gauss' Law, Candy Canes, and the Perfect Null Polygon}\label{sec:ZZCC}
\begin{figure}[t]
  \centering
  \begin{minipage}{0.32\textwidth}
    \centering
\begin{tikzpicture}[scale=1]
    \pgfmathsetmacro{\piVal}{3.141592653589793}
\draw[very thick] (-\piVal/2,0+\piVal/2) -- (0,\piVal/2+\piVal/2) -- (\piVal/2,0+\piVal/2) -- (0,-\piVal/2+\piVal/2) -- cycle;
\draw[very thick] (-\piVal/2,0-\piVal/2) -- (0,\piVal/2-\piVal/2) -- (\piVal/2,0-\piVal/2) -- (0,-\piVal/2-\piVal/2) -- cycle;
\draw[very thick] (-\piVal/2,-\piVal) -- (-\piVal/2,\piVal);
\draw[very thick] (\piVal/2,-\piVal) -- (\piVal/2,\piVal);
\foreach \lw/\op in {3.0/0.1, 2.5/0.2, 1/1} {
  \draw[red, line width=\lw pt, opacity=\op] (0,-\piVal) -- (0,\piVal);
  \draw[blue, line width=\lw pt, opacity=\op] (-\piVal/2,-\piVal) -- (-\piVal/2,\piVal);
}
\draw[red, line width=0pt, ->-=0.27 rotate 0] (0,-\piVal) -- (0,\piVal);
\draw[red, line width=0pt, ->-=0.77 rotate 0] (0,-\piVal) -- (0,\piVal);
\draw[blue, line width=0pt, ->-=0.27 rotate 0] (-\piVal/2,-\piVal) -- (-\piVal/2,\piVal);
\draw[blue, line width=0pt, ->-=0.77 rotate 0] (-\piVal/2,-\piVal) -- (-\piVal/2,\piVal);


  \end{tikzpicture}
  \subcaption{Timelike charges.}\label{fig:antipodalTime}
  \end{minipage}
  \begin{minipage}{0.32\textwidth}
    \centering
\begin{tikzpicture}[scale=1]
    \pgfmathsetmacro{\piVal}{3.141592653589793}
\draw[very thick] (-\piVal/2,0+\piVal/2) -- (0,\piVal/2+\piVal/2) -- (\piVal/2,0+\piVal/2) -- (0,-\piVal/2+\piVal/2) -- cycle;
\draw[very thick] (-\piVal/2,0-\piVal/2) -- (0,\piVal/2-\piVal/2) -- (\piVal/2,0-\piVal/2) -- (0,-\piVal/2-\piVal/2) -- cycle;
\draw[very thick] (-\piVal/2,-\piVal) -- (-\piVal/2,\piVal);
\draw[very thick] (\piVal/2,-\piVal) -- (\piVal/2,\piVal);
\foreach \lw/\op in {3.0/0.1, 2.5/0.2, 1/1} {
  \draw[red, line width=\lw pt, opacity=\op] (\piVal/2,-\piVal) -- (-\piVal/2,0);
  \draw[red, line width=\lw pt, opacity=\op] (\piVal/2,-\piVal+\piVal) -- (-\piVal/2,0+\piVal);
  \draw[blue, line width=\lw pt, opacity=\op] (0,-\piVal) -- (0-\piVal/2,-\piVal+\piVal/2);
  \draw[blue, line width=\lw pt, opacity=\op] (\piVal/2, -\piVal/2) -- (-\piVal/2,\piVal/2);
  \draw[blue, line width=\lw pt, opacity=\op] (0,\piVal) -- (0+\piVal/2,+\piVal-\piVal/2);
}
\draw[red, line width=0pt, ->-=0.53 rotate 0] (\piVal/2,-\piVal) -- (-\piVal/2,0);
\draw[red, line width=0pt, ->-=0.53 rotate 0] (\piVal/2,-\piVal+\piVal) -- (-\piVal/2,0+\piVal);
\draw[blue, line width=0pt, ->-=0.53 rotate 0] (0,-\piVal) -- (0-\piVal/2,-\piVal+\piVal/2);
\draw[blue, line width=0pt, ->-=0.77 rotate 0] (\piVal/2, -\piVal/2) -- (-\piVal/2,\piVal/2);
\draw[blue, line width=0pt, ->-=0.27 rotate 0] (\piVal/2, -\piVal/2) -- (-\piVal/2,\piVal/2);
\draw[blue, line width=0pt, ->-=0.53 rotate 0] (0+\piVal/2,+\piVal-\piVal/2) -- (0,\piVal);


  \end{tikzpicture}
  \subcaption{Lightlike charges.}\label{fig:antipodalNull}
  \end{minipage}
  \begin{minipage}{0.32\textwidth}
    \centering
\begin{tikzpicture}[scale=1]
    \pgfmathsetmacro{\piVal}{3.141592653589793}
\draw[very thick] (-\piVal/2,0+\piVal/2) -- (0,\piVal/2+\piVal/2) -- (\piVal/2,0+\piVal/2) -- (0,-\piVal/2+\piVal/2) -- cycle;
\draw[very thick] (-\piVal/2,0-\piVal/2) -- (0,\piVal/2-\piVal/2) -- (\piVal/2,0-\piVal/2) -- (0,-\piVal/2-\piVal/2) -- cycle;
\draw[very thick] (-\piVal/2,-\piVal) -- (-\piVal/2,\piVal);
\draw[very thick] (\piVal/2,-\piVal) -- (\piVal/2,\piVal);
\foreach \lw/\op in {3.0/0.1, 2.5/0.2, 1/1} {
  \draw[red, line width=\lw pt, opacity=\op] (0,-\piVal)-- (\piVal/2-\piVal/4,-\piVal+\piVal/4);
    \draw[red, line width=\lw pt, opacity=\op] (\piVal/2-\piVal/4,-\piVal+\piVal/4) -- (-\piVal/2+\piVal/4,-\piVal/4);
    \draw[red, line width=\lw pt, opacity=\op] (-\piVal/2+\piVal/4,-\piVal/4) -- (\piVal/4,\piVal/4);
    \draw[red, line width=\lw pt, opacity=\op] (\piVal/2-\piVal/4,\piVal/4) -- (-\piVal/2+\piVal/4,-\piVal/4+\piVal);
    \draw[red, line width=\lw pt, opacity=\op] (-\piVal/2+\piVal/4,-\piVal/4+\piVal) -- (0,\piVal);
    \draw[blue, line width=\lw pt, opacity=\op] (\piVal/2,-\piVal)--(\piVal/4,-\piVal+\piVal/4);
    \draw[blue, line width=\lw pt, opacity=\op] (\piVal/4,-\piVal+\piVal/4) -- (\piVal/2,-\piVal/2);
    \draw[blue, line width=\lw pt, opacity=\op] (-\piVal/2,-\piVal/2) -- (-\piVal/4,-\piVal/4);
    \draw[blue, line width=\lw pt, opacity=\op] (-\piVal/4,-\piVal/4) -- (-\piVal/2,0);
    \draw[blue, line width=\lw pt, opacity=\op] (\piVal/2,0) -- (\piVal/4,\piVal/4);
    \draw[blue, line width=\lw pt, opacity=\op] (\piVal/4,\piVal/4) -- (\piVal/2,\piVal/2);
    \draw[blue, line width=\lw pt, opacity=\op] (-\piVal/2,\piVal/2) -- (-\piVal/2+\piVal/4,\piVal/2+\piVal/4);
    \draw[blue, line width=\lw pt, opacity=\op] (-\piVal/2+\piVal/4,\piVal/2+\piVal/4) -- (-\piVal/2,\piVal);
}
\draw[red, line width=0pt, ->-=0.70 rotate 0] (0,-\piVal)-- (\piVal/2-\piVal/4,-\piVal+\piVal/4);
\draw[red, line width=0pt, ->-=0.38 rotate 0] (\piVal/2-\piVal/4,-\piVal+\piVal/4) -- (-\piVal/2+\piVal/4,-\piVal/4);
\draw[red, line width=0pt, ->-=0.84 rotate 0] (\piVal/2-\piVal/4,-\piVal+\piVal/4) -- (-\piVal/2+\piVal/4,-\piVal/4);
\draw[red, line width=0pt, ->-=0.38 rotate 0] (-\piVal/2+\piVal/4,-\piVal/4) -- (\piVal/4,\piVal/4);
\draw[red, line width=0pt, ->-=0.84 rotate 0] (-\piVal/2+\piVal/4,-\piVal/4) -- (\piVal/4,\piVal/4);
\draw[red, line width=0pt, ->-=0.38 rotate 0] (\piVal/2-\piVal/4,\piVal/4) -- (-\piVal/2+\piVal/4,-\piVal/4+\piVal);
\draw[red, line width=0pt, ->-=0.84 rotate 0] (\piVal/2-\piVal/4,\piVal/4) -- (-\piVal/2+\piVal/4,-\piVal/4+\piVal);
\draw[red, line width=0pt, ->-=0.70 rotate 0] (-\piVal/2+\piVal/4,-\piVal/4+\piVal) -- (0,\piVal);
    \draw[blue, line width=0pt, ->-=0.70 rotate 0] (\piVal/2,-\piVal)--(\piVal/4,-\piVal+\piVal/4);
    \draw[blue, line width=0pt, ->-=0.70 rotate 0] (\piVal/4,-\piVal+\piVal/4) -- (\piVal/2,-\piVal/2);
    \draw[blue, line width=0pt, ->-=0.70 rotate 0] (-\piVal/2,-\piVal/2) -- (-\piVal/4,-\piVal/4);
    \draw[blue, line width=0pt, ->-=0.70 rotate 0] (-\piVal/4,-\piVal/4) -- (-\piVal/2,0);
    \draw[blue, line width=0pt, ->-=0.70 rotate 0] (\piVal/2,0) -- (\piVal/4,\piVal/4);
    \draw[blue, line width=0pt, ->-=0.70 rotate 0] (\piVal/4,\piVal/4) -- (\piVal/2,\piVal/2);
    \draw[blue, line width=0pt, ->-=0.70 rotate 0] (-\piVal/2,\piVal/2) -- (-\piVal/2+\piVal/4,\piVal/2+\piVal/4);
    \draw[blue, line width=0pt, ->-=0.70 rotate 0] (-\piVal/2+\piVal/4,\piVal/2+\piVal/4) -- (-\piVal/2,\piVal);


  \end{tikzpicture}
    \subcaption{Ultraboosted limit.}\label{fig:antipodalSquare}
  \end{minipage}
  \caption{On the cylinder, a charged particle (red) must be compensated by an oppositely charged particle (blue) for consistency with Gauss' law on compact spaces. Left, two static charges sit on the Lorentzian cylinder, and the sink charge is only seen in the first Poincar\'e patch as a sink at spatial infinity $i^0$. Middle, the candy cane configuration arises naturally in the construction of double-twist operators and corresponds to two Wilson lines winding the cylinder antipodally. Right, the ultraboosted limit of timelike defects becomes the ``perfect null polygon,'' describing the repeated creation and annihilation of a pair of charged particles along maximal null trajectories.}
  \label{fig:antipodal}
\end{figure}
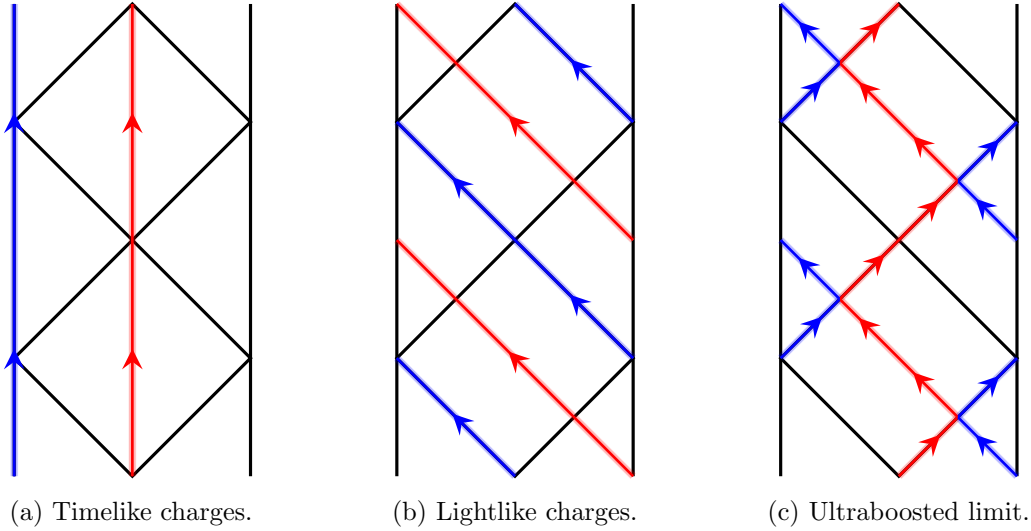

Consider a static charged particle on the Lorentzian cylinder $\bbR \times S^{d-1}$, corresponding to a timelike Wilson line. Since spatial slices are compact, Gauss' law demands that there be another charge on the sphere $S^{d-1}$ for flux lines to end. The simplest remedy is to consider another oppositely charged timelike line running antipodally up the cylinder, see Figure \ref{fig:antipodalTime}. This picture is also familiar from Euclidean CFT: in the defect state-operator correspondence, performing radial quantization around a point on an infinite defect corresponds to two line operators on the cylinder piercing the north and south poles of the $S^{d-1}$. Crucially, the defects carry opposite orientations on the cylinder or (equivalently) the same orientation with equal and opposite charge.

In the case of a null line, the analogous configuration is to have two oppositely charged defects moving up the cylinder antipodally, e.g. a fast moving electron-positron pair, see Figure \ref{fig:antipodalNull}. Such a configuration was already studied in detail in \cite{Alday:2007mf} (see also \cite{Alday:2010ku}), where kinematics can be used to predict the log scaling of anomalous dimensions at high spins, i.e. to explain the leading
\begin{equation}
    \Delta - J \stackrel{J\to\infty}{\sim} f(g) \log J\,.
\end{equation}
scaling of double-twist operators in conformal gauge theories. However, this ``candy cane'' configuration in Figure \ref{fig:antipodalNull} is not actually the ultrarelativistic limit of two charged particles, and the corresponding solutions for field strengths may not necessarily carry the maximal ``geometric'' symmetries of the configuration of lines.

This is easiest to understand by working in embedding space. We introduce an embedding space $\bbR^{2,d}$ with coordinates $\{X^A\}$ and metric $G = \rm{diag}(-1, +1, \dots, +1, -1)$, where the indices $A,B \in \{0,1,\dots, d, d+1\}$. The Lie algebra generators of $\mathfrak{so}(2,d)$ satisfy
\begin{equation}
    [J_{AB}, J_{CD}] = -2i(G_{B[D}J_{C]A} - G_{A[D}J_{C],B})\,,
\end{equation}
and a differential representation is realized by the differential operators $\mathscr{J}_{AB} = -i(X_A \partial_B - X_B \partial_A)$. As standard, our CFT coordinates are identified with points in the projective null-cone
\begin{equation}
    0 = G_{AB}X^A X^B = \eta_{\mu\nu} X^{\mu} X^\nu + (X^d)^2 - (X^{d+1})^2\,.
\end{equation}
More practically, the generators of the conformal group can be identified with
\begin{equation}
    D = J_{d,d+1}\,,\quad
    P_{\mu} = J_{d\mu} + J_{d+1,\mu}\,,\quad
    K_{\mu} = - J_{d\mu} + J_{d+1,\mu}\,,\quad
    M_{\mu\nu} = J_{\mu\nu}\,.
\end{equation}

Now, we have seen many times previously that there is a natural Heisenberg algebra $\mathfrak{h}_{d-2} = \{M_{-i}, K_-, K_i\}$ and its automorphism group $\mathfrak{sl}(2,\bbR)$ embedded inside the null defect algebra. Thus, we should expect that there is some ``Fourier transform'' swapping the role of $M_{-i}$ and $K_i$. In terms of the embedding space generators, these are\footnote{We absorb factors of $F$ into $J$ here and/or set $F = 1/\sqrt{2}$. They can be put back in by replacing $x^2$ in the inversions by $x^2 \mapsto \frac{1}{2F^2} x^2$.}
\begin{equation}
    M_{-i} = - J_{1i} + J_{0i} \,,\quad
    K_{i} = - J_{di} + J_{d+1,i}\,.
\end{equation}
Clearly a rotation in $(10) \leftrightarrow (d,d+1)$ changes $M_{-i}$ and $K_{i}$. A rotation that does this is (restricting to the $\{0,1,d,d+1\}$ coordinates):
\begin{equation}
    R = \begin{pmatrix}
        0 & 0 & 0 & 1 \\
        0 & 0 & 1 & 0 \\
        0 & 1 & 0 & 0 \\
        1 & 0 & 0 & 0
    \end{pmatrix}\,.
\end{equation}
The generators of the null-defect algebra transform as
\begin{equation}
    D \leftrightarrow M_{+-}\,,\quad
    P_- \leftrightarrow -K_{+}\,,\quad
    K_{i} \leftrightarrow M_{-i}\,,\quad
    K_- \leftrightarrow -K_-\,,\quad
    M_{ij} \leftrightarrow M_{ij}\,,
\end{equation}
while the other generators in the conformal algebra satisfy
\begin{equation}
    P_{i} \leftrightarrow M_{+i}\,,\quad P_+ \mapsto -P_+\,.
\end{equation}
On the CFT coordinates, this just descends to a null inversion \cite{Hofman:2008ar}:
\begin{equation}\label{eq:HMBar}
    \tilde{x}^+ = \frac{x^2}{x^-}\,,\quad
    \tilde{x}^- = \frac{1}{x^-}\,,\quad
    \tilde{x}^i = \frac{x^i}{x^-}\,,
\end{equation}
giving an involution mapping the plane $x^- = 0$ to $\scri^+$. Similar arguments can be used to map $x^+ = 0$ to $\scri^+$, swapping the role of $\pm$ in \eqref{eq:HMBar}. We consider the $x^-$ case because it is harder to intuit and will be useful for explaining the ``perfect null polygon'' configuration in Figure \ref{fig:antipodalSquare}.

Now we return to our candy cane configuration in Figure \ref{fig:antipodalNull}. In this case, one line (red) is at $x^+ = 0$ and $x^i = 0$, while the other line (blue) is along $\scri^+$. The symmetries preserving the latter can be computed by using the $x^+ = 0 \leftrightarrow \scri^+$ null inversion. Altogether we have
\begin{equation}
\begin{array}{l c c c c c c c c}
\hphantom{\mathscr{I}^+}x^+ = 0\!:\qquad &\quad
P_- \,,\; & D \,,\; & M_{+-} \,,\; & K_+ \,,\; & M_{ij} \,,\; & M_{-i} \,,\; & K_- \,,\; & K_i \,,\; \\
\hphantom{x^+ = 0}\mathscr{I}^+\!:\qquad &\quad
P_- \,,\; & M_{+-} \,,\; & D \,,\; & K_+ \,,\; & M_{ij} \,,\; & P_i \,,\; & M_{+i} \,,\; & P_+ \,,\;
\end{array}
\end{equation}
meaning that the symmetries preserved (geometrically) by two null lines in the candy cane configuration are:
\begin{equation}
    P_-\,,\quad
    D\,,\quad
    M_{+-}\,,\quad
    K_+\,,\quad
    M_{ij}\,.
\end{equation}
Of course, this matches \cite{Alday:2010ku} (although our argument is slightly different). 

Given our shockwave solution for the one-point function of $F_{\mu\nu}$, we expect that two null Wilson lines together will preserve
\begin{equation}
    P_-\,,\quad
    D\,,\quad
    M_{+-}\,,\quad
    M_{ij}\,.
\end{equation}
A similar argument for a candy cane configuration of two free scalar defects, would suggest that the scalars only preserve $P_-$ and $M_{ij}$ symmetry, since $D \leftrightarrow M_{+-}$ and $M_{+-}$ symmetry is broken in the causal solution. This could have been predicted: the causal scalar $\phi_c \sim (x^+)^{-1}$, and so a source at $\scri^-$ leads to a constant solution in the bulk, breaking $D$-invariance. In principle, only $P_-$ and $M_{ij}$ symmetry would appear in a similar analysis to \cite{Alday:2007mf} to show that the scaling of double twist operators
\begin{equation}
    \Delta - J \stackrel{J\to\infty}{\sim} \text{finite}\,.
\end{equation}
Of course, in the free scalar case this would be an unnecessary exercise.

Somewhat surprisingly, there is a more symmetric configuration than two null lines winding antipodally up the cylinder, obtained from the large boost limit of a charged particle as in Section \ref{sec:NullWilsonLine}. When we introduce an oppositely charged particle on the cylinder and boost the solution, the timelike charges in Figure \ref{fig:antipodalTime} deform to Figure \ref{fig:antipodalSquare}, not the candy cane in Figure \ref{fig:antipodalNull}. The resulting configuration is one big beautiful ``perfect null polygon'' (a null square) that is of ``maximal size.'' There is a neat reinterpretation of this configuration, obtained by moving the charge trajectory in the bulk of spacetime to $\scri^+$ with a null inversion, then the charged particle and anti-particle wrap around the conformal boundary of spacetime. Thus, up to a coordinate transformation: \textit{the ultrarelativistic limit of two timelike lines on the cylinder is a null polygon of maximal size, describing the creation and annihilation of charged particles around the horizon of spacetime.} Clearly it could not be any bigger or smaller without breaking some symmetries; indeed, in the configurations studied in \cite{Alday:2010ku}, the squares are only guaranteed to preserve $D$, $M_{+-}$, and $M_{ij}$. 

What are the symmetries preserved by the perfect null polygon? A standard inversion $x^\mu \mapsto x^\mu/x^2$ maps $\scri^\pm$ to the lightcone, and maps the perfect null polygon to crossing lines through the origin: $x_{(1)}^+=0$ and $x_{(2)}^-=0$ respectively, with all $x_{(i)}^\perp = 0$. Geometrically, the symmetries preserved by the crossing configuration are just the common set of:
\begin{equation}
\begin{array}{l c c c c c c c c}
x^+ = 0\!:\qquad &\quad
P_- \,,\; & D \,,\; & M_{+-} \,,\; & K_+ \,,\; & M_{ij} \,,\; & M_{-i} \,,\; & K_- \,,\; & K_i \,,\; \\
x^- = 0\!:\qquad &\quad
P_+ \,,\; & D \,,\; & M_{+-} \,,\; & K_- \,,\; & M_{ij} \,,\; & M_{+i} \,,\; & K_+ \,,\; & K_i \,,\;\,,
\end{array}
\end{equation}
which is
\begin{equation}
    D\,,\quad 
    M_{+-}\,,\quad 
    K_{+}\,,\quad
    M_{ij}\,,\quad
    K_{-}\,,\quad
    K_{i}\,.
\end{equation}

We can relate this to our original ultraboosted configuration in Figure \ref{fig:antipodalSquare}. We obtained the crossing configuration by concatenating the null-inversion with a standard inversion, which is just \eqref{eq:HMBar}. Therefore, in the original coordinates, the corresponding generators are
\begin{equation}
    D\,,\quad 
    M_{+-}\,,\quad 
    P_{-}\,,\quad
    M_{ij}\,,\quad
    K_{-}\,,\quad
    M_{-i}\,,
\end{equation}
which are precisely the symmetries of the shockwave solution, computed explicitly in Section \ref{sec:NullWilsonLine}, and bigger than symmetries preserved by the lightlike charges in Section \ref{fig:antipodalNull}. In this sense, \textit{null Wilson lines are maximally symmetric: preserving the maximal symmetry allowed by the perfect null polygon and/or the ultraboosted limit of timelike particles on the conformal cylinder}.

\section{Results in (1+1)d}\label{sec:2dPaper}
In this section, we turn to some studies in (1+1)d. While some of our results are similar to $d>2$ dimensions, we use the greatly simplified setting to give a quick classification of (projective) unitary irreducible representations of the null defect algebra. Thereby giving a putative kinematic description of potential defect local operators and their relationship to bulk local operators.

In (1+1)d, the global conformal algebra is just $\mathfrak{conf}(\bbR^{1,1}) = \mathfrak{so}(2,2) \cong \mathfrak{sl}(2,\bbR)_- \times \mathfrak{sl}(2,\bbR)_+$. Bulk local primaries $\calO$ are labelled by conformal weights $(h,\bar{h})$, where $h$ and $\bar{h}$ are independent. Recall that the generators of the two $\mathfrak{sl}(2,\bbR)$'s satisfy the commutation relations:
\begin{equation}
    [J_m, J_n] = i(n-m) J_{m+n}\,,\quad
    [\bar{J}_m, \bar{J}_n] = i(n-m) \bar{J}_{m+n}\,,\quad
    [J_m, \bar{J}_n] = 0\,.
\end{equation}
In (1+1)d, the null-defect algebra is reduced to
\begin{equation}
    \mathfrak{n}_2 = \mathfrak{sl}(2,\bbR)_{-} \times \{\bar{J}_0, \bar{J}_1\}\,.
\end{equation}

To compare to our previous conventions, suppose that we have a conformal primary operator $\calO$ with scaling weight $\Delta$ and $S_{\pm} \calO = i F^2 \ell$, then
\begin{align}\label{eq:DifferentialRep}
    [ J_{-1}, \calO_{h,\bar{h}}(x)] &= i \partial_- \calO_{h,\bar{h}}(x)\,,\nonumber\\
    [ J_0, \calO_{h,\bar{h}}(x)] &= -i (x^{-}\partial_- + h) \calO_{h,\bar{h}}(x)\,, \\
    [ J_1, \calO_{h,\bar{h}}(x)] 
        &= i ((x^-)^2\partial_- + 2hx^-)\calO_{h,\bar{h}}(x)\,,\nonumber
\end{align}
and similarly for the $\bar{J}$'s with $(-,h)\leftrightarrow(+,\bar{h})$, where
\begin{equation}\label{eq:handhbar}
    h := \frac{\Delta + \ell}{2}\,,\quad
    \bar{h} := \frac{\Delta - \ell}{2}\,.
\end{equation}
For future reference, we note that the differential generators satisfy $\mathscr{D}_1 = - x^- (\mathscr{D}_0 + i h)$ and $\bar{\mathscr{D}}_1 = - x^+ (\bar{\mathscr{D}}_0 + i \bar{h})$.


\subsection{Defect Primary Operators}\label{sec:DefectPris}
To study the physics of null defects in a conformal field theory, we might hope to to study defect local operators and defect Hilbert spaces, perhaps finding some form of state-operator correspondence. 

In order to most properly understand a Hilbert space from a physics perspective, we should first specify a quantization scheme i.e. a foliation of spacetime whose leaves support the state space. Let us temporarily put this aside, and focus on more formal algebraic matters: the representations of $\mathfrak{n}_2$. The representations of the $\mathfrak{sl}(2,\bbR)_-$ factor are understood, so we instead focus on the right moving part $\mathfrak{h} := \{\bar{J}_0,\bar{J}_1\}$; this is just the 1d affine group or ``$ax+b$ group.''

If a defect Hilbert space exists, we expect it to decompose into projective unitary irreducible representations of $H$, the exponential of $\mathfrak{h}$. Actually, we should also demand that the representations that appear have energy bounded below for some choice of Hamiltonian. Since we do not yet know what that Hamiltonian should be, we will remain vague on this point. It will not really matter since there are so few representations of $H$. The group $H$ is simply connected, so classifying projective representations is the same as classifying standard linear representations of $H$.\footnote{$\mathfrak{h}$ does have a Virasoro-like extension, which we are choosing to ignore here. Likewise for the Virasoro extension of $\mathfrak{sl}(2,\bbR)$.} A general group element $g(\alpha,\beta) := \exp(i\alpha \bar{J}_0 + i\beta \bar{J}_1)$ acts on the coordinate $\bar{z}$ by:
\begin{equation}
    g(\alpha,\beta)\cdot \bar{z} = \frac{\alpha e^{\alpha} \bar{z}}{(1-e^{\alpha})\beta \bar{z} + \alpha}\,.
\end{equation}
Of course, on the inverted coordinate $\bar{w} = -1/\bar{z}$, $\bar{J}_0$ and $\bar{J}_1$ just act by a dilatation $\bar{w} \mapsto e^{-\alpha} \bar{w}$ and translation $\bar{w} \mapsto \bar{w}+\beta$ respectively, so $H$ is the ``$ax+b$'' group as promised. Thus we can rewrite a general group element as pair $(e^{a}, b)$, with group composition law
\begin{equation}
    (e^{a},b)(e^{a'}, b') =  (e^{a+a'}, b+e^a b')\,.
\end{equation}
The UIRs are computable in a straightforward way by induction (see e.g. \cite{folland2016course}), and come in two types:
\begin{enumerate}
    \item \textbf{1d Characters.} For any $\bar{h} \in \bbR$ there is a 1d  ``scale wave'' character representation of the dilatation subgroup lifted to $H$:
    \begin{equation}\label{eq:1dUIRn1}
        \pi_{\bar{h}}(a,b)v = e^{i\bar{h} a}v\,,\quad \bar{h} \in \bbR\,.
    \end{equation}
    Infinitesimally, these correspond to 1d $\mathfrak{h}$-modules where
    \begin{equation}\label{eq:1dUIRn2}
        \bar{J}_0 \cdot v = -i \bar{h} v\,,\quad
        \bar{J}_1 \cdot v = 0\,.
    \end{equation}
    \item \textbf{Infinite Dimensional Representations.} There also exists two infinite dimensional representations $\pi_{+}$ and $\pi_{-}$, which can be realized on $L^2((0,\infty),\frac{dp}{2\pi})$ and $L^2((-\infty,0),\frac{dp}{2\pi})$ respectively by:
    \begin{equation}\label{eq:infUIRn2}
        [\pi_{\pm}(a,b) f](p) = e^{a/2} e^{i b p} g(e^a p)\,.
    \end{equation}
    Together, $\pi_+ \oplus \pi_-$ just give the usual momentum space representation dual to the (inverted) position coordinate $\bar{w}$. The effect of the scale transformations is to effectively unify all of the positive energy (resp. negative energy) plane waves into one rep $\pi_+$ (resp. $\pi_-$). Fourier transforming to position space, it can be shown that the Lie algebra generators here act as in \eqref{eq:DifferentialRep}, with $\bar{h} = \frac{1}{2}$.
\end{enumerate}
This gives all projective unitary irreducible representations of $H$. Next we consider the decomposition of $\mathfrak{sl}(2,\bbR)_+$ modules into $\mathfrak{h}$-modules. This will give some insight into the bulk-defect OPE, insofar as we can see how multiplets of local operators decompose.

As a brief reminder, it is useful to recall the kinematics of a \emph{timelike} line defect. Consider a bulk primary local operator $\calO_{h,\bar{h}}$ at the origin, it has its pyramid of descendants
\begin{equation}
    V_{h,\bar{h}} = \{J_{-1}^{k} \bar{J}_{-1}^{\bar{k}} \calO_{h,\bar{h}} \,|\, k,\bar{k} \geq 0\}\,.
\end{equation}
If we consider a timelike line defect, we expect operators involving only derivatives normal to the defect to behave like defect primaries: schematically, $\partial_\perp \hat{\calO}(0)$ are killed by $K_\parallel$'s. To see this, note that the symmetry group preserving the defect is the diagonal $\mathfrak{sl}(2,\bbR) \subset \mathfrak{sl}(2,\bbR)_+ \times \mathfrak{sl}(2,\bbR)_-$. A better basis for translation operators is then
\begin{equation}
    J_{-1}^{||} := \tfrac{1}{2}(J_{-1} + \bar{J}_{-1})\,,\quad
    J_{-1}^{\perp} := \tfrac{1}{2}(J_{-1} - \bar{J}_{-1})\,,
\end{equation}
where $J_{-1}^{||}$ generates translations along the line defect and $J_{-1}^{\perp}$ generates translations normal to the defect. Then the space of descendants can be re-written
\begin{equation}
    V_{h,\bar{h}} = \{(J_{-1}^{||})^{k_+} (J_{-1}^{\perp})^{k_-} \calO_{h,\bar{h}} \,|\, k_+,k_- \geq 0\}\,.
\end{equation}
In terms of the defect subalgebra, the bulk algebra descendants involving only normal derivatives to the defect
\begin{equation}
    \hat{\calO}_{h,\bar{h}}^{(k_-)} = (J_{-1}^{\perp})^{k_-} \calO_{h,\bar{h}}\,,
\end{equation}
behave like defect primary operators with 1d conformal weight $\Delta = h + \bar{h} + k_-$.

Now we return to the same exercise for a conformal \textit{null} line defect. First, consider the case that the line only preserves the minimal conformal symmetry, i.e. only the $\mathfrak{sl}(2,\bbR)_-$ factor. Then we have an identical analysis to the timelike case, with 
\begin{equation}
    \hat{\calO}_{h,\bar{h}}^{(k)} = \bar{J}_{-1}^k\calO_{h,\bar{h}}\,,
\end{equation}
a defect primary of conformal weight $h$ for all $k$. If, in addition, $\bar{J}_0$ is preserved, then we would have an additional quantum number and $\hat{\calO}_{h,\bar{h}}^{(k)} = \bar{J}_{-1}^k\calO_{h,\bar{h}}$ looks like a defect primary of weight and charge $(h,\bar{h}+k)$.

The real difference between the timelike and null cases arises when the transverse SCT $\bar{J}_1$ is preserved. In this case, the bulk modules are generally reducible but indecomposable for the defect algebra. Physically, this is undesirable because one could apply $\bar{J}_1$ a finite number of times to any bulk local operator (primary or descendant) to annihilate it, undetectable from just the $J_0$ and $\bar{J}_0$ weights. Such modules are also generally excluded by unitarity.

Let's consider this in more detail, ignoring the $\mathfrak{sl}(2,\bbR)_-$ factor for brevity. Given a lowest weight module of $\mathfrak{sl}(2,\bbR)_+$, with lowest weight $\bar{h}$, we have
\begin{equation}
\bar{J}_0 \cdot \calO_{\bar{h}} = -i \bar{h} \calO_{\bar{h}}\,,\quad
    \bar{J}_1 \cdot \calO_{\bar{h}} = 0\,,
\end{equation}
while all $\bar{J}_{-1}^k \cdot \calO_{\bar{h}}$ form a space of descendants of higher conformal weight
\begin{equation}
    V_{\bar{h}} := \{ \bar{J}_{-1}^k\calO_{\bar{h}} \,|\, k\geq 0\}\,.
\end{equation}
Under $\mathfrak{h}$, this conformal family $V_{\bar{h}}$ is highly reducible. For any $\ell$, the subspace
\begin{equation}
    W_{\bar{h}}^{(\ell)} := \{J_{-1}^{\bar{k}}\calO_{\bar{h}} \,|\, 0 \leq k \leq \ell\}
\end{equation}
is an $(\ell+1)$-dimensional $\mathfrak{h}$-invariant submodule of $V_{\bar{h}}$, giving a filtration:
\begin{equation}
    W_{\bar{h}}^{(0)} \subset W_{\bar{h}}^{(1)} \subset W_{\bar{h}}^{(2)} \subset \cdots \subset V_{\bar{h}}\,.
\end{equation}
Since $\bar{J}_1: W_{\bar{h}}^{(\ell)} \to W_{\bar{h}}^{(\ell-1)}$, they are each reducible but indecomposable, except $W_{\bar{h}}^{(0)}$ which is genuinely irreducible and indecomposable. Here, $W_{\bar{h}}^{(0)}$ is just the 1d module of $\mathfrak{h}$ described in \eqref{eq:1dUIRn2}.\footnote{The two infinite dimensional representations $\pi_{\pm}$ of $\mathfrak{h}$ are directly related to the restriction of the $\bar{h}=\frac{1}{2}$ principal series representation of $\mathfrak{sl}(2,\bbR)$ to $\mathfrak{h}$. This can be seen explicitly by comparing the Fourier transform of $\pi := \pi_+ \oplus \pi_-$ on $L^2(\bbR, d\bar{w})$ to the usual principal series representation on $L^2(\bbR,d\bar{w})$.}

Thus, our general expectation is that defect local primaries, corresponding to states in a putative defect Hilbert space, take the form $\hat{\calO}_{h,\bar{h}}(x^-)$. They should behave like 1d CFT primaries for $\mathfrak{sl}(2,\bbR)_-$, i.e. as in \eqref{eq:DifferentialRep}, while also satisfying
\begin{equation}\label{eq:defectPris}
    [\bar{J}_0, \hat{\calO}_{h,\bar{h}}(x^-)] = -i\bar{h} \hat{\calO}_{h,\bar{h}}(x^-)\,,\quad
    [\bar{J}_1, \hat{\calO}_{h,\bar{h}}(x^-)] = 0\,,
\end{equation}
for all $x^-$. Since every $\bar{J}_{-1}$ descendant in an $\mathfrak{sl}(2,\bbR)_+$ module lives in a reducible but indecomposable representation, we further expect that $\bar{J}_{-1}$ descendants of bulk operators $\bar{J}_{-1}^{\bar{k}}\calO_{h,\bar{h}}$ should decouple from defect dynamics at least in the context of unitary theories. In summary, if a null defect preserves $\bar{J}_0$ it effectively adds an extra internal quantum number for defect primaries and the bulk-defect OPE; if the defect also preserves $\bar{J}_1$, it effectively decouples bulk transverse descendants from the defect. 

Thus far the discussion has been fairly technical, but we can gain intuition for reducible but indecomposable representations in a different but perhaps more familiar setting. For ordinary bulk conformal primaries saturating or violating the unitarity bound, it is possible to have descendants that are themselves primaries. As mentioned in Section \ref{sec:WLKinematics}, a prototypical example is the gauge field $A$, with dimension 1 and spin 1 (a 1-form). When $d>2$ this violates the unitarity bound, but the field strength descendant  $F = dA$ is a primary and annihilated by all special conformal transformations. The field strength also satisfies (saturates) unitarity bounds in 3 and 4 dimensions; the saturation of the unitarity bound is equivalent to the Bianchi identity $dF=0$. Thus the multiplet of the gauge field is clearly reducible but indecomposable. In the context of gauge theories, we know that in order to have a unitary and gauge-invariant theory we must remove the states associated with the gauge field that are not the field strength. That is, we consider only local operators associated with the irreducible submodule with $F$ as its primary. Analogously for the null defect, we can only allow local operators associated in the irreducible submodule \(W_{\bar{h}}^{(0)}\) to participate in the defect CFT dynamics.

Finally, we return to positive energy constraints. For bulk CFT operators $\calO_{h,\bar{h}}$, both $h$ and $\bar{h}$ are bounded below $h,\bar{h} \geq 0$. This follows from the non-negativity of the bulk conformal Hamiltonian  $H_{\mathrm{LM}}$ which implies $\Delta = h + \bar{h} \geq 0$ \cite{Luscher:1974ez, Mack:1975je}. In the case of a lightlike conformal defect, in any dimension, two other natural Hamiltonians are the conformal lightcone Hamiltonians $H = J_{-1} + J_{1}$ and $\bar{H} = \bar{J}_{-1} + \bar{J}_{1}$, which generate translations parallel and perpendicular to the defect on the Lorentzian cylinder. For a defect primary $\hat{\calO}_{h,\bar{h}}$, requiring $H \geq 0$ only enforces the usual 1d defect unitarity boundary $h \geq 0$, while allowing $\bar{h}$ to be completely arbitrary. This is not necessarily a problem, as defect CFT unitarity bounds are typically weaker than the bulk CFT unitarity bounds. Additionally demanding that $\bar{H} \geq 0$ is at least as strong as $H_{\mathrm{LM}} = H + \bar{H} \geq 0$. In the bulk, demanding that $H \geq 0$ and $\bar{H} \geq 0$ independently gives equivalent unitarity bounds to $H_{\mathrm{LM}} \geq 0$. However, it is not obvious that this observation is useful for the defect since the defect does not create an eigenstate of $\bar{H}$; as a result, we refrain from making any sweeping statements about positivity conditions.

\subsection{Ward Identities for Correlators}\label{eq:BulkWardIdentity2d}
Separate from the Hilbert space discussion of the previous subsections, we can also study the kinematics of correlation functions in the presence of a null defect. We do so by explicitly solve the Ward identities for correlators in the presence of the null defect. These are abstract kinematic $n$-point functions, not a particular time ordered or causal correlation function.

\paragraph{1 Bulk Primary Operator.} An abstract bulk one-point function takes the form:
\begin{equation}
    \expval*{\calO_{h,\bar{h}}(x^+,x^-)}_{\rm{D}} = f(h,\bar{h}; x^+, x^-)\,.
\end{equation}
As this is an invariant correlator, it must satisfy the Ward identities $\mathscr{D}_{J} \expval{\calO_{h,\bar{h}}(x^+,x^-)} = 0$ for all generator $J$. Th $J_{-1}$ symmetry means that $f$ does not depend on $x^-$. $J_0$ then gives
\begin{equation}
    0 = \mathscr{D}_{J_0} f(h,\bar{h}; x^+) = h f(h,\bar{h}; x^+)\,,
\end{equation}
from which we determine that either $h = 0$ or else $f = 0$ trivially. For the non-trivial case of $h=0$ the operator is necessarily anti-chiral. $J_1$ does not add any additional constraints. 

So far we have what we would expect by analogy to the timelike or spacelike defect case. For a timelike line defect in (1+1)d, the only operators which get a non-trivial one-point function are the scalars
\begin{equation}
    \expval*{\calO_{h,h}(x^0,x^1)}_{\rm{D}} = \frac{c}{(x^1)^{2h}}\,.
\end{equation}
Here we have already seen that the only bulk operators which can get a one-point function are the anti-chiral operators $\calO_{0,\bar{h}}$.

Now we can consider the effects of $\bar{J}_0$ and $\bar{J}_1$. The $\bar{J}_0$ Ward identity is
\begin{equation}
    0 = \mathscr{D}_{\bar{J}_0} f(\bar{h};x^{+}) = x^{+} f'(\bar{h};x^{+}) + \bar{h} f(\bar{h};x^+)\,.
\end{equation}
For generic $\bar h$, the solution is
\begin{equation}
    f(\bar{h};x^+) = c(\bar{h}) |x^+|^{-\bar{h}}\,.
\end{equation}
When $h = n+1 \in \bbZ_{>0}$, corresponding to a (possibly higher-spin) anti-chiral bulk current operator $J(x^-,x^+)$, the ODE also admits delta-function solutions $\propto c_n \delta^{(n)}(x^+)$. This correctly recovers the expected one-point function contact terms we'd expect for higher-spin currents.

The $\bar{J}_1$ Ward identity is
\begin{equation}
    0 = \mathscr{D}_{\bar{J}_1} f(\bar{h};x^{+}) = (x^{+})^2 f'(\bar{h};x^{+}) + 2\bar{h}x^+ f(\bar{h};x^+)\,.
\end{equation}
For generic $\bar h$, the solution $c(\bar{h}) |x^+|^{-\bar{h}}$ does not satisfy the equation unless $c = 0$. However, the contact terms from higher-spin currents do satisfy the Ward identity and thus remain as solutions.

In summary, for bulk one-point functions we have the following: if the null defect only preserves the left moving $\mathfrak{sl}(2,\bbR)_-$, only anti-chiral bulk primary operators have a non-trivial one-point function. By locality, operators with conformal dimension $(0,\bar{h})$ are necessarily bulk currents (or anti-chiral fermions) in unitary CFTs. If $\bar{J}_0$ symmetry is preserved, the correlation functions generically take the form
\begin{equation}
    \expval*{\calO_{h,\bar{h}}(x^0,x^1)}_{\rm{D}} = \delta_{h,0}\frac{c}{|x^+|^{\bar{h}}}\,.
\end{equation}
When there is maximal $\mathfrak{n}_2$ symmetry, all one-point functions vanish except possibly some contact terms for higher-spin currents.

\paragraph{2 Bulk Primary Operators.}  We can consider the constraints on a 2 point correlator in the same way. A general two point function in 2d takes the form
\begin{equation}
    \expval*{\calO_{h_1,\bar{h}_1}(x^+_1,x^-_1)\calO_{h_2,\bar{h}_2}(x^+_2,x^-_2)}_{\rm{D}} = f(h_i,\bar{h}_i; x_{i}^{\pm})\,.
\end{equation}
Going forward, we will ignore the non-generic contact term solutions in $\delta(x^+)$. Unlike higher dimensions, $\delta(x^+)$ is just a contact term on the defect, not a whole shockwave plane, so it is inherently less interesting. As noted below \eqref{eq:handhbar}, the $J_1$ Ward identity can be re-written as
\begin{align}
    0 
        &= -(\mathscr{D}_1^{(1)} + \mathscr{D}^{(2)}_1)f(h_i;x_i^-)\\
        &= (x_1^- (\mathscr{D}_0^{(1)}+ih_1) + x_2^- (\mathscr{D}_0^{(2)}+ih_2)) f(h_i; x_i^-)\,,
\end{align}
and similarly for $\bar{J}_1$. If we introduce $h_{\pm} := h_2 \pm h_1$ and $\delta^{\pm} := x^-_2 \pm x_1^-$, then this is equivalent to
\begin{equation}
    0
        = \left(\frac{1}{2}\left((\delta^-)^2-(\delta^+)^2\right)\partial_{\delta^+}+h_-\delta^--\delta^+(\mathscr{D}_0^{(1)} + \mathscr{D}_0^{(2)})\right) f(h_{\pm}; \delta^{\pm})\,.\label{eq:J1WardSimple}
\end{equation}
Thus, after using the Ward identities for $J_0$, the Ward identities for $J_1$ greatly simplify.

The $J_{-1}$ Ward identity says that $0 = (\partial_-^{(1)}+\partial_-^{(2)})f$ which means that $f$ only depends on the combination $\delta^- = x_2^{-} - x^-_1$. The $J_0$ Ward identity then implies that
\begin{equation}
    f(h_i,\bar{h}_i; x_i^+,\delta^-) = c(h_i,\bar{h}_i; x_i^+)|\delta^-|^{-h_+}\,.
\end{equation}
Now we turn to the $J_1$ Ward identity. From \eqref{eq:J1WardSimple}, the previous solution must also solve
\begin{equation}
    0 = h_- \delta^- c(h_i,\bar{h}_i; x_i^+)|\delta^-|^{-h_+}\,,
\end{equation}
which is satisfied only when $h_- = 0$. Together, the left-moving Ward identities constrain the two-point function to the form
\begin{equation}
    \expval*{\calO_{h_1,\bar{h}_1}(x^+_1,x^-_1)\calO_{h_2,\bar{h}_2}(x^+_2,x^-_2)}_{\rm{D}} = \delta_{h_1,h_2}\frac{c(\bar{h}_i; x_i^+)}{|\delta^-|^{h_1 + h_2}}\,.
\end{equation}

Now we can consider the right-moving algebra. The $\bar{J}_0$ Ward identity is just
\begin{equation}
    0 = (\bar{h}_+ + x^+_1 \partial_{+}^{(1)} + x^+_2 \partial_{+}^{(2)})c(\bar{h}_{\pm}; x^+_i)\,,
\end{equation}
and the solution can be written in the form
\begin{equation}
    c(\bar{h}_{\pm}; x_1^+,x_2^+) = \abs{x_1^+ x_2^+}^{-\tfrac{\bar{h}_+}{2}}\tilde{g}\left(\bar{h}_\pm;
    \abs{\frac{x_1^+}{x_2^+}}, \mathrm{sign}\, {x_1^+}, \mathrm{sign}\, {x_2^+}\right)\,.
\end{equation}
We expect $\tilde{g}$ to depend discontinuously on the sign of $x_1^+/x_2^+$. In particular, we expect different answers if the two local operators are on the same side before the defect, same side after the defect, or opposite sides of the defect.

If in addition, we also have $\bar{J}_1$ symmetry, then the previous solution must also satisfy
\begin{equation}
    0 
        = \left(\frac{1}{2}\left((\epsilon^-)^2-(\epsilon^+)^2\right)\partial_{\epsilon^+}+\bar{h}_-\epsilon^-\right) c(\bar{h}_{\pm}; \epsilon^{\pm})
\end{equation}
where we have introduced $\epsilon^{\pm} := x_2^+ \pm x_1^+$. A straightforward manipulation gives:
\begin{equation}
    \expval*{\calO_{h_1,\bar{h}_1}(x^+_1,x^-_1)\calO_{h_2,\bar{h}_2}(x^+_2,x^-_2)}_{\rm{D}} = \frac{\delta_{h_1,h_2}}{|x_2^--x_1^-|^{h_1 + h_2}}
    \frac{\tilde{g}\left(\bar{h}_\pm;\mathrm{sign}\, {x_1^+}, \mathrm{sign}\, {x_2^+}\right)}{|x_2^+-x_1^+|^{\bar{h}_1+\bar{h}_2}}\abs{\frac{x_1^+}{x_2^+}}^{\bar{h}_2 - \bar{h}_1}\,.
\end{equation}
Based on whether one wants time-ordered, anti-time-ordered, Wightman, etc. correlators, the expressions above can be adjusted with $i\epsilon$ prescriptions without much trouble.

\paragraph{Defect Correlation Functions.} We can also comment on defect correlation functions. Based on the discussion in Section \ref{sec:DefectPris}, we expect defect primaries to behave like 1d CFT primaries and also satisfy the additional conditions in \eqref{eq:defectPris}. Thus we expect defect correlators to essentially behave like 1d CFT correlation functions with an additional abelian quantum number $\bar{J}_0$, for example
\begin{equation}
    \expval*{\hat{\calO}_{h,\bar{h}}(x^-)} = 0\,,
\end{equation}
and 
\begin{equation}
    \expval*{\hat{\calO}_{h_1,\bar{h}_1}(x^-_1)\hat{\calO}_{h_2,\bar{h}_2}(x^-_2)} = \frac{c\,\delta_{h_1, h_2} \delta_{\bar{h}_1 \bar{h}_2}}{(x_1^- - x_2^-)^{h_1+h_2}}\,,
\end{equation}
etc.

The real question, as alluded to at the end of Section \ref{sec:DefectPris}, is what the correct positivity condition is for $\bar{J}_0$. For example, if it turns out that we should have $\bar{h} \geq 0$ for defect local operators, then no defect correlation function will ever depend on $\bar{h}$, all $\bar{h}_i$ can effectively be set to zero, and the situation is no different from having zero right-moving symmetry at all. However, if $\bar{h}$ is only constrained to be ``not too negative,'' say with $h + \bar{h} \geq 0$, then we can admit far more interesting sets of correlation functions. For example, where $\bar{J}_0$ invariance is violated by bulk local operators, but absorbed/compensated by defect local operators. We leave further considerations of positivity conditions and consequences for defect correlators to future works.



\appendix
\section{Conformal Algebra and Lightcone Coordinates}\label{app:Conventions}
In this appendix, we record our conventions for the conformal algebra and, importantly, our choices for lightcone coordinates. 

\subsection{Conformal Algebra}
We will write the commutation relations of the Minkowski conformal algebra $\mathfrak{so}(2,d)$ as:
\begin{align}
    [ D, P_{\mu}] &= +i P_{\mu}\,,\\
    [ D, K_{\mu}] &= -i K_{\mu}\,,\\
    [ K_{\mu}, P_{\nu}] &= -2i(\eta_{\mu\nu}  D +  M_{\mu\nu})\,,\\
    [ M_{\mu\nu}, P_{\rho}] &= i(\eta_{\mu\rho} P_{\nu} - \eta_{\nu\rho}  P_{\mu})\,,\\
    [ M_{\mu\nu}, K_{\rho}] &= i(\eta_{\mu\rho} K_{\nu} - \eta_{\nu\rho}  K_{\mu})\,,\\
    [ M_{\mu\nu}, M_{\rho\lambda}] &= -2i(\eta_{\nu[\lambda} M_{\rho]\mu}-\eta_{\mu[\lambda} M_{\rho]\nu})\,.
\end{align}
Greek letters $\mu,\nu = 0,1,\dots,d-1$ and the Minkowski metric is $\eta_{\mu\nu} = \mathrm{diag}(-1,+1,\dots,+1)$. On a conformal primary operator $\calO_I(x)$, the generators act by:\footnote{In our conventions, the abstract generator acts by minus the differential operator
\begin{equation}
    [Q,\calO_I(x)] := -\mathscr{D}_Q\calO_I(x)
\end{equation}
so that the commutation relations of the abstract generators and the differential operators are the same (without a minus sign the Lie algebras would have opposite structure).}
\begin{align}
    [ D, \calO_I(x)] &= i (x\cdot\partial + \Delta) \calO_I(x)\\
    [ P_{\mu}, \calO_I(x)] &= i \partial_{\mu} \calO_I(x)\,,\\
    [ K_{\mu}, \calO_I(x)] &= i(x^2 \partial_{\mu} - 2 x_{\mu} x\cdot \partial - 2 \Delta x_{\mu})\calO_I(x)+2{(S_{\mu\nu})_I}^J x^{\nu} \calO_J(x)\,,\\
    [ M_{\mu\nu}, \calO_I(x)] &= i(x_{\mu} \partial_{\nu} - x_{\nu} \partial_{\mu}) \calO_I(x)-{(S_{\mu\nu})_I}^J\calO_J(x)\,.
\end{align}
Here $S_{\mu\nu}$ is the appropriate spin-representation matrix for $\calO_I(x)$. For the vector representation it is given by:
\begin{equation}
    {(S_{\mu\nu})_\rho}^\lambda = -i (\eta_{\mu\rho}\delta_\nu^\lambda - \eta_{\nu\rho}\delta_\mu^\lambda)\,.
\end{equation}

\subsection{Lightcone Coordinates}
It will be extremely useful for us to switch to lightcone coordinates. We record these conventions and other useful relations here:
\begin{alignat}{3}
    x^{\pm} &= \frac{1}{\sqrt{2}F}(x^0 \pm x^1)\,,\qquad 
        && \partial_{x^{\pm}} = \frac{F}{\sqrt{2}} (\partial_{x^0} \pm \partial_{x^1})\,,\qquad
        && T_{\mu} = \eta_{\mu\nu} T^\nu\,,\\
    x^0 &= \frac{F}{\sqrt{2}}(x^+ + x^-)\,,\qquad 
        && \partial_{0} = \frac{1}{\sqrt{2}F}(\partial_+ + \partial_-)\,,\qquad
        && \partial_{{\pm}} x^{\pm} = 1\,,\\  
    x^1 &= \frac{F}{\sqrt{2}}(x^+ - x^-)\,,\qquad 
        && \partial_{1} = \frac{1}{\sqrt{2}F}(\partial_+ - \partial_-)\,,\qquad
        && \partial_{{\pm}} x^{\mp} = 0\,.
\end{alignat}
We have left the symbol $F$ in for ease of comparison to other references. 

We introduce $A,B = +,-,2,\dots, d-1$ for the lightcone coordinates and $i,j = 2,\dots,d-1$ for the shared transverse coordinates to the relevant light plane. We will also denote these directions collectively as $x^\perp$. $x_\perp = x^\perp$, a vector quantity, is distinctly different from the transverse radial distance $|x_\perp| = |x^\perp| = \sqrt{x_i x^i}$, both of which appear in expressions. The $0,1$-components of the metric change to $\pm$ components and the metric transforms as:
\begin{equation}
    \tilde{\eta}_{AB} = \eta_{\mu\nu} \frac{d x^\mu}{d\tilde{x}^C}\frac{d x^\nu}{d\tilde{x}^D}\,,\qquad 
        \tilde{\eta}_{AB} = 
            F^2 \begin{pmatrix}
                0 & -1 \\
                -1 & 0 
            \end{pmatrix}_{AB}\,,\qquad
        \tilde{\eta}^{AB} = 
            \frac{1}{F^2}\begin{pmatrix}
                0 & -1 \\
                -1 & 0 
            \end{pmatrix}^{AB}\,. 
\end{equation} 
Or, in terms of the line element:
\begin{equation}
    ds^2 
        = -(dx^0)^2+(dx^1)^2+(dx^\perp)^2 
        = -2F^2 dx^{+} dx^{-} + (dx^\perp)^2\,.
\end{equation}

More generally, any tensor, e.g. with two components like $\tilde{T}_{AB}$, will satisfy $\tilde{T}_{ij} = T_{ij}$ and
\begin{align}
    \tilde{T}_{\pm i} &= \frac{F}{\sqrt{2}}(T_{0i}\pm T_{1i})\,,\\
    \tilde{T}_{i\pm} &= \frac{F}{\sqrt{2}}(T_{i0}\pm T_{i1})\,,\\
    \tilde{T}_{\pm_1 \pm_2} &= \frac{F^2}{2}(T_{00}\pm_2 T_{01} \pm_1 T_{10} \pm_1 \pm_2 T_{11})\,,
\end{align}
etc. If a tensor is symmetric/anti-symmetric, it will remain that way since its a property of the abstract object, not the presentation. 

Using the metric, we can raise and lower indices. The $i,j$ indices are raised/lowered freely. The $\pm$ indices swap signs and collect Jacobians, e.g.
\begin{equation}
    \tilde{T}_{\pm i} = - F^2 \tilde{T}^{\mp i}\,,\qquad
    \tilde{T}_{\pm_1 \pm_2} = F^4 \tilde{T}^{\mp_1 \mp_2}\,.
\end{equation}
And also:
\begin{equation}\label{eq:SIdentities}
    {(S_{+-})_-}^\lambda = i F^2 \delta_-^\lambda\,,\quad
    {(S_{-i})_-}^\lambda = 0\,,\quad
    {(S_{ij})_-}^\lambda = 0\,.
\end{equation}
For the traceless symmetric spin-$\ell$ representation
\begin{equation}
    (S_{+-}^{(\ell)})_{-\dots-}^{\lambda_1\cdots\lambda_\ell} = iF^2\ell \delta^{\lambda_1}_-\cdots \delta^{\lambda_\ell}_-\,.
\end{equation}

\subsection{Lightcone Conformal Algebra}
For lightcone calculations, a better basis for the conformal algebra uses the generators:
\begin{align}
    P_{\pm} 
        &= \frac{F}{\sqrt{2}}(P_0 \pm P_1)\,,\\
    K_{\pm} 
        &= \frac{F}{\sqrt{2}}(K_0 \pm K_1)\,,\\
    M_{\pm i} 
        &= \frac{F}{\sqrt{2}}(M_{0i} \pm M_{1i})\\
    M_{+-} &= - M_{-+} = -F^2 M_{01}\,.
\end{align}
The non-zero commutators in the light-cone coordinates involving the new $\pm$-coordinates are:
\begin{alignat}{2}
    [ D, P_{\pm}] &= +i P_{\pm}\,,        &\quad
    [ D, K_{\pm}] &= -i K_{\pm}\,,\nonumber\\
    [ K_{\pm}, P_{\mp}] &= -2i(-F^2  D \pm M_{+-})\,, &\quad
    [ P_{\pm}, K_{i}] &= -2i M_{\pm i}\,,\nonumber\\
    [ K_{\pm}, P_{i}] &= -2i M_{\pm i}\,, &\quad
    [ M_{\pm i}, P_j] &= -i \delta_{ij} P_{\pm}\,,\nonumber\\
    [ M_{+ -}, P_{\pm}] &= \pm iF^2 P_{\pm}\,, &\quad
    [ M_{\pm i}, K_j] &= -i \delta_{ij} K_{\pm}\,,\\
    [ M_{\pm i}, K_{\mp}] &= -i F^2 K_{i}\,, &\quad
    [ M_{+ -}, K_{\pm}] &= \pm iF^2 K_{\pm}\,,\nonumber\\
    [ M_{\pm i}, M_{jk}] &= -i (\delta_{ij}M_{\pm k}-\delta_{ik} M_{\pm j})\,, &\quad
    [ M_{+ i}, M_{- j}] &= -iF^2M_{ij} + i\delta_{ij}M_{+-}\,,\nonumber\\
    [ M_{+ -}, M_{\pm i}] &= \pm iF^2 M_{\pm i}\,. & &\nonumber
\end{alignat}
Note that $[K_{\pm},P_{\pm}]=0$ in these coordinates, and they \textit{do not} commute if they have different signs. Note: some authors define things so that $[K_{\pm},P_{\mp}] = 0$.


\section{Details of Calculations}\label{app:Calculations}

\subsection{Details for Example: (3+1)d Free Scalar + \texorpdfstring{$\phi$}{phi}}\label{app:freeScalarCalcs}
This appendix details the calculations in Section \ref{sec:ExampleFreeScalar}. The action is
\begin{equation}
    S = \frac{1}{2}\int d^{4} x \, (\partial_\mu \phi)^2 + h  \int dx^-\, \phi(0,x^-,0)
\end{equation}
with equation of motion and source
\begin{equation}
    \partial^2 \phi 
        = h \delta(x^+)\delta^2(x^\perp) =: J(x^+,x^\perp)\,.
\end{equation}

\subsubsection{Causal Solution and \texorpdfstring{$\Theta(\scarex)$}{Theta(``x+'')}}
It is helpful to see the solution to this classical PDE in some detail as there are some subtleties presaging subsequent examples. The retarded Green's function is
\begin{gather}
    G_R(x) = -\frac{\delta(x^0 - r)}{4\pi \sqrt{r^2 + \delta^2}}\,,
\end{gather}
where $r^2 = x_i x^i$. In our conventions, it satisfies
\begin{equation}
    \Box G_R(x) = \partial^2 G_R(x) = \delta^{4}(x)\,.
\end{equation}
We find the field configuration by computing:
\begin{align}
    \phi(x^0,x^1, x^\perp) 
        &= \int d^{2}\tilde{x}^\perp d\tilde{x}^1 d\tilde{x}^0\, G_R(x^0-\tilde{x}^0,x^1-\tilde{x}^1,x^\perp-\tilde{x}^\perp) J(\tilde{x}^+,\tilde{x}^\perp)\\
        &= -\frac{\sqrt{2} h F}{4\pi} \int d\tilde{x}^1 d\tilde{x}^0\,
        \frac{\delta((x^0-\tilde{x}^0) - \sqrt{(x^1 - \tilde{x}^1)^2 + x_\perp^2})}{\sqrt{(x^1 - \tilde{x}^1)^2 + x_\perp^2 + \delta^2}} \delta(\tilde{x}^0 + \tilde{x}^1)\\
        &= -\frac{\sqrt{2} h F}{4\pi} \int d\tilde{x}^0\,
        \frac{\delta((x^0-\tilde{x}^0) - \sqrt{(x^1 + \tilde{x}^0)^2 + x_\perp^2})}{\sqrt{(x^1 + \tilde{x}^0)^2 + x_\perp^2 + \delta^2}}\,.
\end{align}

The support of the final delta function is
\begin{gather}
    x^0-\tilde{x}^{0*} - \sqrt{ (x^1 + \tilde{x}^{0*})^2 + x_\perp^2} = 0\,,\\
    \Longrightarrow \quad \tilde{x}^{0*} = -\frac{-(x^0)^2 + (x^1)^2 + (x^\perp)^2}{2(x^0+x^1)} = -\frac{x_\mu x^\mu}{2\sqrt{2}Fx^+}\,,
\end{gather}
and tells us ``when the source kicks in.'' i.e. $\phi$ should only be supported on the region
\begin{gather}
    x^0 - x^{0*} = x^0 + \frac{-(x^0)^2+(x^1)^2+(x^\perp)^2}{2(x^0 + x^1)} > 0\,,
\end{gather}
because only these points are contained in the future null cone of the source. In lightcone coordinates, this region is equivalent to:
\begin{equation}
    \frac{F}{\sqrt{2}} x^+ + \frac{x_\perp^2}{2\sqrt{2}F x^+} > 0 \quad \Longrightarrow \quad x^+ > 0\,.
\end{equation}
Thus we will ensure that our final solution is multiplied by the appropriate Heaviside $\Theta$-function to match this support.

For real numbers $x, y \in \bbR$, we note that $\Theta(xy^2) \neq \Theta(x)$. The two expressions are equal everywhere except along the curve $y = 0$. Consequently, we define $x^+$ with scare quotes to be the quantity:
\begin{equation}
    \scarex := \frac{F}{\sqrt{2}} x^+ + \frac{x_\perp^2}{2\sqrt{2}F x^+} = x^0 -x^{0*}
\end{equation}
which is positive when $x^+ > 0$ but with different zeroes. As we will verify below, it is important not to ``simplify'' expressions involving $\scarex$ when dealing with/expecting distributional quantities. We have dedicated the separate Appendix \ref{app:xpIdentities} entirely to discussion of $\scarex$ and manipulations/identities involving it.

Using the above, and specifically $x^0 + x^1 > 0$, we can complete the final integration:
\begin{equation}
    \phi(x^0,x^1, x^\perp) 
        = -\frac{\sqrt{2} h F}{4\pi} \Theta(x^0 - x^{0*}) \int d\tilde{x}^0\, \left(\frac{(x^0+x^1)^2+x_\perp^2}{2(x^0+x^1)^2}\right)\frac{\delta(\tilde{x}^0 - \tilde{x}^{0*})}{\sqrt{(x^1 + \tilde{x}^0)^2+x_\perp^2 + \delta^2}}\,,
\end{equation}
which gives
\begin{align}
    \phi(x^0,x^1, x^\perp) 
        &= -\frac{\sqrt{2} h F}{4\pi}\, \frac{1}{x^0+x^1}  \frac{\Theta(x^0 - x^{0*})}{\sqrt{1 + \frac{\delta^2}{\left(\scarex\right)^2}}}\,,\\
    \phi(x^+,x^-, x^\perp) 
        &=  -\frac{h}{4\pi x^+} \,  \Theta\left(\scarex\right)\,.
\end{align}
We can check that the final answer satisfies the appropriate Laplace equation:
\begin{equation}\label{eq:LaplaceCheck}
    \partial^2 \phi(x^+,x^-,x^\perp) 
        = h \delta(x^+) \delta^2(x^\perp) = J(x^+,x^\perp)\,.
\end{equation}
This is justified in detail in Appendix \ref{app:correctTheta}.

\subsection{\texorpdfstring{$\scarex$}{``x+''} and Its Discontents}\label{app:xpIdentities}
In the main text, we make repeated use of the object $\scarex$, defined as
\begin{equation}
    \scarex := \frac{F}{\sqrt{2}} x^+ + \frac{x_\perp^2}{2\sqrt{2}F x^+}\,.
\end{equation}
It appears naturally in Section \ref{sec:ExampleFreeScalar} and Appendix \ref{app:freeScalarCalcs} in studying the support of the delta-function in the retarded propagator, and is crucial for ensuring that field profiles actually satisfy the equations of motion, as in \eqref{eq:LaplaceCheck}. Consequently, it appears in the entire rest of the text as the solution to the Laplacian with null source. It is helpful to understand precisely what $\scarex$ is, and identities involving it, as the quantity is rather natural.

First, it is easy to see that $\scarex > 0$ when $x^+ > 0$. The two only differ by their behaviour on the lightplane $x^+ = 0$. As we will see, $\scarex$ arises in the retarded propagator to delete the points with $x^+ = 0$ but $x_\perp^2 \neq 0$.

Consider a light signal emitted from a point $y$, the light signal spreads out and passes through all points $x = (x^+,x^-,x^\perp)$ on the forward null cone of $y$, i.e. those points satisfying
\begin{equation}\label{eq:nullCone}
    0 = (x-y)^2 = -2F^2(x^+-y^+)(x^- -y^-)+(x_\perp-y_\perp)^2\,,\quad
    x^0 - y^0 > 0\,.
\end{equation}
A point on a null line defect necessarily has $y=(0,y^-,0)$, and so, when $x^+ \neq 0$, a point on its forward null-cone satisfies $y^- = x^- - x_\perp^2/(2F^2x^+)$ and $x^+ + x^- > y^-$. Substituting the first expression into the second, the inequality is $y^-$ independent and becomes
\begin{equation}
    \scarex > 0\,.
\end{equation}
Since $y^-$ can be arbitrary, any point satisfying just $\scarex > 0$ is in the future null cone of \textit{some} point on the defect. Now we consider what happens when $x^+ = 0$, then points in the future null cone of $y$ with $x^+ = 0$ necessarily also have $x_\perp^2 = 0$. This leaves just the line of points $x^-$ with $x^- > y^-$, i.e. points ``further along the defect.'' Together, we see that the future null cone excludes the $x^+ = 0$ regions with $x_\perp^2 \neq 0$, as promised.

\begin{figure*}
    \centering
    \newcommand{\eps}{1}      
    \newcommand{\xmin}{-2}    
    \newcommand{\xmax}{2}     
    \newcommand{\ymax}{3}     
    %
    \begin{minipage}{0.32\textwidth}
        \centering
        \begin{tikzpicture}[scale=1]
            \fill[gray!30] (\xmin,\eps) rectangle (\xmax,\ymax);
            \draw[->, thick] (\xmin,0) -- (\xmax,0) node[below right] {$x^\perp$};
            \draw[->, thick] (0,0) -- (0,\ymax) node[above] {$x^+$};
            \draw[thick, dashed] (\xmin,\eps) -- (\xmax,\eps) node[right] {$\epsilon$};
        \end{tikzpicture}
        \subcaption{$\Theta(x^+ - \epsilon)$}\label{fig:xplus-a}
    \end{minipage}
    \hfill
    \begin{minipage}{0.32\textwidth}
        \centering
        \begin{tikzpicture}[scale=1]
            \coordinate (C) at (0,\eps);
            \def\radius{\eps}
            \fill[gray!30, even odd rule]
                (\xmin,0) rectangle (\xmax,\ymax)
                (C) circle (\radius);
            \draw[thick, dashed] (C) circle (\radius);
            \draw[->, thick] (\xmin,0) -- (\xmax,0) node[below right] {$x^\perp$};
            \draw[->, thick] (0,0) -- (0,\ymax) node[above] {$x^+$};
            \fill (C) circle (1pt) node[right] {$\!(0,\tfrac{\epsilon}{2})$};
        \end{tikzpicture}
        \subcaption{$\Theta(\scarex - \epsilon)$}\label{fig:xplus-b}
    \end{minipage}
    \hfill
    \begin{minipage}{0.32\textwidth}
        \centering
        \begin{tikzpicture}[scale=1]
            \pgfmathsetmacro{\yleft}{\eps * abs(\xmin)+0.3}
            \pgfmathsetmacro{\yright}{\eps * abs(\xmax)+0.3}
            %
            \fill[gray!30]
                (0,0.3) --
                (\xmin,\yleft) --
                (\xmin,\ymax) --
                (\xmax,\ymax) --
                (\xmax,\yright) -- cycle;
            %
            \draw[->, thick] (\xmin,0) -- (\xmax,0) node[below right] {$x^\perp$};
            \draw[->, thick] (0,0.0) -- (0,\ymax) node[above] {$x^+$};
            %
            \draw[thick,dashed,domain=\xmin:0] plot ({\x},{0.3+\eps*abs(\x)});
    \draw[thick,dashed,domain=0:\xmax] plot ({\x},{0.3+\eps*abs(\x)});
        \end{tikzpicture}
        \subcaption{$\Theta(x^+ - \epsilon\,|x^\perp|)$}\label{fig:xplus-c}
    \end{minipage}
    \caption{Supports of three $\Theta$-functions. All three are supported in the upper half space, with different behaviour for the boundaries as $\epsilon \to 0$. Figures \ref{fig:xplus-b} and \ref{fig:xplus-c} give the same limit $\Theta(\scarex)$ as $\epsilon \to 0$, and appear in studying null defects.}
    \label{fig:xplusfigures}
\end{figure*}
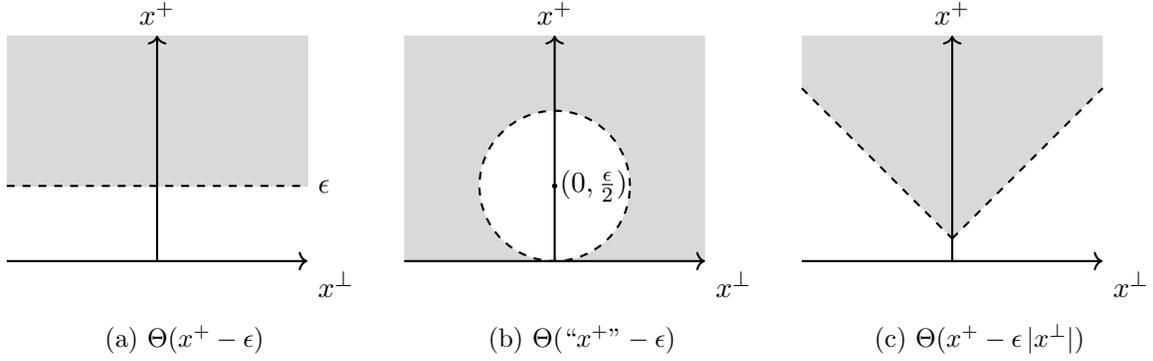

The deletion of the $x_\perp^2 \neq 0$ parts of the shockwave plane can be realized different ways in regulators. As a warmup, consider the support of $\Theta(x^+-\epsilon)$, this is the upper-half plane $x^+ > \epsilon$, see Figure \ref{fig:xplus-a}. In the limit as $\epsilon \to 0$, the discontinuity of the step function approaches $x^+=0$ at the same rate in $\epsilon$. 

Now consider $\Theta(\scarex - \epsilon)$, for any $\epsilon$ it is supported on the entire upper half space $x^+>0$ outside the cylinder $x_\perp^2 + F^2 (\sqrt{2} x^+ - \epsilon)^2 = \epsilon^2$, see Figure \ref{fig:xplus-b}. In the limit as $\epsilon \to 0$, the boundary of the $\Theta$ function that actually changes is just the cylinder shrinking towards the origin in the $(x^\perp x^+)$-plane, leading to different singular behaviour.

Instead of $\Theta(\scarex - \epsilon)$, we can consider
\begin{equation}\label{eq:spacelikeWedge}
    \Theta\left(\tfrac{x^0 + u x^1}{\sqrt{1-u^2}}-|x^\perp|\right)\,,
\end{equation}
which appears in studying e.g. boosts of spacelike sources, see Section \ref{sec:limitsOfDefects}. Substituting $u = 1-\epsilon$, then for any fixed $x^-$, the support is a wedge in the $(x^\perp x^+)$-plane, which sharply touches the points $(x^+=0,x^\perp = 0)$ as $\epsilon \to 0$. This is depicted in Figure \ref{fig:xplus-c}.

In the main text, our regulator for $\Theta(\scarex)$ is chosen to be:
\begin{gather}
    \Theta(\scarex) := \lim_{\delta \to 0}\frac{\Theta(x^+)}{\sqrt{1 + \delta^2/(\scarex)^2}}\,.
\end{gather}
It can be considered as a variation of $\Theta(\scarex - \epsilon)$ presented above, with correct singular behaviour in the limit as $\delta \to 0$. For any finite $\delta$, it is supported on the \textit{entire} $x^+ > 0$ plane, but the function rapidly decays inside the cylinder in Figure \ref{fig:xplus-b}. In Appendix \ref{app:correctTheta} we show that this regulator makes the free scalar satisfy the equations of motion.

In Appendix \ref{app:wedgeApprox} we show how the spacelike/wedge-regularized $\Theta$-function in \eqref{eq:spacelikeWedge} relates to $\Theta(\scarex)$. Miraculously, this leads to an extra factor of $\frac{1}{2}$ in $\delta$-function identities which are relevant in matching timelike and spacelike ultrarelativistic limits.

\subsubsection{Scare Theta Identities}\label{app:correctTheta}
In order to confirm that our answer for the free-scalar makes sense, we should check if it satisfies the equations of motion. In order to do that, we must introduce a regularization for $\Theta(\scarex)$. In practice, we define
\begin{gather}
    \Theta(\scarex) := \lim_{\delta \to 0}\frac{\Theta(x^+)}{\sqrt{1 + \delta^2/(\scarex)^2}}\,.
\end{gather}

Let us prove that the scalar solution solves the equations of motion by using this regulated expression. Consider
\begin{equation}
    \partial^2 \left(\frac{-h}{4\pi x^+} \Theta(\scarex)\right)
        = \lim_{\delta \to 0} \frac{16 h F^2}{\pi} \Theta(x^+) \frac{x^+ \delta^2 (x_\perp^4 + F^2(x_\perp^2 - 4 \delta^2)(x^+)^2 - 2F^4 (x^+)^4)}{(x_\perp^4+(4F^2x_\perp^2 + 8 F^2\delta^2)(x^+)^2 +4F^4(x^+)^4))^{5/2}}\,.
\end{equation}
We claim that this is a delta-function. To see this, consider the integral against a test function $f(x^+, x^2, x^3)$; switch to polar coordinates in the transverse directions, and rescale $r \mapsto \delta r$ and $x^+ \mapsto \tfrac{\delta}{\sqrt{2}F} x^+$, then:
\begin{align}
    \int &dx^+ dx^2 dx^3\, \partial^2 \left(\frac{-h}{4\pi x^+} \Theta(x^+)\right)\, f(x^+, x^2, x^3)\nonumber\\
    &= \frac{4 h}{\pi}\lim_{\delta \to 0} \int dx^+ dr d\theta\, \Theta(x^+) \frac{x^+ r(2r^4+(r^2-4)(x^+)^2-(x^+)^4)}{(r^4+2(2+r^2)(x^+)^2+(x^+)^4)^{5/2}} f\left(\tfrac{\delta x^+}{\sqrt{2}F},\delta r \cos\theta,\delta r\sin\theta\right)\,.
\end{align}
In the limit as $\delta \to 0$, the result localizes entirely on $f(0,0,0)$, which confirms that it is proportional to a $\delta$-function. Now we just need to find the normalization; as it turns out:
\begin{equation}
    8h f(0,0,0) \int_0^{\infty} dx^+ \int_0^{\infty} dr\, \frac{x^+ r(2r^4+(r^2-4)(x^+)^2-(x^+)^4)}{(r^4+2(2+r^2)(x^+)^2+(x^+)^4)^{5/2}} = h f(0,0,0)\,.
\end{equation}
Thus we have the promised result:
\begin{equation}
    \partial^2 \left(\frac{-1}{4\pi x^+}\Theta(\scarex)\right) = h \delta(x^+)\delta^2(x^\perp)\,.
\end{equation}


\subsubsection{\texorpdfstring{$\Theta(x^+ - \epsilon\,|x^\perp|)$}{Theta(x+)-eps|xperp|} Limits to \texorpdfstring{$\Theta(\scarex)/2$}{Theta(``x+'')/2}}\label{app:wedgeApprox}
In this section we will prove that the wedge-regularized $\Theta$-function, arising from boosted limits of spacelike source functions, limits to $\Theta(\scarex)/2$. The wedge-regularized $\Theta$-function is
\begin{equation}
    \Theta\left(\tfrac{x^0 + u x^1}{\sqrt{1-u^2}}-|x^\perp|\right) 
        = \Theta\left(\tfrac{x^0 + (1-\epsilon)x^1}{\sqrt{\epsilon(2-\epsilon)}} - |x^\perp|\right)\,,
\end{equation}
where $u = 1-\epsilon$. More specifically, we will actually prove that
\begin{equation}
    \lim_{\epsilon \to 0} \partial^2 \frac{\Theta(x^+ - \epsilon |x^\perp|)}{4\pi x^+} = -\frac{1}{2}\delta(x^+) \delta(x^1) \delta(x^2)\,.
\end{equation}

To this end, we first note that $x^-$ contributes to the argument of the $\Theta$ function at subleading orders in $\epsilon$
\begin{equation}
    \frac{x^0 + (1-\epsilon)x^1}{\sqrt{\epsilon(2-\epsilon)}} - |x^\perp| = \frac{F}{\sqrt{\epsilon}} x^+ - |x^\perp| + \frac{F}{4}(2x^- - x^+)\sqrt{\epsilon} + \dots\,,
\end{equation}
so we will ignore $x^-$ and only consider the first two terms. So now we study
\begin{equation}
    \lim_{\epsilon \to 0} \partial^2 \frac{1}{4\pi x^+}\Theta\left(\tfrac{F}{\sqrt{\epsilon}}x^+ - |x^\perp|\right)
    = \lim_{\epsilon \to 0} \frac{1}{4\pi x^+} \nabla^2 \Theta\left(\tfrac{F}{\sqrt{\epsilon}}x^+ - |x^\perp|\right)
    \,,
\end{equation}
where $\nabla^2$ is the Laplacian in the $(x^2,x^3)$ coordinates.

We can integrate the kernel above against a test function $f(x^+,x^2,x^3)$ and switch to polar coordinates in the transverse directions:
\begin{equation}
    \int \frac{dx^+}{4\pi x^+} \int dr d\theta\, \partial_r \left(r \partial_r \Theta\left(\tfrac{F}{\sqrt{\epsilon}}x^+ - r\right)\right) f(x^+,r\cos\theta,r\sin\theta)\,.
\end{equation}
As done many times before, we can rescale $x^+ \mapsto \frac{\sqrt{\epsilon}}{F} x^+$. We also integrate by parts in $r$ and take $\partial_r\Theta$ to get:
\begin{align}
    \int \frac{dx^+}{4\pi x^+} &\int dr d\theta\, r \delta(x^+ - r) \partial_r f(\tfrac{\sqrt{\epsilon}}{F}x^+,r\cos\theta,r\sin\theta)\\
        &= \frac{1}{4\pi} \int dr d\theta \, \partial_r f(\tfrac{\sqrt{\epsilon}}{F}r,r\cos\theta,r\sin\theta)\,.
\end{align}
We return to the limit as $\epsilon \to 0$, in this case the argument in $x^+$ is set to $0$. Since we have a total derivative in $r$, the integral only depends on the value of $f$ at infinity and at $0$. However, test functions go to $0$ at infinity, and thus the result only depends on $r$ at $0$. Consequently, all $\theta$ dependence also disappears, giving us a factor of $2\pi$, and thus
\begin{equation}
    \lim_{\epsilon \to 0} \partial^2 \frac{\Theta(x^+ - \epsilon |x^\perp|)}{4\pi x^+} = -\frac{1}{2}\delta(x^+) \delta(x^2) \delta(x^3)\,.
\end{equation}

\subsection{Ultra-Relativistic Limits}\label{app:URLimits}
In this appendix we briefly consider some useful ultra-relativistic limits that appear from limits of timelike or spacelike solutions. We first define the following quantities that appear in problems involving timelike ($R_u$) and spacelike ($T_u$) defects:
\begin{align}
    R_u^2 
        &:= (ux^0 + x^1)^2 + (1-u^2) x_\perp^2\,,\\
    T_u^2
        &:= (x^0 + u x^1)^2 - (1-u^2) x_\perp^2\,.\label{eq:Tu2}
\end{align}

Relevant to timelike limits of field strengths and potentials, it can be shown that \cite{Jackiw:1991ck, Aichelburg:1970dh}:
\begin{gather}
\lim_{u\to 1} \frac{1-u^2}{R_u^3} 
        = \frac{2}{x_\perp^2}\delta(x^0+x^1)
        = \frac{\sqrt{2}}{F} \frac{\delta(x^+)}{x_\perp^2}
        \,,\label{eq:FirstURIdentity}\\
    \lim_{u\to 1} \frac{1}{R_u} 
        = -\log{(\mu^2 x_\perp^2)}\delta(x^0 + x^1) + \frac{1}{|x^0 + x^1|}
        = -\sqrt{2}F\log{(\mu^2 x_\perp^2)}\delta(x^+) + \frac{\sqrt{2}F}{|x^+|}
        \,.\label{eq:SecondURIdentity}
\end{gather}
The first identity is proven as follows: replace $u = 1-\epsilon$ and integrate the quantity against a test-function in $x^+$, then rescale $x^+$ by $\sqrt{\epsilon}$, i.e.
\begin{align}
    \lim_{u\to 1} \int dx^+ \, \frac{1-u^2}{R_u^3} f(x^+) 
        &= \lim_{\epsilon \to 0} \int dx^+ \frac{2\epsilon^{3/2} - \epsilon^{5/2}}{(\epsilon (2x_\perp^2 + 2 F^2 (x^+)^2) + O(\epsilon^{>1}))^{3/2}} f(\sqrt{\epsilon}x^+)\\
        &= f(0) \int dx^+\, \frac{2}{(2x_\perp^2+2F^2 (x^+)^2)^{3/2}}\\
        &= f(0) \frac{\sqrt{2}}{F x_\perp^2}\,.
\end{align}
And thus the identity follows. The second identity can be obtained by noting that the regulated quantity $R_u$ satisfies
\begin{equation}
    \frac{\partial}{\partial|x_\perp|^2}\frac{1}{R_u} = - \frac{1-u^2}{2R_u^3}\,.
\end{equation}
Then \eqref{eq:SecondURIdentity} follows from \eqref{eq:FirstURIdentity} if the limit as $u\to 1$ and derivative in the transverse directions commute.

Identical results can be proven for the ultrarelativistic limit coming from a spacelike source, with some subtleties. Specifically, $T_u$ has a relative sign between the two terms in \eqref{eq:Tu2} and, correspondingly, should always come dressed with an appropriate theta function to avoid divergences (see also Appendix \ref{app:wedgeApprox}). To derive a useful ultrarelativistic limit of $1/T_u$, we first consider:
\begin{equation}
    \frac{\partial}{\partial |x_\perp|^2} \left[\frac{1}{T_u}\Theta\left(\tfrac{x^0 + u x^1}{\sqrt{1-u^2}}-|x^\perp|\right) \right]
        = \frac{1-u^2}{2T_u^{3}}\Theta\left(\tfrac{x^0 + u x^1}{\sqrt{1-u^2}}-|x^\perp|\right)- \frac{1}{2|x_\perp|T_u}\Theta'\left(\tfrac{x^0 + u x^1}{\sqrt{1-u^2}}-|x^\perp|\right)\,.
\end{equation}
As explained in Appendix \ref{app:wedgeApprox}, we can replace the $\Theta$-function by the simpler $\Theta(\tfrac{F}{\sqrt{\epsilon}}x^+ - |x^\perp|)$ (note: this is not necessary). Now we proceed as in the previous derivation: we take $u=1-\epsilon$, integrate against a test-function in $x^+$, and rescale $x^+$ by $\tfrac{\sqrt{\epsilon}}{F}x^+$, to get:
\begin{equation}
    \frac{1}{2 \sqrt{2}F} \int dx^+ \left(\frac{\Theta(x^+-|x_\perp|)}{((x^+)^2-x_\perp^2)^{3/2}}-\frac{\Theta'(x^+-|x_\perp|)}{((x^+)^2-x_\perp^2)^{1/2}}\right)f(\tfrac{\sqrt{\epsilon}}{F}x^+)\,.
\end{equation}
In the limit as $\epsilon \to 0$, we have $\delta(x^+)$ -- so we must just finish the integral to obtain the normalization. While each of the terms is divergent on their own, together the divergences cancel. A straightforward integration by parts gives the final answer
\begin{equation}
    \frac{\partial}{\partial |x_\perp|^2} \left[\frac{1}{T_u}\Theta\left(\tfrac{x^0 + u x^1}{\sqrt{1-u^2}}-|x^\perp|\right) \right] = -\frac{1}{2\sqrt{2}F x_\perp^2} \delta(x^+)\,.
\end{equation}
We can now try and compute
\begin{equation}
    \lim_{u \to 1} \frac{1}{T_u}\Theta\left(\tfrac{x^0 + u x^1}{\sqrt{1-u^2}}-|x^\perp|\right)\,.
\end{equation}
As before, the ``naive limit'' gives the term $\tfrac{1}{\sqrt{2}F|x^+|}\tfrac{\Theta(\scarex)}{2}$ while the shockwave term escapes the naive limit. However, we can insert an $\alpha \log(\mu^2 x_\perp^2) \delta(x^+)$ term by hand and tune $\alpha$ by assuming the derivative commutes with the limit. The final answer is
\begin{equation}\label{eq:ThirdURIdentity}
    \lim_{u \to 1} \frac{1}{T_u}\Theta\left(\tfrac{x^0 + u x^1}{\sqrt{1-u^2}}-|x^\perp|\right)
    = -\frac{1}{2\sqrt{2}F}\log(\mu^2x_\perp^2)\delta(x^+)+\frac{1}{\sqrt{2}F|x^+|}\frac{\Theta(\scarex)}{2}\,.
\end{equation}

\subsection{Details for Ward Identities and Tariff Functions}
This appendix details the calculations in Section \ref{sec:WITariffs}.

\subsubsection{One Point Function of a Scalar}\label{app:scalarCalc}
We can consider the form of a general scalar one-point function in the presence of a null defect by explicitly solving the Ward identities. The form of the solution, crucially, depends on what subalgebra of the conformal group survives in the presence of our defect. Here we consider four cases, we consider the one-point function of a scalar primary operator of scaling dimension $\delta$ when: the defect is a pinning field defect of dimension $\Delta = 1$, $\Delta = 2$, and generic $\Delta$, as in Section \ref{sec:ScalarPrimary}; and the case that the defect preserves the maximal symmetry algebra $\mathfrak{n}_d$.

We consider an ansatz for the one-point function of a scalar primary of dimension $\delta$,
\begin{equation}
    \expval*{\calO(x^+,x^-, x^\perp)}_{\rm{D}} = f(\delta; x^+,x^-, x^\perp)\,.
\end{equation}
In all cases, the defect preserves $P_-$ and $M_{ij}$, which implies that the $f$ does not depend on $x^-$ and only depends on $x^\perp$ through $|x^\perp|$. As mentioned in Section \ref{sec:ScalarPrimary}, a pinning defect made from the scalar primary with dimension $\Delta$ always preserves the chiral scaling
\begin{equation}
    J' = D - \frac{(\Delta-1)}{F^2} M_{+-}\,.
\end{equation}
Solving the $J'$ Ward identities by straightforward manipulations gives a solution of the form
\begin{equation}\label{eq:regularSoln}
    f(\delta;x^+,|x^\perp|) = \frac{1}{|x_\perp|^\delta} \tilde{f}\left(\frac{x^+}{|x_\perp|^{2-\Delta}}\right)\,,
\end{equation}
for some unknown function $\tilde{f}$. With an eye towards some of our particular examples, we note that there can also be distributional solutions, e.g. of the form
\begin{equation}
    f_{\mathrm{s}}(\delta;x^+,|x^\perp|) = \delta^{(n)}(x^+)h_n^{\Delta}(\delta; |x_\perp|)\,,
\end{equation}
where $h_n$ is a homogeneous function of scaling weight $(2-\Delta)(n+1)-\delta$. Normally we would ignore such $\delta$-function correlators, but we have already seen that shockwave solutions are common, so will consider $\delta(x^+)$-dependent solutions (we ignore derivatives of shockwaves for simplicity).

For any scalar pinning field defect, we still expect $K_i$, $K_-$, and $M_{-i}$ symmetry to be preserved, i.e. the Heisenberg subalgebra. We start with $K_-$, for a non-trivial one-point function, the non-shockwave solution must satisfy:
\begin{equation}
    (\Delta-1)(x^+)^2 \tilde{f}'\left(\frac{x^+}{|x_\perp|^{2-\Delta}}\right) = 0\,.
\end{equation}
When $\Delta = 1$, this is automatically satisfied, and when $\Delta \neq 1$, it already restricts the form of the one-point function to
\begin{equation}
    f_{\Delta \neq 1}(\delta;|x_\perp|) = \frac{c}{|x_\perp|^\delta}\,,
\end{equation}
for some constant $c$. For the shockwave solution, the $K_-$ Ward identity immediately vanishes from terms proportional to $x^+ \delta(x^+)$.

Now we turn to the $K_i$ and $M_{-i}$ constraints.  When $\Delta = 1$, the $K_i$ Ward identities force the non-distributional solution to 
\begin{equation}
    f_{\Delta = 1}(\delta;x^+) = \frac{c}{|x^+|^{\delta}}\,,
\end{equation}
i.e. the dependence on the transverse directions vanishes entirely. When $\Delta \neq 1$, they make the non-shockwave one-point function vanish identically -- only shockwave solutions are allowed. For all $\Delta$, the $K_i$ Ward identities force the shockwave solutions to have $\delta = \Delta$
\begin{equation}\label{eq:abstractShockwave}
    f_{\mathrm{s}}(\Delta;x^+,|x^\perp|) = \delta(x^+)h(\Delta; |x_\perp|)\,.
\end{equation}
The solutions automatically satisfy the $M_{-i}$ Ward identities. This is clear by inspection, but also because $P_-$ and $K_i$ invariance imply $M_{-i}$ invariance by \eqref{eq:outerAutomorphism}. In the $\Delta \neq 2$ case, we have exhausted all of the Ward identities, and can admit potentially interesting solutions such as
\begin{equation}
    f_{\mathrm{s}}(\Delta;x^+,|x^\perp|) = \delta(x^+)\frac{c}{|x_\perp|^{2\Delta-2}}\,, \quad \Delta \neq 2\,,
\end{equation}
which is the obvious generalization of the 4d scalar expression \eqref{eq:scalarShockwave}. When $\Delta = 2$, \eqref{eq:abstractShockwave} should also preserve $K_+ \sim J_1$, and the only solution is a pure contact term localized on the defect itself.

If we can push cases further to maximal symmetry then, when $\delta=1$, adding either of $M_{+-}$ or $K_+$ makes one-point functions of non-identity scalar operators trivial.


\bibliographystyle{JHEP}
\bibliography{mono.bib} 

\end{document}